\documentclass[a4paper]{article}
\usepackage{amsmath}
\usepackage{amsfonts}
\usepackage{amssymb}
\usepackage{wasysym}
\usepackage{amsthm}
\usepackage[most]{tcolorbox}
\usepackage{xcolor}
\usepackage[pdftex]{pict2e}

\tcbset{boxrule=0.6pt,colback=white!97!red}

\usepackage{array}

\usepackage{tikz}
\usepackage{tikz-cd}
\usetikzlibrary{arrows}
\usetikzlibrary{arrows.meta}
\usetikzlibrary{positioning}
\usetikzlibrary{shapes,snakes}
\usetikzlibrary{fit}
\usetikzlibrary{decorations.pathmorphing,decorations.pathreplacing,decorations.markings}
\usetikzlibrary{calc}

\usepackage{authblk}

\def\myblue{white!40!blue}
\def\mygreen{black!30!green}
\def\myred{black!30!red}

\usepackage{longtable}
\usepackage{empheq}
\usepackage{ytableau}
\usepackage{url}
\usepackage{dsfont}

\usepackage[utf8]{inputenc}

\def\resp{\begin{tikzpicture}[scale=0.35]
		\draw[-stealth] (-0.5,-0.5) to[out=45,in=270] (-0.15,0) to[out=90,in=315] (-0.5,0.5);
		\draw[-stealth] (0.5,-0.5) to[out=135,in=270] (0.15,0) to[out=90,in=225] (0.5,0.5);
	    \end{tikzpicture}}
    
\def\respexp{\begin{tikzpicture}[scale=0.35]
		\draw[-stealth] (-0.5,-0.5) to[out=45,in=270] (-0.2,0) to[out=90,in=315] (-0.5,0.5);
		\draw[-stealth] (0.5,-0.5) to[out=135,in=270] (0.2,0) to[out=90,in=225] (0.5,0.5);
		\draw[fill=\mygreen] (-0.2,-0.07) circle (0.1) (0.2,-0.07) circle (0.1);
\end{tikzpicture}}

\def\resm{\begin{tikzpicture}[scale=0.35]
		\draw[-stealth] (-0.5,-0.5) -- (0.5,0.5);
		\draw[-stealth] (0.5,-0.5) -- (-0.5,0.5);
		\draw[fill=burgundy] (0,0) circle (0.1);
	    \end{tikzpicture}}

\def\smallresp{)(}

\def\smallresm{\begin{tikzpicture}[scale=0.2]
		\draw (-0.5,-0.5) -- (0.5,0.5);
		\draw (0.5,-0.5) -- (-0.5,0.5);
		\draw[fill=burgundy] (0,0) circle (0.15);
	    \end{tikzpicture}}

\def\be{\begin{eqnarray}}
\def\ee{\end{eqnarray}}
\def\nn{\nonumber}

\def\ba{\begin{equation}\begin{aligned}}
\def\ea{\end{aligned}\end{equation}}

\def\vth{\vartheta}

\def\MF{D}

\def\MOY{\Gamma}

\def\dag{\dagger}
\def\IQ{\mathbb{Q}}
\def\lm{\limits}
\def\p{\partial}
\def\Lra{\Longrightarrow}

\def\KhR{{\rm KhR}}
\def\Ker{{\rm Ker}}
\def\Im{{\rm Im}}

\def\CM {{\cal M}}
\def\CL {{\cal L}}
\def\CR {{\cal R}}
\def\fD{\mathfrak{D}}

\usepackage[hidelinks,colorlinks=true,unicode]{hyperref}
\hypersetup{
	linkcolor=blue,
	citecolor=blue,
	filecolor=magenta,
	urlcolor=blue,
}
\usepackage[left=2cm,right=2cm,top=2cm,bottom=2cm,bindingoffset=0cm]{geometry}

\definecolor{burgundy}{rgb}{0.5, 0.0, 0.13}
\def\myblue{white!40!blue}

\numberwithin{equation}{section}

\usepackage[natbib=true, backend = bibtex, style=numeric-comp, sorting=none,maxbibnames=99]{biblatex}
\date{}

\begin{document}
	
\title{\bf Reductions in Khovanov--Rozansky operator formalism}

\author[1,2,3]{{\bf D. Galakhov}\thanks{\href{mailto:galakhov@itep.ru}{galakhov@itep.ru}}}
\author[1,2,3]{{\bf E. Lanina}\thanks{\href{mailto:lanina.en@phystech.edu}{lanina.en@phystech.edu}}}
\author[1,2,3]{{\bf A. Morozov}\thanks{\href{mailto:morozov@itep.ru}{morozov@itep.ru}}}

\vspace{5cm}

\affil[1]{Moscow Institute of Physics and Technology, 141700, Dolgoprudny, Russia}

\affil[2]{Institute for Information Transmission Problems, 127051, Moscow, Russia}
\affil[3]{NRC "Kurchatov Institute", 123182, Moscow, Russia\footnote{former Institute for Theoretical and Experimental Physics, 117218, Moscow, Russia}}

\renewcommand\Affilfont{\itshape\small}

\maketitle

\vspace{-7cm}

\begin{center}
	\hfill MIPT/TH-11/26\\
	\hfill ITEP/TH-11/26\\
	\hfill IITP/TH-11/26
\end{center}

\vspace{4.5cm}

\begin{abstract}
	{
	Sophisticated Khovanov--Rozansky (KhR) description of knot invariants in the fundamental representation
	can be reformulated in terms of bicomplex with a simple physical meaning.
	Namely, the counterintuitive matrix factorization is substituted by simple operators $D$,
	locally constructed for every MOY resolution of a link diagram,
	which becomes nilpotent when the diagram has no external lines.
	Operators for different resolutions are related by equally simple conjugations $\chi^{(\pm)}$.
	The  KhR procedure then splits in two steps --- defining ``vertical''  cohomologies of $D$,
	which are associated with particular resolutions and will be put at vertices of the {\it hypercube},
	and conjugations $\chi^{(\pm)}$, that define morphisms along its edges.
	As usual, standard combinations of morphisms are nilpotent, and one can define ``horizontal'' cohomologies ---
	which are then combined into Poincar\'e polynomial, called KhR polynomial in application to links.
	This construction remains {\it global} in the sense that resulting cohomologies depend on the entire
	link diagram, but all its building blocks, including the operators and morphisms
	are {\it local} in the sense that they are defined for its particular  vertices.
	Sometimes, this allows simple {\it local reductions}, allowing to eliminate or change particular 
	vertices or sets of those.
	Along with the obvious case of Reidemeister equivalencies this happens also for 
	antiparallel-lock tangles, what is responsible for simplification of bipartite calculus. In the $N=2$ and arbitrary $N$ bipartite cases, one can also provide {\it global reductions} transferring the local construction of the KhR double-complex to the global construction of the Khovanov(-like) single-complex.
}
\end{abstract}

%

\tableofcontents



\section{Introduction}

In this paper, we explain how to simplify the Khovanov--Rozansky operator formalism~\cite{2508.05191}. In particular, we show how the bipartite Khovanov--Rozansky calculus~\cite{2506.08721} arise from this operator formalism. 

The Khovanov--Rozansky polynomials are believed to be observables in the refined Chern--Simons theory. It was shown~\cite{aganagic2015knot} that the partition function of the refined Chern--Simons theory on a 3d sphere equals to the partition function of the refined topological string. The Khovanov--Rozasky homology theory has lots of other implications in theoretical and mathematical physics. In particular, the Khovanov--Rozansky polynomials can be formulated in terms of integers which capture the spectrum of BPS states in the string Hilbert space~\cite{gukov2005khovanov}. They also determine the topological amplitudes of strings. Moreover, there is an approach to Khovanov--Rozansky cohomology of links based on counting the solutions of specific elliptic partial differential equations arising in 4d and 5d gauge theories~\cite{witten2011fivebranes,witten2012khovanov,gaiotto2012knot}.     

Let us describe first the construction of the Jones polynomial~\cite{jones2005jones,jones1987hecke,jones1985polynomial} being Wilson loop in the 3d Chern--Simons theory~\cite{Chern-Simons} with the $SU(2)$ gauge group with gauge fields taken in the fundamental representation~\cite{Witten}. In the Wilson loop, we take the integration contour being tied in an arbitrary knot $\cal K$ and also consider its generalization to a link $\cal L$. Due to the topological nature of the theory, the Jones polynomial turns out to be a link invariant called quantum link invariant having non-perturbative answer obtainable because of the hidden structure of the quantum group $U_q(\mathfrak{sl}_2)$. The corresponding method of computation is called the Reshetikhin--Turaev (RT) approach~\cite{Reshetikhin,guadagnini1990chern2,reshetikhin1990ribbon,turaev1990yang,reshetikhin1991invariants} in which each crossing is associated to a quantum $\cal R$-matrix. In the Jones case, this $\cal R$-matrix and its inverse (in the fundamental representation) can be chosen in a simple way~\cite{Kauff,dolotin2013introduction}:
\ba \label{Krule} 
{\cal R}^{ij}_{kl} \ &= \  \epsilon^{ij}\epsilon_{kl} -q\cdot \delta^i_k \delta^j_l\,, \\
({\cal R}^{-1})^{ij}_{kl} \ &= \  \epsilon^{ij}\epsilon_{kl} -q^{-1}\cdot \delta^i_k \delta^j_l\,,
\ea
and depicted as shown in Fig.\,\ref{fig:Kauff}. Here, we also add the value of the Jones polynomial for the unknot. Thus, the computation of the Jones polynomial reduces just to combinatorial cycle calculus, and instead of {\it local} $\cal R$-matrix technique, we arrive just to {\it global} planar approach. 
\begin{figure}[h!]
	\setlength{\unitlength}{0.9pt}
	
	\begin{picture}(100,200)(-200,-100)
		{\footnotesize
			\put(0,70){
				
				\put(-60,-20){\line(1,1){40}}
				\put(-20,-20){\line(-1,1){18}}
				\put(-60,20){\line(1,-1){18}}

				\put(10,-2){\mbox{$=$}}
				
				\qbezier(30,20)(50,0)(70,20)
				\qbezier(30,-20)(50,0)(70,-20)
				
				\put(85,-2){\mbox{$- \ \ \ q $}}
				
				\qbezier(125,20)(145,0)(125,-20)
				\qbezier(150,20)(130,0)(150,-20)
				
				\put(-68,22){\mbox{$i$}}  \put(-15,22){\mbox{$j$}}  \put(-68,-28){\mbox{$k$}}  \put(-15,-28){\mbox{$l$}}
				
%
			}
			
			\put(0,-20){
				
				\put(-60,-20){\line(1,1){18}}
				\put(-20,-20){\line(-1,1){40}}
				\put(-37.5,2){\line(1,1){18}}
				
				\put(-68,22){\mbox{$i$}}  \put(-15,22){\mbox{$j$}}  \put(-68,-28){\mbox{$k$}}  \put(-15,-28){\mbox{$l$}}
				
				\put(10,-2){\mbox{$=$}}
				
				\qbezier(30,20)(50,0)(70,20)
				\qbezier(30,-20)(50,0)(70,-20)
				
				\put(85,-2){\mbox{$- \ \ \ q^{-1} $}}
				
				\qbezier(125,20)(145,0)(125,-20)
				\qbezier(150,20)(130,0)(150,-20)
				
%
			}
			
			\put(60,-20){
				
				\put(-100,-70){\circle{30}}
				\put(-75,-72.5){\mbox{$=\ \ \ D_2 \ = \ q+q^{-1}$}}
			}
		}
	\end{picture}
	\caption{\footnotesize The Kauffman bracket --- the planar decomposition
		of the $\mathcal{R}$-matrix vertex for the fundamental representation of $U_q(\mathfrak{sl}_2)$.
		In this case ($N=2$), the conjugate of the fundamental representation is isomorphic to it,
		thus, tangles in the picture have no orientation. 
	}
	\label{fig:Kauff}
\end{figure}

The generalization to the $SU(N)$ case is provided by the HOMFLY polynomial~\cite{freyd1985new,przytycki1988invariants}. In this case, $\cal R$-matrices for the quantum universal enveloping algebra $U_q(\mathfrak{sl}_N)$ arise, but they are not so simple. Still, one can split the $\cal R$-matrix into the sum of the identity component and the projector to the antisymmetric representation called MOY vertex~\cite{MOY}, see~\eqref{R}. Then, resolutions of a link are some MOY graphs satisfying MOY relations~\eqref{MOYloc_II}--\eqref{MOYloc_VI}, but there is still infinite amount of types of end-point MOY graphs. However, in our recent works~\cite{ALM,ALM2,ALM3}, we have shown that for a specific family of bipartite links, these end-point graphs are just cycles as in the Jones case. 

Then, one can consider the Khovanov--Rozansky (KhR) polynomials being $t$-deformation (categorification) of the HOMFLY polynomials and ask if there is a cycle calculus too. In this paper, we show that the answer is positive and as simple as in the case of the Khovanov polynomial being in turn the categorification of the Jones polynomial. 

This fact provides an example of reduction of the KhR operator formalism that allows to transfer from the initial bicomplex structure to the monocomplex on $N$-dimensional $q$-graded vector spaces. Another example is the equivalence of the Khovanov cycle calculus and the KhR operator technique for $N=2$.

We also demonstrate the reductions of the KhR bicomplex itself. This bicomplex contains vertical and horizontal morphisms. These morphisms and their cohomologies can be simplified, and such reductions are called vertical and horizontal reductions correspondingly.  

%

\bigskip


\section{Plan}


In this section, we provide the organization and brief review of this paper. At first, we sketch the content of Section~\ref{sec:prerequisites}. We use a formalism of differential operators in odd (Grassman) variables implicit in many works on the subject since the original \cite{Kauff,khovanov2000categorification,KRI},
and made fully explicit in recent \cite{2508.05191}.

\medskip

\noindent $\bullet$
Each vertex $v$ of a link/knot diagram $\cal L$\footnote{We denote closed diagrams (knots and links) by curly letters and open diagrams (tangles) by straight letters.} can be {\it resolved} in different ways.
In the conventional RT approach \cite{MMM1}, these are provided by distinguishing of ${\cal R}$-matrices ${\cal R}$ and ${\cal R}^{-1}$.
A more interesting option is a decomposition of $\cal R$-matrices over representations \cite{MMM2} based on expansion of the tensor product of representations $\lambda$, $\lambda'$ into irreducible components $\lambda''$:
\be
\lambda\otimes \lambda' = \bigoplus_{\lambda''} C_{\lambda,\lambda'}^{\lambda''}\, \lambda''\,.
\ee
For the fundamental representation $F$, there are just two items at the r.h.s.,
and one can substitute this decomposition by shifting from symmetric representation to unity \cite{dolotin2014introduction,AnoM}
\be
F\otimes F = I \oplus A
\ee
where $A$ is the first antisymmetric representation, often denoted by a Young diagram $[1,1]$.
This last resolution, sometimes called MOY \cite{MOY}, is used in the present text. At the level of $\CR$-matrices, we have the decomposition~\eqref{R}.

\medskip

\noindent $\bullet$
Two-fold resolution of any vertex $v$ leads to $2^{\#(v)}$ resolutions $\Gamma$ of a diagram $\cal L$, where $\#(v)$ is a number of vertices in $\cal L$. These resolutions can be organized in a 2-hypercube, see the example below:
\begin{equation}\label{Hopf-ex-HOMFLY}
	\begin{aligned}
		\begin{array}{c}
			\begin{tikzpicture}[scale=0.7]
				\begin{scope}[shift={(0,1)}]
					\draw[thick] (0.5,-0.5) to[out=90,in=270] (-0.5,0.5);
					\draw[white, line width = 1.5mm] (-0.5,-0.5) to[out=90,in=270] (0.5,0.5);
					\draw[thick] (-0.5,-0.5) to[out=90,in=270] (0.5,0.5);
				\end{scope}
				\draw[thick,-stealth] (0.5,-0.5) to[out=90,in=270] (-0.5,0.5) -- (-0.5,0.6);
				\draw[white, line width = 1.5mm] (-0.5,-0.5) to[out=90,in=270] (0.5,0.5);
				\draw[thick,-stealth] (-0.5,-0.5) to[out=90,in=270] (0.5,0.5) -- (0.5,0.6);
				\draw[thick] (-0.5,1.5) to[out=90,in=90] (-1.2,1.5) -- (-1.2,-0.5) to[out=270,in=270] (-0.5,-0.5);
				\begin{scope}[xscale=-1]
					\draw[thick] (-0.5,1.5) to[out=90,in=90] (-1.2,1.5) -- (-1.2,-0.5) to[out=270,in=270] (-0.5,-0.5);
				\end{scope}
				\node[above] at (0,0) {$\scriptstyle 1$};
				\node[above] at (0,1) {$\scriptstyle 2$};
			\end{tikzpicture}
		\end{array}&=\left[\begin{array}{c}
			\begin{tikzpicture}
				\node(A) at (0,0) {
					$\begin{tikzpicture}[scale=0.5]
						\draw[thick] (0.5,-0.5) to[out=90,in=270] (0.3,0) to[out=90,in=270] (0.5,0.5) -- (0.5,0.6);
						\draw[thick] (-0.5,-0.5) to[out=90,in=270] (-0.3,0) to[out=90,in=270] (-0.5,0.5) -- (-0.5,0.6);
						\begin{scope}[shift={(0,1)}]
							\draw[thick] (0.5,-0.5) to[out=90,in=270] (0.3,0) to[out=90,in=270] (0.5,0.5);
							\draw[thick] (-0.5,-0.5) to[out=90,in=270] (-0.3,0) to[out=90,in=270] (-0.5,0.5);
						\end{scope}
						\draw[thick, postaction={decorate},decoration={markings, mark= at position 0.6 with {\arrow{stealth}}}] (-0.5,1.5) to[out=90,in=90] (-1.2,1.5) -- (-1.2,-0.5) to[out=270,in=270] (-0.5,-0.5);
						\begin{scope}[xscale=-1]
							\draw[thick, postaction={decorate},decoration={markings, mark= at position 0.6 with {\arrow{stealth}}}] (-0.5,1.5) to[out=90,in=90] (-1.2,1.5) -- (-1.2,-0.5) to[out=270,in=270] (-0.5,-0.5);
						\end{scope}
					\end{tikzpicture}_{\,00}$};
				\node(B) at (3,0.7) {$\begin{tikzpicture}[scale=0.5]
						\draw[thick] (0.5,-0.5) to[out=90,in=270] (-0.5,0.5);
						\draw[thick] (-0.5,-0.5) to[out=90,in=270] (0.5,0.5);
						\begin{scope}[shift={(0,1)}]
							\draw[thick] (0.5,-0.5) to[out=90,in=270] (0.3,0) to[out=90,in=270] (0.5,0.5);
							\draw[thick] (-0.5,-0.5) to[out=90,in=270] (-0.3,0) to[out=90,in=270] (-0.5,0.5);
						\end{scope}
						\draw[thick, postaction={decorate},decoration={markings, mark= at position 0.6 with {\arrow{stealth}}}] (-0.5,1.5) to[out=90,in=90] (-1.2,1.5) -- (-1.2,-0.5) to[out=270,in=270] (-0.5,-0.5);
						\begin{scope}[xscale=-1]
							\draw[thick, postaction={decorate},decoration={markings, mark= at position 0.6 with {\arrow{stealth}}}] (-0.5,1.5) to[out=90,in=90] (-1.2,1.5) -- (-1.2,-0.5) to[out=270,in=270] (-0.5,-0.5);
						\end{scope}
						\draw[fill=\myred] (0,0) circle (0.13);
					\end{tikzpicture}_{\,10}$};
				\node(C) at (3,-0.7) {$\begin{tikzpicture}[scale=0.5]
						\draw[thick] (0.5,-0.5) to[out=90,in=270] (0.3,0) to[out=90,in=270] (0.5,0.5);
						\draw[thick] (-0.5,-0.5) to[out=90,in=270] (-0.3,0) to[out=90,in=270] (-0.5,0.5);
						\begin{scope}[shift={(0,1)}]
							\draw[thick] (0.5,-0.5) to[out=90,in=270] (-0.5,0.5);
							\draw[thick] (-0.5,-0.5) to[out=90,in=270] (0.5,0.5);
						\end{scope}
						\draw[thick, postaction={decorate},decoration={markings, mark= at position 0.6 with {\arrow{stealth}}}] (-0.5,1.5) to[out=90,in=90] (-1.2,1.5) -- (-1.2,-0.5) to[out=270,in=270] (-0.5,-0.5);
						\begin{scope}[xscale=-1]
							\draw[thick, postaction={decorate},decoration={markings, mark= at position 0.6 with {\arrow{stealth}}}] (-0.5,1.5) to[out=90,in=90] (-1.2,1.5) -- (-1.2,-0.5) to[out=270,in=270] (-0.5,-0.5);
						\end{scope}
						\draw[fill=\myred] (0,1) circle (0.13);
					\end{tikzpicture}_{\,01}$};
				\node(D) at (6,0) {$\begin{tikzpicture}[scale=0.5]
						\draw[thick,-stealth] (0.5,-0.5) to[out=90,in=270] (-0.5,0.5) -- (-0.5,0.6);
						\draw[thick,-stealth] (-0.5,-0.5) to[out=90,in=270] (0.5,0.5) -- (0.5,0.6);
						\begin{scope}[shift={(0,1)}]
							\draw[thick] (0.5,-0.5) to[out=90,in=270] (-0.5,0.5);
							\draw[thick] (-0.5,-0.5) to[out=90,in=270] (0.5,0.5);
						\end{scope}
						\draw[thick, postaction={decorate},decoration={markings, mark= at position 0.6 with {\arrow{stealth}}}] (-0.5,1.5) to[out=90,in=90] (-1.2,1.5) -- (-1.2,-0.5) to[out=270,in=270] (-0.5,-0.5);
						\begin{scope}[xscale=-1]
							\draw[thick, postaction={decorate},decoration={markings, mark= at position 0.6 with {\arrow{stealth}}}] (-0.5,1.5) to[out=90,in=90] (-1.2,1.5) -- (-1.2,-0.5) to[out=270,in=270] (-0.5,-0.5);
						\end{scope}
						\draw[fill=\myred] (0,0) circle (0.13) (0,1) circle (0.13);
					\end{tikzpicture}_{\,11}$};
				\path (A) edge[-] (B) (A) edge[-] (C) (B) edge[-] (D) (C) edge[-] (D);
			\end{tikzpicture}
		\end{array}\right]\,.
	\end{aligned}
\end{equation}

\medskip

\noindent $\bullet$
Ascribe a simply-constructed nilpotent operator\footnote{We usually omit hats over operators.} $\hat D_\Gamma$
to any {\it resolution} $\Gamma$ of a diagram $L\,$:
\be
\hat D_\Gamma^2
= 0 \ \ \ \ \ \ \ \text {for a closed $\CL$ with no external edges.}
\ee

\medskip

\noindent $\bullet$
Two resolutions $\resp\,$ and $\resm\,$, differing at a single vertex $v$, can be related by {\it homomorphisms} $\chi_v^{(\pm)}$, acting on the operators:
\ba\label{mu-D-commut}
\chi_v^{(+)} \cdot \hat D^{\smallresp}_v = \hat D^{\smallresm}_v \cdot \chi_v^{(+)}\,, \\
\hat D^{\smallresp}_v \cdot \chi_v^{(-)} = \chi_v^{(-)} \cdot \hat D^{\smallresm}_v\,. 
\ea
This is an analogue of the cut-and-join morphism in the Kauffman theory \cite{dolotin2013introduction}.

\medskip

\noindent $\bullet$
From linear combinations of these morphisms one can construct new nilpotent differentials $\hat {\cal D}_i$,
acting between cohomologies of different $\hat D_\Gamma$.
\be
\bigoplus_{\Gamma\in S_i} H^*(\hat D_\Gamma) \ \overset{\hat{\cal D}_i}{\longrightarrow}\  \bigoplus_{\Gamma\in S_{i+1}}  H^*(\hat D_\Gamma)
\ee
where $S_i$ and $S_{i+1}$ are adjacent ``slices''  of the hypercube of resolutions of a diagram $\CL$.

\medskip

\noindent $\bullet$
Poincar\'e polynomial (the KhR superpolynomial) $\KhR^\CL$ is made from  cohomologies of $\tilde{\mathfrak{D}}_i$:
\be
\KhR^\CL(A,q,T) = \sum_{i} t^i \cdot {\rm dim}_q\, H^i(\tilde{\mathfrak{D}}_i)\,.
\ee
The central point is the construction of operators $\hat D_\Gamma$,
realizing/substituting the matrix factorization idea of \cite{KRI}.
The main simplification is that they can be associated with diagrams with external legs (i.e. with {\it tangles}),
and then for any diagram $\Gamma$, they are just sums over the operators for particular vertices:
\be
\hat D_\Gamma=\sum_{{\smallresp}\, \in \Gamma} \hat D^{\smallresp}_v + \sum_{{\smallresm}\, \in \Gamma} \hat D^{\smallresm}_v\,.
\ee
What guarantees nilpotency for closed diagrams is the special condition that
\be\label{D^2-gen}
\hat D_\Gamma^2 = w(\text{outgoing edges}) - w(\text{incoming edges})
\ee
for some function $w$.
Actually, for fundamental representations of $\mathfrak{sl}_N$ in \cite{KRI} the choice is $w = x^{N+1}$.

\medskip

\noindent The last point of the construction is the specification of its  building blocks: the operators for a single vertex.

\medskip

\noindent $\bullet$
With each vertex $v$, we associate a pair of odd operators $\hat D^{\smallresp}_v$, $\hat D^{\smallresm}_v$, differing by a resolution $\resp\,$, $\resm\,$.
They are linear in two odd parameters $\theta_v,\eta_{\,v}$ and derivatives $\theta^\dagger_v= \cfrac{\p}{\p \theta_v}$,
$\eta^\dagger_v= \cfrac{\p}{\p \eta_v}$ and are functions (not operators) of four even parameters:
$x_1,x_2$, associated with two outgoing legs
and $x_3,x_4$ associated with two incoming legs:
\be\label{Dresp-Dresm-gen}
\begin{aligned}
    \begin{array}{c}
    	\begin{tikzpicture}[scale=0.65]
    		\draw[thick, -stealth] (-0.5,-0.5) to[out=45,in=270] (-0.2,0) to[out=90,in=315] (-0.5,0.5);
    		\draw[thick, -stealth] (0.5,-0.5) to[out=135,in=270] (0.2,0) to[out=90,in=225] (0.5,0.5);
    		\draw[fill=\mygreen] (-0.2,0) circle (0.08) (0.2,0) circle (0.08);
    		\node[above left] at (-0.5,0.5) {$\scriptstyle x_1$};
    		\node[above right] at (0.5,0.5) {$\scriptstyle x_2$};
    		\node[below right] at (0.5,-0.5) {$\scriptstyle x_4$};
    		\node[below left] at (-0.5,-0.5) {$\scriptstyle x_3$};
    		\node[left] at (-0.2,0) {$\scriptstyle \theta_v$};
    		\node[right] at (0.2,0) {$\scriptstyle \eta_v$};
    	\end{tikzpicture}
    \end{array}:\;\hat D^{\smallresp}_v &= A^{\smallresp}_v\left[\begin{array}{cc} x_1 & x_2 \\ x_3 & x_4 \end{array}\right]\theta_v +
B^{\smallresp}_v\left[\begin{array}{cc} x_1 & x_2 \\ x_3 & x_4 \end{array}\right]\theta^\dag_v
+ C^{\smallresp}_v\left[\begin{array}{cc} x_1 & x_2 \\ x_3 & x_4 \end{array}\right]\eta_v +
E^{\smallresp}_v\left[\begin{array}{cc} x_1 & x_2 \\ x_3 & x_4 \end{array}\right]\eta^\dag_v\,, \\
\begin{array}{c}
	\begin{tikzpicture}[scale=0.65]
		\draw[thick, -stealth] (-0.5,-0.5) -- (0.5,0.5);
		\draw[thick, -stealth] (0.5,-0.5) -- (-0.5,0.5);
		\draw[fill=\myred] (0,0) circle (0.1);
		\node[above left] at (-0.5,0.5) {$\scriptstyle x_1$};
		\node[above right] at (0.5,0.5) {$\scriptstyle x_2$};
		\node[below right] at (0.5,-0.5) {$\scriptstyle x_4$};
		\node[below left] at (-0.5,-0.5) {$\scriptstyle x_3$};
		\node[left] at (-0.1,0) {$\scriptstyle \theta_v$};
		\node[right] at (0.1,0) {$\scriptstyle \eta_v$};
	\end{tikzpicture}
\end{array}:\; \hat D^{\smallresm}_v &= A^{\smallresm}_v\left[\begin{array}{cc} x_1 & x_2 \\ x_3 & x_4 \end{array}\right]\theta_v +
B^{\smallresm}_v\left[\begin{array}{cc} x_1 & x_2 \\ x_3 & x_4 \end{array}\right]\theta^\dag_v
+ C^{\smallresm}_v\left[\begin{array}{cc} x_1 & x_2 \\ x_3 & x_4 \end{array}\right]\eta_v +
E^{\smallresm}_v\left[\begin{array}{cc} x_1 & x_2 \\ x_3 & x_4 \end{array}\right]\eta^\dag_v\,.
\end{aligned}
\ee
The concrete choice of functions $A,\,B,\,C,\,E$ is specified in Section \ref{sec:vert-morph} below.

\medskip

\noindent $\bullet$ Then, the morphisms $\chi_v^{(\pm)}$ are fixed by the homomorphism conditions~\eqref{mu-D-commut} and the requirements that it should be a local even operator of the fixed degree \cite{2508.05191}, see Section~\ref{sec:hor-morphisms}. The relation~\eqref{mu-D-commut} for $\chi^{(+)}_v$ can be pictorially sketched as follows:
\begin{equation}\label{MFhomo}
	\begin{array}{c}
		\begin{tikzpicture}
			\node(A) at (0,0) {$\begin{array}{c}
					\begin{tikzpicture}[scale=0.65]
						\draw[thick, -stealth] (-0.5,-0.5) to[out=45,in=270] (-0.2,0) to[out=90,in=315] (-0.5,0.5);
						\draw[thick, -stealth] (0.5,-0.5) to[out=135,in=270] (0.2,0) to[out=90,in=225] (0.5,0.5);
						\draw[fill=\mygreen] (-0.2,0) circle (0.08) (0.2,0) circle (0.08);
						\node[above left] at (-0.5,0.5) {$\scriptstyle x_1$};
						\node[above right] at (0.5,0.5) {$\scriptstyle x_2$};
						\node[below right] at (0.5,-0.5) {$\scriptstyle x_4$};
						\node[below left] at (-0.5,-0.5) {$\scriptstyle x_3$};
						\node[left] at (-0.2,0) {$\scriptstyle \theta_v$};
						\node[right] at (0.2,0) {$\scriptstyle \eta_v$};
					\end{tikzpicture}
				\end{array}$};
			\node(B) at (4,0) {$\begin{array}{c}
					\begin{tikzpicture}[scale=0.65]
						\draw[thick, -stealth] (-0.5,-0.5) -- (0.5,0.5);
						\draw[thick, -stealth] (0.5,-0.5) -- (-0.5,0.5);
						\draw[fill=\myred] (0,0) circle (0.1);
						\node[above left] at (-0.5,0.5) {$\scriptstyle x_1$};
						\node[above right] at (0.5,0.5) {$\scriptstyle x_2$};
						\node[below right] at (0.5,-0.5) {$\scriptstyle x_4$};
						\node[below left] at (-0.5,-0.5) {$\scriptstyle x_3$};
						\node[left] at (-0.1,0) {$\scriptstyle \theta_v$};
						\node[right] at (0.1,0) {$\scriptstyle \eta_v$};
					\end{tikzpicture}
				\end{array}$};
			\node(C) at (0,-3) {$\begin{array}{c}
					\begin{tikzpicture}[scale=0.65]
						\draw[thick, -stealth] (-0.5,-0.5) to[out=45,in=270] (-0.2,0) to[out=90,in=315] (-0.5,0.5);
						\draw[thick, -stealth] (0.5,-0.5) to[out=135,in=270] (0.2,0) to[out=90,in=225] (0.5,0.5);
						\draw[fill=\mygreen] (-0.2,0) circle (0.08) (0.2,0) circle (0.08);
						\node[above left] at (-0.5,0.5) {$\scriptstyle x_1$};
						\node[above right] at (0.5,0.5) {$\scriptstyle x_2$};
						\node[below right] at (0.5,-0.5) {$\scriptstyle x_4$};
						\node[below left] at (-0.5,-0.5) {$\scriptstyle x_3$};
						\node[left] at (-0.2,0) {$\scriptstyle \theta_v$};
						\node[right] at (0.2,0) {$\scriptstyle \eta_v$};
					\end{tikzpicture}
				\end{array}$};
			\node(D) at (4,-3) {$\begin{array}{c}
					\begin{tikzpicture}[scale=0.65]
						\draw[thick, -stealth] (-0.5,-0.5) -- (0.5,0.5);
						\draw[thick, -stealth] (0.5,-0.5) -- (-0.5,0.5);
						\draw[fill=\myred] (0,0) circle (0.1);
						\node[above left] at (-0.5,0.5) {$\scriptstyle x_1$};
						\node[above right] at (0.5,0.5) {$\scriptstyle x_2$};
						\node[below right] at (0.5,-0.5) {$\scriptstyle x_4$};
						\node[below left] at (-0.5,-0.5) {$\scriptstyle x_3$};
						\node[left] at (-0.1,0) {$\scriptstyle \theta_v$};
						\node[right] at (0.1,0) {$\scriptstyle \eta_v$};
					\end{tikzpicture}
				\end{array}$};
			\draw[-stealth] ([shift={(0,0.1)}]A.east) -- ([shift={(0,0.1)}]B.west) node[pos=0.5,above] {$\scriptstyle \chi_v^{(+)}$};
			\draw[-stealth] ([shift={(0,0.1)}]C.east) -- ([shift={(0,0.1)}]D.west) node[pos=0.5,above] {$\scriptstyle \chi_v^{(+)}$};
			\draw[-stealth] ([shift={(-0.1,0)}]A.south) -- ([shift={(-0.1,0)}]C.north);
			\draw[-stealth] ([shift={(-0.1,0)}]B.south) -- ([shift={(-0.1,0)}]D.north);
			\node at (-0.5,-1.5) {$\scriptstyle \hat D^{\smallresp}_v$};
			\node at (3.5,-1.5) {$\scriptstyle \hat D^{\smallresm}_v$};
		\end{tikzpicture}
	\end{array}
\end{equation}
what reveals the structure of a bicomplex. For the morphism $\chi^{(-)}_v$, there is an analogous diagram. Keeping in mind this picture, we call $\hat D_v^{\smallresp}$ and $\hat D_v^{\smallresm}$ {\it vertical} morphisms and $\chi_v^{(\pm)}$ {\it horizontal} morphisms.

%
%
\bigskip

\noindent In this article, we study reductions of complexes for the Khovanov--Rozansky cohomologies. In addition, the notion of reduction splits into three cases --- the vertical reduction of cohomologies of operators $\hat D_\Gamma$, the horizontal reduction of resulting complexes of these reduced cohomologies and, in the $N=2$ and the bipartite cases --- the afterwards reduction from bicomplexes to Khovanov(-like) monocomplexes. 
The vertical reductions are first demonstrated in Section~\ref{sec:examples} with the use of the examples, see Table~\ref{tab:examples-1}. These examples are the simplest ones that illustrate some of the general reductions from Table~\ref{tab:isom}, see Sections~\ref{sec:lin_red}--\ref{sec:box-2}. 

\begin{table}[h!]
	\centering
\begin{tabular}{|c|m{6.5cm}|}
	\hline 
	\vspace{5pt}
	Example & \vspace{5pt} Key phenomenon at this level \\
	\hline 
	{\centering $\begin{array}{c}\begin{tikzpicture}
		\draw[thick,postaction={decorate},decoration={markings, mark= at position 0.5 with {\arrow{stealth}}}] (0,0) circle (0.5);
		\node[right] at (0.5,0) {$\scriptstyle x$};
		\node[left] at (-0.51,0) {$\scriptstyle y$};
		\node[above] at (0,0.5) {$\scriptstyle \theta_1$};
		\node[below] at (0,-0.5) {$\scriptstyle \theta_2$};
		\draw[fill=\mygreen] (0,0.5) circle (0.06);
		\draw[fill=\mygreen] (0,-0.5) circle (0.06);
	\end{tikzpicture}\end{array} \cong \begin{array}{c}
	\begin{tikzpicture}
		\draw[thick,postaction={decorate},decoration={markings, mark= at position 0.5 with {\arrow{stealth}}}] (0,0) circle (0.5);
		\node[right] at (0.5,0) {$\scriptstyle x$};
		\node[left] at (-0.51,0) {$\scriptstyle x$};
		\node[above] at (0,0.5) {$\scriptstyle \theta_1$};
		\draw[fill=\mygreen] (0,0.5) circle (0.06);
	\end{tikzpicture}
	\end{array}$} & {\centering Linear reduction holds, i.e. one can delete, for example, $\theta_2$ and equalize $x=y$.}  \\ [1.5ex]
	\hline 
	\vspace{5pt}
	$\begin{array}{c}\begin{tikzpicture}[scale=0.8]
		\draw[thick,-stealth] (0.5,-0.5) to[out=90,in=270] (-0.5,0.5);
		\draw[thick,-stealth] (-0.5,-0.5) to[out=90,in=270] (0.5,0.5);
		\draw[thick] (-0.5,0.5) to[out=90,in=90] (-0.8,0.5) -- (-0.8,-0.5) to[out=270,in=270] (-0.5,-0.5);
		\begin{scope}[xscale=-1]
			\draw[thick] (-0.5,0.5) to[out=90,in=90] (-0.8,0.5) -- (-0.8,-0.5) to[out=270,in=270] (-0.5,-0.5);
		\end{scope}
		\node[right] at (-0.5,-0.5) {$\scriptstyle x$};
		\node[right] at (-0.5,0.5) {$\scriptstyle x$};
		\node[left] at (0.5,-0.5) {$\scriptstyle z$};
		\node[left] at (0.5,0.5) {$\scriptstyle z$};
		\node[left] at (-0.15,0) {$\scriptstyle \theta_1$};
		\node[right] at (0.25,0) {$\scriptstyle \theta_2$};
		\draw[fill=\myred] (0,0) circle (0.1);
	\end{tikzpicture}\end{array}\cong \begin{array}{c}
	\begin{tikzpicture}
	\draw[thick,postaction={decorate},decoration={markings, mark= at position 0.5 with {\arrow{stealth}}}] (0,0) circle (0.5);
	\node[right] at (0.5,0) {$\scriptstyle x$};
	\node[left] at (-0.51,0) {$\scriptstyle x$};
	\node[above] at (0,0.5) {$\scriptstyle \theta_1$};
	\draw[fill=\mygreen] (0,0.5) circle (0.06);
	\end{tikzpicture}
\end{array} \otimes \sum\limits_{j=0}^{N-2} a_j z^j \theta_2$ & \vspace{5pt} MOY I holds, i.e. the cohomology corresponding to the figure-eight MOY graph splits into the tensor product $V_N\otimes V_{N-1}$ with $\dim_q V_x = [x]\,.$ \\ [1.5ex]
	\hline 
	\vspace{5pt}
	{$\begin{array}{c}\begin{tikzpicture}[scale=1.0]
		\draw[thick,-stealth] (0.5,-0.5) to[out=90,in=270] (-0.5,0.5);
		\draw[thick,-stealth] (-0.5,-0.5) to[out=90,in=270] (0.5,0.5);
		\draw[thick] (-0.5,0.5) to[out=90,in=90] (-0.8,0.5) -- (-0.8,-0.5) to[out=270,in=270] (-0.5,-0.5);
		\begin{scope}[xscale=-1]
			\draw[thick] (-0.5,0.5) to[out=90,in=90] (-0.8,0.5) -- (-0.8,-0.5) to[out=270,in=270] (-0.5,-0.5);
		\end{scope}
		\node[right] at (-0.5,-0.5) {$\scriptstyle x_4$};
		\node[right] at (-0.5,0.5) {$\scriptstyle x_1$};
		\node[left] at (0.5,-0.5) {$\scriptstyle x_3$};
		\node[left] at (0.5,0.5) {$\scriptstyle x_2$};
		\node[left] at (-0.15,0) {$\scriptstyle \theta_1$};
		\node[left] at (-0.85,0) {$\scriptstyle \theta_3$};
		\node[right] at (0.25,0) {$\scriptstyle \theta_2$};
		\node[right] at (0.85,0) {$\scriptstyle \theta_4$};
		\draw[fill=\myred] (0,0) circle (0.07);
		\draw[fill=\mygreen] (-0.8,0) circle (0.06);
		\draw[fill=\mygreen] (0.8,0) circle (0.06);
	\end{tikzpicture}\end{array}\cong \left\{\begin{array}{c}
	\begin{tikzpicture}[scale=0.9]
		\draw[thick,-stealth] (0.5,-0.5) to[out=90,in=270] (-0.5,0.5);
		\draw[dashed,thick,-stealth] (-0.5,-0.5) to[out=90,in=270] (0.5,0.5);
		\draw[thick] (-0.5,0.5) to[out=90,in=90] (-0.8,0.5) -- (-0.8,-0.5) to[out=270,in=270] (-0.5,-0.5);
		\begin{scope}[xscale=-1]
			\draw[thick] (-0.5,0.5) to[out=90,in=90] (-0.8,0.5) -- (-0.8,-0.5) to[out=270,in=270] (-0.5,-0.5);
		\end{scope}
		\node[right] at (-0.5,-0.5) {$\scriptstyle x_2$};
		\node[right] at (-0.5,0.5) {$\scriptstyle x_1$};
		\node[left] at (0.5,-0.5) {$\scriptstyle x_1$};
		\node[left] at (0.5,0.5) {$\scriptstyle x_2$};
		\node[left] at (-0.85,0) {$\scriptstyle \theta_3$};
		\node[right] at (0.85,0) {$\scriptstyle \theta_4$};
		\draw[fill=\mygreen] (-0.8,0) circle (0.06);
		\draw[fill=\mygreen] (0.8,0) circle (0.06);
	\end{tikzpicture} \\
		\begin{array}{c}\begin{tikzpicture}[scale=0.8]
		\draw[thick,postaction={decorate},decoration={markings, mark= at position 0.5 with {\arrow{stealth}}}] (0,0) circle (0.5);
		\node[right] at (0.5,0) {$\scriptstyle x_1$};
		\node[left] at (-0.51,0) {$\scriptstyle x_1$};
		\node[above] at (0,0.5) {$\scriptstyle \theta_3$};
		\draw[fill=\mygreen] (0,0.5) circle (0.06);
	\end{tikzpicture}\end{array}\otimes \begin{array}{c}\begin{tikzpicture}[scale=0.8]
	\draw[thick,postaction={decorate},decoration={markings, mark= at position 0.5 with {\arrow{stealth}}}] (0,0) circle (0.5);
	\node[right] at (0.5,0) {$\scriptstyle x_2$};
	\node[left] at (-0.51,0) {$\scriptstyle x_2$};
	\node[above] at (0,0.5) {$\scriptstyle \theta_4$};
	\draw[fill=\mygreen] (0,0.5) circle (0.06);
	\end{tikzpicture}\end{array} \\
	\hline
	\psi_-(x,x) = \psi_+(x,x)
	\end{array}\right\}$} & \vspace{5pt} Quadratic reduction holds, i.e. the cohomology of the shown figure-eight MOY graph splits into two cohomologies corresponding to the r.h.s. MOY graphs subject to the relation on these cohomologies. \\ [1.5ex]
	\hline 
	\vspace{5pt}
	$\begin{array}{c}\begin{tikzpicture}[scale=0.7]
		\draw[thick,-stealth] (0.5,-0.5) to[out=90,in=270] (-0.5,0.5) -- (-0.5,0.6);
		\draw[thick,-stealth] (-0.5,-0.5) to[out=90,in=270] (0.5,0.5) -- (0.5,0.6);
		\begin{scope}[shift={(0,1)}]
			\draw[thick] (0.5,-0.5) to[out=90,in=270] (-0.5,0.5);
			\draw[thick] (-0.5,-0.5) to[out=90,in=270] (0.5,0.5);
		\end{scope}
		\draw[thick, postaction={decorate},decoration={markings, mark= at position 0.6 with {\arrow{stealth}}}] (-0.5,1.5) to[out=90,in=90] (-1.2,1.5) -- (-1.2,-0.5) to[out=270,in=270] (-0.5,-0.5);
		\begin{scope}[xscale=-1]
			\draw[thick, postaction={decorate},decoration={markings, mark= at position 0.6 with {\arrow{stealth}}}] (-0.5,1.5) to[out=90,in=90] (-1.2,1.5) -- (-1.2,-0.5) to[out=270,in=270] (-0.5,-0.5);
		\end{scope}
		\draw[fill=\myred] (0,0) circle (0.1) (0,1) circle (0.1);
		\node[left] at (-0.5,-0.5) {$\scriptstyle x_1$};
		\node[right] at (0.5,-0.5) {$\scriptstyle x_2$};
		\node[left] at (-0.5,0.5) {$\scriptstyle x_3$};
		\node[right] at (0.5,0.5) {$\scriptstyle x_4$};
		\node[left] at (-0.5,1.5) {$\scriptstyle x_1$};
		\node[right] at (0.5,1.5) {$\scriptstyle x_2$};
		\node[left] at (-0.3,0) {$\scriptstyle \theta_1$};
		\node[right] at (0.3,0) {$\scriptstyle \theta_2$};
		\node[left] at (-0.3,1) {$\scriptstyle \theta_3$};
		\node[right] at (0.3,1) {$\scriptstyle \theta_4$};
	\end{tikzpicture}\end{array}\cong \; \begin{array}{c}
	\begin{tikzpicture}[scale=0.8]
	\draw[thick,-stealth] (0.5,-0.5) to[out=90,in=270] (-0.5,0.5);
	\draw[thick,-stealth] (-0.5,-0.5) to[out=90,in=270] (0.5,0.5);
	\draw[thick] (-0.5,0.5) to[out=90,in=90] (-0.8,0.5) -- (-0.8,-0.5) to[out=270,in=270] (-0.5,-0.5);
	\begin{scope}[xscale=-1]
		\draw[thick] (-0.5,0.5) to[out=90,in=90] (-0.8,0.5) -- (-0.8,-0.5) to[out=270,in=270] (-0.5,-0.5);
	\end{scope}
	\node[right] at (-0.55,-0.5) {$\scriptstyle x_1$};
	\node[right] at (-0.55,0.5) {$\scriptstyle x_1$};
	\node[left] at (0.55,-0.5) {$\scriptstyle x_2$};
	\node[left] at (0.55,0.5) {$\scriptstyle x_2$};
	\node[left] at (-0.15,0) {$\scriptstyle \theta_1$};
	\node[right] at (0.25,0) {$\scriptstyle \theta_2$};
	\draw[fill=\myred] (0,0) circle (0.1);
	\end{tikzpicture}
\end{array} \otimes V_{2}\,,\quad \dim_q V_{2} = [2]$ & \vspace{5pt} Quadratic reduction in its degenerate form holds. Namely, it turns out that the differentials $D_\pm$ are such that the cohomology splits into a direct sum of two isomorphic cohomologies differing by the factor $(x_3 - x_4)$ so that one can separate the $V_2$ space. This fact also allows to validate the categorified MOY relation~\eqref{MOYcat-loc_IV}. \\ [1.5ex]
	\hline
\end{tabular}\caption{\footnotesize Examples from Section~\ref{sec:examples} and the outcomes. By the equivalence of pictures we mean the isomorphism of cohomologies of the corresponding operators. These examples are then lifted to the generic isomorphisms from Table~\ref{tab:isom}.}\label{tab:examples-1}
\end{table}

\begin{table}[h!]
	\centering
\begin{tabular}{|m{7.5cm}|m{9.5cm}|}
	\hline
	\vspace{5pt}
	Schematic isomorphism & Differentials and isomorphism \\
	\hline 
	\vspace{5pt}
	\begin{center}
		\quad \quad Linear reduction \vspace{3pt} \newline
		$\begin{array}{c}
			\begin{tikzpicture}[scale=0.7]
				\draw[thick, postaction={decorate},decoration={markings, mark = at position 0.25 with {\arrowreversed{stealth}}, mark = at position 0.75 with {\arrowreversed{stealth}}}] (0,0) to[out=45,in=180] (0.8,0.5) to[out=0,in=90] (1.2,0) to[out=270,in=0] (0.8,-0.5) to[out=180,in=315] (0,0);
				\draw[fill=\myblue] (0,0) circle (0.5);
				\node[white] at (0,0) {$\scriptstyle G$};
				\node[right] at (1,0.5) {$\scriptstyle x$};
				\node[right] at (1,-0.5) {$\scriptstyle y$};
				\node[right] at (1.25,0) {$\scriptstyle \theta$};
				\draw[fill=\mygreen] (1.2,0) circle (0.1);
			\end{tikzpicture}
		\end{array}\cong\begin{array}{c}
			\begin{tikzpicture}[scale=0.7]
				\draw[thick, postaction={decorate},decoration={markings, mark = at position 0.4 with {\arrowreversed{stealth}}}] (0,0) to[out=45,in=180] (0.8,0.5) to[out=0,in=90] (1.2,0) to[out=270,in=0] (0.8,-0.5) to[out=180,in=315] (0,0);
				\draw[fill=\myblue] (0,0) circle (0.5);
				\node[white] at (0,0) {$\scriptstyle G$};
				\node[right] at (1.2,0) {$\scriptstyle x$};
			\end{tikzpicture}
		\end{array}$
	\end{center}
	 & 
	 \vspace{5pt}
	 {\centering \small
	 	Linear reduction allows to eliminate 2-valent vertex. Namely, if we start from the differential $D=d_G(x,y)+\pi_{xy}\theta+\underline{(x-y)}\theta^{\dagger}$, then the following isomorphism holds: $H^*(D) \cong H^*(d_G(x,x))$. The reduction is called linear because of the underlined linear in even variables multiplier in the differential.}
	 	\vspace{5pt}
	  \\
	\hline 
	\vspace{5pt}
	\begin{center}
		\quad \quad Categorified MOY I relation \vspace{3pt} \newline
		$\begin{array}{c}
			\begin{tikzpicture}[scale=0.5]
				\draw[thick] (0.5,0.5) to[out=0,in=0] (0.5,-0.5);
				\draw[thick,postaction={decorate},decoration={markings, mark= at position 0.9 with {\arrow{stealth}}}] (0.5,-0.5) to[out=180,in=300] (-0.7,0.7);
				\draw[thick, postaction={decorate},decoration={markings, mark= at position 0.8 with {\arrow{stealth}}}] (-0.7,-0.7) to[out=60,in=180] (0.5,0.5);
				\draw[fill=burgundy] (-0.3,0) circle (0.1);
				\draw[fill=\myblue, even odd rule] (0.4,0) circle (1.3) (0.4,0) circle (1.8);
				\node[right] at (-0.7,0.7) {$\scriptstyle x$};
				\node[left] at (-1.4,0) {$\scriptstyle G$};
				\node[right] at (-0.7,-0.7) {$\scriptstyle y$};
				\node[above] at (0.5,0.5) {$\scriptstyle z$};
				\node[below] at (0.5,-0.5) {$\scriptstyle z$};
			\end{tikzpicture}
		\end{array}\cong\begin{array}{c}
			\begin{tikzpicture}[scale=0.5]
				\draw[thick,postaction={decorate},decoration={markings, mark= at position 0.9 with {\arrow{stealth}}}] (-0.7,-0.7) to[out=60,in=270] (-0.4,0) to[out=90,in=300] (-0.7,0.7);
				\draw[fill=\myblue, even odd rule] (0.4,0) circle (1.3) (0.4,0) circle (1.8);
				\node[right] at (-0.7,0.7) {$\scriptstyle x$};
				\node[right] at (-0.7,-0.7) {$\scriptstyle x$};
				\node[left] at (-1.4,0) {$\scriptstyle G$};
			\end{tikzpicture}
		\end{array}\otimes V_{N-1}$
	\end{center}
	 &\vspace{5pt} {\centering \small With the use of the categorified MOY I relation, one can eliminate a loop with 4-valent vertex. Namely, at the beginning we have the differential $D = d_G(x,y) + u_1\left[\begin{array}{cc} x & z \\ y & z \end{array}\right]\theta_1 + u_2\left[\begin{array}{cc} x & z \\ y & z \end{array}\right] \theta_2 +(x-y)\theta_1^{\dagger}+z(x-y)\theta_2^{\dagger}$, but actually we can reduce the cohomology as follows: $H^*(D) \cong H^*(d_G(x,x))\otimes V_{N-1}$.}\vspace{5pt} \\
	\hline
	\vspace{5pt}
	\begin{center}
		\quad \quad Quadratic reduction \vspace{1pt} \newline
	$\begin{array}{c}
		\begin{tikzpicture}[scale=0.97]
			\node[left] at (-1,0) {$\scriptstyle G$};
			\draw[fill=\myblue] (0,0) circle (1);
			\draw[fill=white] (0,0) circle (0.7);
			\draw[thick, postaction={decorate},decoration={markings, mark= at position 0.75 with {\arrow{stealth}}, mark= at position 0.25 with {\arrow{stealth}}}] (-0.5,-0.5) -- (0.5,0.5) node[left,pos=0.3] {$\scriptstyle x_4$} node[right,pos=0.7] {$\scriptstyle x_2$};
			\draw[thick, postaction={decorate},decoration={markings, mark= at position 0.75 with {\arrow{stealth}}, mark= at position 0.25 with {\arrow{stealth}}}] (0.5,-0.5) -- (-0.5,0.5) node[right,pos=0.3] {$\scriptstyle x_3$} node[left,pos=0.7] {$\scriptstyle x_1$};
			\draw[fill=\myred] (0,0) circle (0.07);
		\end{tikzpicture}
	\end{array} \cong \left\{\begin{array}{c} 
	\Gamma_+=\begin{array}{c}
		\begin{tikzpicture}[scale=0.97]
			\draw[thick, postaction={decorate},decoration={markings, mark= at position 0.75 with {\arrow{stealth}}, mark= at position 0.25 with {\arrow{stealth}}}] (-0.5,-0.5) to[out=45,in=315] node[left,pos=0.5,shift={(0.1,0)}] {$\scriptstyle x_1$} (-0.5,0.5);
			\draw[thick, postaction={decorate},decoration={markings, mark= at position 0.75 with {\arrow{stealth}}, mark= at position 0.25 with {\arrow{stealth}}}] (0.5,-0.5) to[out=135,in=225] node[right,pos=0.5,shift={(-0.1,0)}] {$\scriptstyle x_2$} (0.5,0.5);
			\begin{scope}[rotate=45]
				\node at (1.1,0) {$\scriptstyle G$};
			\end{scope}
			\draw[fill=\myblue, even odd rule] (0,0) circle (0.7)  (0,0) circle (0.9);
		\end{tikzpicture}
	\end{array} \\
	\Gamma_-=\begin{array}{c}
		\begin{tikzpicture}[scale=0.97]
			\draw[dashed,thick, postaction={decorate},decoration={markings, mark= at position 0.75 with {\arrow{stealth}}, mark= at position 0.25 with {\arrow{stealth}}}] (-0.5,-0.5) -- (0.5,0.5) node[left,pos=0.3] {$\scriptstyle x_2$} node[right,pos=0.7] {$\scriptstyle x_2$};
			\draw[thick, postaction={decorate},decoration={markings, mark= at position 0.75 with {\arrow{stealth}}, mark= at position 0.25 with {\arrow{stealth}}}] (0.5,-0.5) -- (-0.5,0.5) node[right,pos=0.3] {$\scriptstyle x_1$} node[left,pos=0.7] {$\scriptstyle x_1$};
			\begin{scope}[rotate=45]
				\node at (1.1,0) {$\scriptstyle G$};
			\end{scope}
			\draw[fill=\myblue, even odd rule] (0,0) circle (0.7)  (0,0) circle (0.9);
		\end{tikzpicture}
	\end{array} \\
	\hline 
	\psi_-(x,x) = \psi_+(x,x)
	\end{array}\right\}$
\end{center}
	 & \vspace{5pt}{\centering \small Quadratic reduction allows to simplify the cohomology for a MOY graph with 4-valent vertices. Given the differential $D=d_G\left[\begin{array}{cc} x_1 & x_2 \\ x_4 & x_3 \end{array}\right]+u_1\theta_1+s_1\theta_1^{\dagger}+u_2\theta_2+s_2\theta_2^{\dagger}$ initially, where $s_1=x_1+x_2-x_3-x_4$, $s_2=x_1x_2-x_3x_4$, we can reduce its cohomology: $H^*(D) \cong \left\{\left(\begin{array}{c}
	 	H^*\left(d_+(x_1,x_2)\right) \\
	 	H^*(d_-(x_1,x_2))
	 \end{array}\right)\;\Bigg|\;\begin{array}{c}
	 	\xi_\pm (x_1,x_2)\in d_\pm(x_1,x_2) \\
	 	\xi_+(x,x)=\xi_-(x,x)
	 \end{array}\right\}$, where $d_+(x_1,x_2)=d_G\left[\begin{array}{cc} x_1 & x_2 \\ x_1 & x_2 \end{array}\right]$ and $d_-(x_1,x_2)=d_G\left[\begin{array}{cc} x_1 & x_2 \\ x_2 & x_1 \end{array}\right]$, if $d_+(x_1,x_2) \neq d_-(x_1,x_2)$. The reduction is called quadratic because it is provided with respect to $s_1$ and the quadratic polynomial $s_2$.}\vspace{5pt}
	\\
	\hline
	\vspace{5pt}
	\begin{center}
		\quad \quad Categorified MOY II relation \vspace{3pt} \newline
		$\begin{array}{c}
			\begin{tikzpicture}[scale=0.7]
				\node[left] at (-1.3,0) {$\scriptstyle G$};
				\begin{scope}[shift={(0,1)}]
					\draw[thick,-stealth] (-0.5,-0.5) to[out=90,in=270] (0.5,0.5);
					\draw[thick,-stealth] (0.5,-0.5) to[out=90,in=270] (-0.5,0.5);
					\draw[fill=\myred] (0,0) circle (0.1);
				\end{scope}
				\draw[thick,-stealth] (-0.5,-0.5) to[out=90,in=270] (0.5,0.5) -- (0.5,0.6);
				\draw[thick,-stealth] (0.5,-0.5) to[out=90,in=270] (-0.5,0.5) -- (-0.5,0.6);
				\draw[fill=\myred] (0,0) circle (0.1);
				\node[left] at (-0.25,-0.1) {$\scriptstyle x_4$};
				\node[right] at (0.25,-0.1) {$\scriptstyle x_3$};
				\node[left] at (-0.5,0.5) {$\scriptstyle y$};
				\node[right] at (0.5,0.5) {$\scriptstyle z$};
				\node[left] at (-0.25,1.0) {$\scriptstyle x_1$};
				\node[right] at (0.25,1.0) {$\scriptstyle x_2$};
				\draw[fill=\myblue, even odd rule] (0,0.5) circle (1.1)  (0,0.5) circle (1.4);
			\end{tikzpicture}
		\end{array}\cong \begin{array}{c}
			\begin{tikzpicture}[scale=0.7]
				\node[left] at (-1,0) {$\scriptstyle G$};
				\draw[thick,-stealth] (-0.5,-0.5) to[out=90,in=270] (0.5,0.5);
				\draw[thick,-stealth] (0.5,-0.5) to[out=90,in=270] (-0.5,0.5);
				\draw[fill=\myred] (0,0) circle (0.1);
				\node[left] at (-0.6,-0.8) {$\scriptstyle x_4$};
				\node[right] at (0.6,-0.8) {$\scriptstyle x_3$};
				\node[left] at (-0.6,0.8) {$\scriptstyle x_1$};
				\node[right] at (0.6,0.8) {$\scriptstyle x_2$};
				\draw[fill=\myblue, even odd rule] (0,0) circle (0.725)  (0,0) circle (1.0);
			\end{tikzpicture}
		\end{array}\otimes V_2$
			\end{center} & \vspace{5pt}{\centering \small Using the categorified MOY II relation, one can eliminate multiple parallel connected 4-valent vertices. Namely, given the differential $D=d_G\left[\begin{array}{cc} x_1 & x_2 \\ x_4 & x_3 \end{array}\right]+u_1\left[\begin{array}{cc} x_1 & x_2 \\ y & z \end{array}\right]\theta_1+\underline{(x_1 + x_2 - y - z)}\theta_1^{\dagger}+u_2\left[\begin{array}{cc} x_1 & x_2 \\ y & z \end{array}\right]\theta_2+\underline{(x_1 x_2 - y z)}\theta_2^{\dagger}+ u_1\left[\begin{array}{cc} y & z \\ x_4 & x_3 \end{array}\right]\theta_3 + (y + z - x_3 - x_4)\theta_3^\dagger + u_2\left[\begin{array}{cc} y & z \\ x_4 & x_3 \end{array}\right]\theta_4 + (y z - x_3 x_4)\theta_4^\dagger$, we can reduce its cohomology: $H^*(D)\cong V_2 \otimes H^*(\tilde D)$, where $\tilde D = d_G\left[\begin{array}{cc} x_1 & x_2 \\ x_4 & x_3 \end{array}\right]+ u_1\left[\begin{array}{cc} x_1 & x_2 \\ x_4 & x_3 \end{array}\right]\theta_3 + (x_1 + x_2 - x_3 - x_4)\theta_3^\dagger + u_2\left[\begin{array}{cc} x_1 & x_2 \\ x_4 & x_3 \end{array}\right]\theta_4 + (x_1 x_2 - x_3 x_4)\theta_4^\dagger$.}\vspace{5pt} \\
		\hline
		\vspace{5pt}
		\begin{center}
			\quad \quad Categorified MOY III relation \vspace{3pt} \newline
			$\begin{array}{c}
				\begin{tikzpicture}[scale=0.7]
					\node[left] at (-1.5,0) {$\scriptstyle G$};
					\draw[thick, postaction={decorate},decoration={markings, mark= at position 0.9 with {\arrow{stealth}}}] (0,0.4) to[out=0,in=120] (1,-0.7);
					\draw[thick, postaction={decorate},decoration={markings, mark= at position 0.9 with {\arrow{stealth}}}] (0,-0.4) to[out=180,in=300] (-1,0.7);
					\draw[thick, -stealth] (-1,-0.7) to[out=60,in=180] (0,0.4) -- (0.1,0.4);
					\draw[thick, -stealth] (1,0.7) to[out=240,in=0] (0,-0.4) -- (-0.1,-0.4);
					\draw[fill=burgundy] (-0.63,0) circle (0.1) (0.63,0) circle (0.1);
					\draw[fill=\myblue, even odd rule] (0,0) circle (1.2) (0,0) circle (1.5);
					\node[right] at (-1,0.7) {$\scriptstyle x_1$};
					\node[left] at (1,-0.7) {$\scriptstyle x_2$};
					\node[left] at (1,0.7) {$\scriptstyle x_3$};
					\node[right] at (-1,-0.7) {$\scriptstyle x_4$};
					\node[above] at (0,0.4) {$\scriptstyle x_5$};
					\node[below] at (0,-0.4) {$\scriptstyle x_6$};
				\end{tikzpicture}
			\end{array}\cong \begin{array}{c}
			\begin{tikzpicture}[scale=0.7]
				\node[left] at (-1.5,0) {$\scriptstyle G$};
				\draw[thick, postaction={decorate},decoration={markings, mark= at position 0.9 with {\arrow{stealth}}}] (-1,-0.7) to[out=60,in=120] (1,-0.7);
				\draw[thick, postaction={decorate},decoration={markings, mark= at position 0.9 with {\arrow{stealth}}}] (1,0.7) to[out=240,in=300] (-1,0.7);
				\draw[fill=\myblue, even odd rule] (0,0) circle (1.2) (0,0) circle (1.5);
				\node[right] at (-1,0.7) {$\scriptstyle x_1$};
				\node[left] at (1,-0.7) {$\scriptstyle x_2$};
				\node[left] at (1,0.7) {$\scriptstyle x_1$};
				\node[right] at (-1,-0.7) {$\scriptstyle x_2$};
			\end{tikzpicture}
			\end{array} \oplus \left( \begin{array}{c}
			\begin{tikzpicture}[scale=0.7]
				\node[left] at (-1.5,0) {$\scriptstyle G$};
				\draw[thick, postaction={decorate},decoration={markings, mark= at position 0.9 with {\arrow{stealth}}}] (-1,-0.7) to[out=60,in=300] (-1,0.7);
				\draw[thick, postaction={decorate},decoration={markings, mark= at position 0.9 with {\arrow{stealth}}}] (1,0.7) to[out=240,in=120] (1,-0.7);
				\draw[fill=\myblue, even odd rule] (0,0) circle (1.2) (0,0) circle (1.5);
				\node[right] at (-1,0.7) {$\scriptstyle x_1$};
				\node[left] at (1,-0.7) {$\scriptstyle x_2$};
				\node[left] at (1,0.7) {$\scriptstyle x_2$};
				\node[right] at (-1,-0.7) {$\scriptstyle x_1$};
			\end{tikzpicture}
			\end{array}\otimes V_{N-2} \right)$
		\end{center} & {\centering \footnotesize Using the categorified MOY III relation, one can eliminate multiple antiparallel connected 4-valent vertices. Namely, given the differential $\MF=d_G\left[\begin{array}{cc}
			x_1 & x_3\\
			x_4 & x_2\\
		\end{array}\right]+u_1\left[\begin{array}{cc}
			x_1 & x_5\\
			x_4 & x_6\\
		\end{array}\right]\theta_1+(x_1 + x_5 - x_4 - x_6)\theta_1^{\dagger}+u_2\left[\begin{array}{cc}
			x_1 & x_5\\
			x_4 & x_6\\
		\end{array}\right]\theta_2+(x_1 x_5 - x_4 x_6)\theta_2^{\dagger}+u_1\left[\begin{array}{cc}
			x_2 & x_6\\
			x_3 & x_5\\
		\end{array}\right]\theta_3+(x_2 + x_6 - x_3 - x_5)\theta_3^{\dagger}+u_2\left[\begin{array}{cc}
			x_2 & x_6\\
			x_3 & x_5\\
		\end{array}\right]\theta_4+(x_2 x_6 - x_3 x_5)\theta_4^{\dagger}$, we can reduce its cohomology: $H^*(D)\cong H^*(d_-) \oplus \left( H^*(d_+) \otimes V_{N-2} \right)$, where $d_+ :=d_G\left[\begin{array}{cc}
		x_1 & x_2\\
		x_1 & x_2\\
		\end{array}\right]$ and $d_-:=d_G\left[\begin{array}{cc}
		x_1 & x_1\\
		x_2 & x_2\\
		\end{array}\right]$.} \\
		\hline
\end{tabular}\caption{\footnotesize Reductions proved in Section~\ref{sec:vert-red}. Here blue rings and disks labeled $G$ denote an arbitrary MOY graph so that the whole diagram is closed.}\label{tab:isom}
\end{table}

Then, in Section~\ref{sec:IR}, we demonstrate the horizontal reduction by the example of the Reidemeister I invariance. Section~\ref{sec:ex} provides examples of calculation of the Khovanov--Rozansky polynomials. These examples either illustrate straightforward calculation of the Khovanov--Rozansky polynomials via our differential operators approach or also demonstrates the use of different reductions. Then, we demonstrate that sometimes complicated KhR operator formalism can be substituted to Khovanov(-like) cycle calculus in Section~\ref{sec:Kh-like}. Namely, in Section~\ref{sec:N=2}, we prove that the local operator formalism for the Khovanov--Rozansky polynomials at $N=2$  reduces to the global technique suggested by Khovanov~\cite{khovanov2000categorification}. Section~\ref{sec:bip-red} shows that for bipartite links, one can use the relatively simple analagous cycle calculus from~\cite{2506.08721}.

\newpage 

\section{Prerequisites}\label{sec:prerequisites}

In this section, we introduce basics of constructions of the HOMFLY and the KhR polynomials (by our differential operators approach).

\subsection{HOMFLY polynomials from MOY diagrams}\label{sec:HOMFLY}

The HOMFLY polynomial is calculated in the Reshetikhin--Turaev approach using $\mathcal{R}$-matrices for the universal enveloping algebra $U_q(\mathfrak{sl}_N)$, which can be decomposed (in the fundamental representation) as follows:
\begin{equation}\label{R}
	\begin{aligned}
		&\qquad {\rm type} \ 0 \quad {\rm type} \ 1 \qquad \qquad \qquad \qquad \qquad \qquad \qquad {\rm type} \ 1 \quad {\rm type} \ 0
		\\
		\begin{array}{c}
			\begin{tikzpicture}[scale=0.5]
				\draw[thick,-stealth] (-0.5,-0.5) -- (0.5,0.5);
				\draw[thick,-stealth] (0.5,-0.5) -- (-0.5,0.5);
				\draw[fill=black] (0,0) circle (0.15);
				\node[left] at (-0.5,-0.5) {$\scriptstyle k$};
				\node[right] at (0.5,-0.5) {$\scriptstyle l$};
				\node[left] at (-0.5,0.5) {$\scriptstyle i$};
				\node[right] at (0.5,0.5) {$\scriptstyle j$};
			\end{tikzpicture}
		\end{array}:=\begin{array}{c}
			\begin{tikzpicture}[scale=0.5]
				\draw[thick,-stealth] (-0.5,-0.5) -- (0.5,0.5);
				\draw[thick,-stealth] (-0.1,0.1) -- (-0.5,0.5);
				\draw[thick] (0.5,-0.5) -- (0.1,-0.1);
			\end{tikzpicture}
		\end{array}&=\ q\begin{array}{c}
			\begin{tikzpicture}[scale=0.5]
				\draw[thick,-stealth] (-0.5,-0.5) to[out=45,in=270] (-0.15,0) to[out=90,in=315] (-0.5,0.5);
				\draw[thick,-stealth] (0.5,-0.5) to[out=135,in=270] (0.15,0) to[out=90,in=225] (0.5,0.5);
			\end{tikzpicture}
		\end{array}-\begin{array}{c}
			\begin{tikzpicture}[scale=0.5]
				\draw[thick,-stealth] (-0.5,-0.5) -- (0.5,0.5);
				\draw[thick,-stealth] (0.5,-0.5) -- (-0.5,0.5);
				\draw[fill=\myred] (0,0) circle (0.15);
			\end{tikzpicture}
		\end{array},\qquad
		\begin{array}{c}
			\begin{tikzpicture}[scale=0.5]
				\draw[thick,-stealth] (-0.5,-0.5) -- (0.5,0.5);
				\draw[thick,-stealth] (0.5,-0.5) -- (-0.5,0.5);
				\draw[fill=white] (0,0) circle (0.15);
				\node[left] at (-0.5,-0.5) {$\scriptstyle k$};
				\node[right] at (0.5,-0.5) {$\scriptstyle l$};
				\node[left] at (-0.5,0.5) {$\scriptstyle i$};
				\node[right] at (0.5,0.5) {$\scriptstyle j$};
			\end{tikzpicture}
		\end{array}:=\begin{array}{c}
			\begin{tikzpicture}[scale=0.5, xscale=-1]
				\draw[thick,-stealth] (-0.5,-0.5) -- (0.5,0.5);
				\draw[thick,-stealth] (-0.1,0.1) -- (-0.5,0.5);
				\draw[thick] (0.5,-0.5) -- (0.1,-0.1);
			\end{tikzpicture}
		\end{array} = \ q^{-1}\begin{array}{c}
			\begin{tikzpicture}[scale=0.5]
				\draw[thick,-stealth] (-0.5,-0.5) to[out=45,in=270] (-0.15,0) to[out=90,in=315] (-0.5,0.5);
				\draw[thick,-stealth] (0.5,-0.5) to[out=135,in=270] (0.15,0) to[out=90,in=225] (0.5,0.5);
			\end{tikzpicture}
		\end{array}-\begin{array}{c}
			\begin{tikzpicture}[scale=0.5]
				\draw[thick,-stealth] (-0.5,-0.5) -- (0.5,0.5);
				\draw[thick,-stealth] (0.5,-0.5) -- (-0.5,0.5);
				\draw[fill=\myred] (0,0) circle (0.15);
			\end{tikzpicture}
		\end{array}\,.
		\\
		\mathcal{R}^{ij}_{kl} &= \ q \cdot \delta^i_k \delta^j_l \ - \ {\cal M}_{kl}^{ij}\,, \qquad \qquad \qquad \quad \left(\mathcal{R}^{-1}\right)^{ij}_{kl}  \ = \ q^{-1} \cdot \delta^i_k \delta^j_l \ - \ {\cal M}_{kl}^{ij}\,.
	\end{aligned}
\end{equation}
The explicit non-zero elements of the operator ${\cal M}_{kl}^{ij}$ are
\begin{equation}
\begin{aligned}
	&\CM^{ji}_{ij} = -1\,,\quad \CM^{ij}_{ji} = -1\,,\quad \CM^{ij}_{ij} = q^{-1}\,,\quad \CM^{ji}_{ji} = q\,,\quad 1\leq i < j \leq N\,. \\
	&\begin{array}{c}
		\begin{tikzpicture}[scale=0.5]
			\draw[thick,-stealth,dashed] (-0.5,-0.5) -- (0.5,0.5);
			\draw[thick,-stealth] (0.5,-0.5) -- (-0.5,0.5);
		\end{tikzpicture}
	\end{array}\quad \quad \ \ \ \ \ \begin{array}{c}
		\begin{tikzpicture}[scale=0.5]
			\draw[thick,-stealth,dashed] (-0.5,-0.5) -- (0.5,0.5);
			\draw[thick,-stealth] (0.5,-0.5) -- (-0.5,0.5);
		\end{tikzpicture}
	\end{array}\quad \quad \ \ \ \ \ \begin{array}{c}
		\begin{tikzpicture}[scale=0.5]
			\draw[thick,-stealth] (-0.5,-0.5) to[out=45,in=270] (-0.15,0) to[out=90,in=315] (-0.5,0.5);
			\draw[thick,-stealth] (0.5,-0.5) to[out=135,in=270] (0.15,0) to[out=90,in=225] (0.5,0.5);
		\end{tikzpicture}
	\end{array}\quad \quad \ \ \ \ \ \begin{array}{c}
	\begin{tikzpicture}[scale=0.5]
		\draw[thick,-stealth] (-0.5,-0.5) to[out=45,in=270] (-0.15,0) to[out=90,in=315] (-0.5,0.5);
		\draw[thick,-stealth] (0.5,-0.5) to[out=135,in=270] (0.15,0) to[out=90,in=225] (0.5,0.5);
	\end{tikzpicture}
	\end{array}
\end{aligned}
\end{equation}
Note that subtracting the two relations~\eqref{R}, one gets the so-called HOMFLY {\it skein relation}:
\begin{equation}\label{skein}
\begin{aligned}
	\begin{array}{c}
		\begin{tikzpicture}[scale=0.5]
			\draw[thick,-stealth] (-0.5,-0.5) -- (0.5,0.5);
			\draw[thick,-stealth] (-0.1,0.1) -- (-0.5,0.5);
			\draw[thick] (0.5,-0.5) -- (0.1,-0.1);
		\end{tikzpicture}
	\end{array}&-\begin{array}{c}
	\begin{tikzpicture}[scale=0.5, xscale=-1]
		\draw[thick,-stealth] (-0.5,-0.5) -- (0.5,0.5);
		\draw[thick,-stealth] (-0.1,0.1) -- (-0.5,0.5);
		\draw[thick] (0.5,-0.5) -- (0.1,-0.1);
	\end{tikzpicture}
	\end{array}=(q-q^{-1}) \begin{array}{c}
	\begin{tikzpicture}[scale=0.5]
		\draw[thick,-stealth] (-0.5,-0.5) to[out=45,in=270] (-0.15,0) to[out=90,in=315] (-0.5,0.5);
		\draw[thick,-stealth] (0.5,-0.5) to[out=135,in=270] (0.15,0) to[out=90,in=225] (0.5,0.5);
	\end{tikzpicture}
	\end{array} \\
	{\cal R}^{ij}_{kl} & - ({\cal R}^{-1})^{ij}_{kl} = (q-q^{-1}) \delta^i_k \delta^j_l
\end{aligned}	
\end{equation}
The red vertex is called the {\it MOY vertex}, and we call all the resolutions of a link as {\it MOY graphs} or {\it MOY diagrams}. MOY diagrams satisfy {\it MOY relations}:
\begin{subequations}
	\begin{align}
		\label{MOYloc_II}&\begin{array}{c}
			\begin{tikzpicture}[scale=0.8]
				\draw[thick,postaction={decorate},decoration={markings, mark= at position 0.5 with {\arrow{stealth}}}] (0,0) circle (0.35);
			\end{tikzpicture}
		\end{array}= \ [N]\,, \\
		\label{MOYloc_III}&\begin{array}{c}
			\begin{tikzpicture}[scale=0.65]
				\draw[thick,stealth-,postaction={decorate},decoration={markings, mark= at position 0.7 with {\arrow{stealth}}}] (0,1) -- (0,0.5) to[out=180,in=90] (-0.5,0) to[out=270,in=180] (0,-0.5) -- (0,-1);
				\draw[fill=\myred] (-0.1,0.5) to[out=90,in=180] (0,0.6) to[out=0,in=90] (0.6,0) to[out=270,in=0] (0,-0.6) to[out=180,in=270] (-0.1,-0.5) to[out=90,in=180] (0,-0.4) to[out=0,in=270] (0.4,0) to[out=90,in=0] (0,0.4) to[out=180,in=270] (-0.1,0.5);
			\end{tikzpicture}
		\end{array}=\begin{array}{c}
			\begin{tikzpicture}
				\draw[thick, -stealth] (0,-0.5) -- (0,0.5);
			\end{tikzpicture}
		\end{array}\cdot \ [N-1]\,,\\
		\label{MOYloc_IV}&\begin{array}{c}
			\begin{tikzpicture}[scale=0.65]
				\begin{scope}[shift={(0,1)}]
					\draw[thick,-stealth] (-0.5,-0.5) to[out=90,in=270] (0.5,0.5);
					\draw[thick,-stealth] (0.5,-0.5) to[out=90,in=270] (-0.5,0.5);
					\draw[fill=\myred] (0,0) circle (0.1);
				\end{scope}
				\draw[thick,-stealth] (-0.5,-0.5) to[out=90,in=270] (0.5,0.5) -- (0.5,0.6);
				\draw[thick,-stealth] (0.5,-0.5) to[out=90,in=270] (-0.5,0.5) -- (-0.5,0.6);
				\draw[fill=\myred] (0,0) circle (0.1);
				\node[left] at (-0.5,-0.5) {$\scriptstyle i'$};
				\node[right] at (0.5,-0.5) {$\scriptstyle j'$};
				\node[left] at (-0.5,0.5) {$\scriptstyle k$};
				\node[right] at (0.5,0.5) {$\scriptstyle l$};
				\node[left] at (-0.5,1.5) {$\scriptstyle i$};
				\node[right] at (0.5,1.5) {$\scriptstyle j$};
			\end{tikzpicture}
		\end{array} = \begin{array}{c}
			\begin{tikzpicture}[scale=0.7]
				\draw[thick,-stealth] (-0.5,-0.5) to[out=90,in=270] (0.5,0.5);
				\draw[thick,-stealth] (0.5,-0.5) to[out=90,in=270] (-0.5,0.5);
				\draw[fill=\myred] (0,0) circle (0.1);
				\node[left] at (-0.5,-0.5) {$\scriptstyle i'$};
				\node[right] at (0.5,-0.5) {$\scriptstyle j'$};
				\node[left] at (-0.5,0.5) {$\scriptstyle i$};
				\node[right] at (0.5,0.5) {$\scriptstyle j$};
			\end{tikzpicture}
		\end{array}\cdot \ [2]\,,\\
		&\label{MOYloc_V}\begin{array}{c}
			\begin{tikzpicture}[scale=0.65]
				\draw[thick,postaction={decorate},decoration={markings, mark= at position 0.7 with {\arrow{stealth}}}] (-1,-1) -- (-0.5,-0.5);
				\draw[thick,postaction={decorate},decoration={markings, mark= at position 0.7 with {\arrow{stealth}}}] (1,1) -- (0.5,0.5);
				\draw[thick,postaction={decorate},decoration={markings, mark= at position 0.7 with {\arrow{stealth}}}] (-0.5,0.5) -- (-1,1);
				\draw[thick,postaction={decorate},decoration={markings, mark= at position 0.7 with {\arrow{stealth}}}] (0.5,-0.5) -- (1,-1);
				\draw[thick,postaction={decorate},decoration={markings, mark= at position 0.7 with {\arrow{stealth}}}] (-0.5,0.5) -- (0.5,0.5);
				\draw[thick,postaction={decorate},decoration={markings, mark= at position 0.7 with {\arrow{stealth}}}] (0.5,-0.5) -- (-0.5,-0.5);
				\begin{scope}[shift={(-0.5,0)}]
					\draw[fill=\myred] (0.1,0.5) to[out=90,in=0] (0,0.6) to[out=180,in=90] (-0.1,0.5) -- (-0.1,-0.5) to[out=270,in=180] (0,-0.6) to[out=0,in=270] (0.1,-0.5) -- (0.1,0.5);
				\end{scope}
				\begin{scope}[shift={(0.5,0)}]
					\draw[fill=\myred] (0.1,0.5) to[out=90,in=0] (0,0.6) to[out=180,in=90] (-0.1,0.5) -- (-0.1,-0.5) to[out=270,in=180] (0,-0.6) to[out=0,in=270] (0.1,-0.5) -- (0.1,0.5);
				\end{scope}
			\end{tikzpicture}
		\end{array}\;=\;\begin{array}{c}
			\begin{tikzpicture}
				\draw[thick, stealth-] (-0.5,0.5) to[out=315,in=225] (0.5,0.5);
				\draw[thick, stealth-] (0.5,-0.5) to[out=135,in=45] (-0.5,-0.5);
			\end{tikzpicture}
		\end{array} + \ \left(\begin{array}{c}
			\begin{tikzpicture}
				\draw[thick, stealth-] (-0.5,0.5) to[out=315,in=45] (-0.5,-0.5);
				\draw[thick, stealth-] (0.5,-0.5) to[out=135,in=225] (0.5,0.5);
			\end{tikzpicture}
		\end{array} \cdot \ [N-2]\right)\,,\\
		&\label{MOYloc_VI}\begin{array}{c}
			\begin{tikzpicture}[scale=0.55]
				\draw[thick,-stealth] (0,0) to[out=90,in=225] (1.5,1.5) to[out=45,in=270] (2,3);
				\draw[thick,-stealth] (2,0) to[out=90,in=315] (1.5,1.5) to[out=135,in=270] (0,3);
				\draw[thick,-stealth] (1,0) to[out=90,in=270] (0,1.5) to[out=90,in=270] (1,3);
				\draw[fill=\myred] (1.5,1.5) circle (0.13) (0.45,0.78) circle (0.13) (0.45,2.22) circle (0.13);
			\end{tikzpicture}
		\end{array}+ \ \begin{array}{c}
			\begin{tikzpicture}[scale=0.6]
				\draw[thick,-stealth] (0,0) -- (0,3);
				\draw[thick,-stealth] (1,0) to[out=90,in=270] (2,3);
				\draw[thick,-stealth] (2,0) to[out=90,in=270] (1,3);
				\draw[fill=\myred] (1.5,1.5) circle (0.13);
			\end{tikzpicture}
		\end{array}= \ \begin{array}{c}
			\begin{tikzpicture}[scale=0.6,xscale=-1]
				\draw[thick,-stealth] (0,0) to[out=90,in=225] (1.5,1.5) to[out=45,in=270] (2,3);
				\draw[thick,-stealth] (2,0) to[out=90,in=315] (1.5,1.5) to[out=135,in=270] (0,3);
				\draw[thick,-stealth] (1,0) to[out=90,in=270] (0,1.5) to[out=90,in=270] (1,3);
				\draw[fill=\myred] (1.5,1.5) circle (0.13) (0.45,0.78) circle (0.13) (0.45,2.22) circle (0.13);
			\end{tikzpicture}
		\end{array}+ \ \begin{array}{c}
			\begin{tikzpicture}[scale=0.6,xscale=-1]
				\draw[thick,-stealth] (0,0) -- (0,3);
				\draw[thick,-stealth] (1,0) to[out=90,in=270] (2,3);
				\draw[thick,-stealth] (2,0) to[out=90,in=270] (1,3);
				\draw[fill=\myred] (1.5,1.5) circle (0.13);
			\end{tikzpicture}
		\end{array}\,.
	\end{align}
\end{subequations}
The first one is just a consequence of quantum grading needed to produce quantum knot invariance. The MOY relations~\eqref{MOYloc_III},~\eqref{MOYloc_IV},~\eqref{MOYloc_VI} are consequences of Reidemeister invariance. The MOY relation~\eqref{MOYloc_V} follows from the HOMFLY skein relation~\eqref{skein} --- the same property of the fundamental $\CR$-matrix decomposability into symmetric and antisymmetric parts. For example, for the l.h.s. of the relation~\eqref{MOYloc_IV}, we have
\begin{equation}
\begin{aligned}
		{\cal M}^{ij}_{kl}\CM^{kl}_{i'j'} &= \left(q \delta^i_k \delta^j_l - \CR^{ij}_{kl}\right) \left(q^{-1} \delta^k_{i'} \delta^l_{j'} - (\CR^{-1})^{kl}_{i'j'}\right) = \\
		&= \underbrace{\delta^i_{i'}\delta^j_{j'} - q(\CR^{-1})^{ij}_{i'j'}}_{q \CM^{ij}_{i'j'}} \underbrace{- q^{-1} \CR^{ij}_{i'j'} + \overbrace{\CR^{ij}_{kl}(\CR^{-1})^{kl}_{i'j'}}^{\delta^i_{i'}\delta^j_{j'}}}_{q^{-1}\CM^{ij}_{i'j'}} = [2]\cdot\CM^{ij}_{i'j'}\,,
\end{aligned}
\end{equation}
and we arrive to the r.h.s. of~\eqref{MOYloc_IV}. Here, we use the consequence of the II Reidemeister move: $\CR^{ij}_{kl}(\CR^{-1})^{kl}_{i'j'}=\delta^i_{i'}\delta^j_{j'}$. 

It is also important to note that the projector to the antisymmetric representation $[1,1]$ is
\begin{equation}
	\hat {\cal P}_{[1,1]} = \frac{q\cdot \mathds{1} - \hat \CR }{[2]} = \frac{\hat \CM}{[2]}
\end{equation}
so that properties~\eqref{MOYloc_II}--\eqref{MOYloc_VI} become obvious. For example, the relation~\eqref{MOYloc_IV} is just the manifestation of a projector definition: $\hat {\cal P}_{[1,1]}^2 = \hat {\cal P}_{[1,1]}$.

\medskip

The HOMFLY polynomial (in the fundamental representation) for a link $\cal L\,$:
\begin{equation}
	H^{\cal L} = A^{-(n_\bullet - n_\circ)} \sum_{r\, \in \, {\rm resolutions}}^{2^n} (-1)^{\nu(r)} q^{\mu_\bullet(r) - \mu_\circ(r)} H^r
\end{equation}
where $n_\bullet$ and $n_\circ$ are the numbers of vertices in~\eqref{R}, $n = n_\bullet + n_\circ$ is the total number of crossings, $\nu(r)$ is the number of MOY vertices in a resolution $r$, $\mu_\bullet$ and $\mu_\circ$ are the numbers of $\,\resp\, $ resolutions coming from $\bullet$ and $\circ$ vertices respectively. 

We provide the example:
\begin{equation}
	\begin{aligned}
		\begin{array}{c}
			\begin{tikzpicture}[scale=0.55]
				\begin{scope}[shift={(0,1)}]
					\draw[thick] (0.5,-0.5) to[out=90,in=270] (-0.5,0.5);
					\draw[white, line width = 1.5mm] (-0.5,-0.5) to[out=90,in=270] (0.5,0.5);
					\draw[thick] (-0.5,-0.5) to[out=90,in=270] (0.5,0.5);
				\end{scope}
				\draw[thick,-stealth] (0.5,-0.5) to[out=90,in=270] (-0.5,0.5) -- (-0.5,0.6);
				\draw[white, line width = 1.5mm] (-0.5,-0.5) to[out=90,in=270] (0.5,0.5);
				\draw[thick,-stealth] (-0.5,-0.5) to[out=90,in=270] (0.5,0.5) -- (0.5,0.6);
				\draw[thick] (-0.5,1.5) to[out=90,in=90] (-1.2,1.5) -- (-1.2,-0.5) to[out=270,in=270] (-0.5,-0.5);
				\begin{scope}[xscale=-1]
					\draw[thick] (-0.5,1.5) to[out=90,in=90] (-1.2,1.5) -- (-1.2,-0.5) to[out=270,in=270] (-0.5,-0.5);
				\end{scope}
			\end{tikzpicture}
		\end{array}&=q^2\begin{array}{c}
			\begin{tikzpicture}[scale=0.55]
				\draw[thick] (0.5,-0.5) to[out=90,in=270] (0.3,0) to[out=90,in=270] (0.5,0.5) -- (0.5,0.6);
				\draw[thick] (-0.5,-0.5) to[out=90,in=270] (-0.3,0) to[out=90,in=270] (-0.5,0.5) -- (-0.5,0.6);
				\begin{scope}[shift={(0,1)}]
					\draw[thick] (0.5,-0.5) to[out=90,in=270] (0.3,0) to[out=90,in=270] (0.5,0.5);
					\draw[thick] (-0.5,-0.5) to[out=90,in=270] (-0.3,0) to[out=90,in=270] (-0.5,0.5);
				\end{scope}
				\draw[thick, postaction={decorate},decoration={markings, mark= at position 0.6 with {\arrow{stealth}}}] (-0.5,1.5) to[out=90,in=90] (-1.2,1.5) -- (-1.2,-0.5) to[out=270,in=270] (-0.5,-0.5);
				\begin{scope}[xscale=-1]
					\draw[thick, postaction={decorate},decoration={markings, mark= at position 0.6 with {\arrow{stealth}}}] (-0.5,1.5) to[out=90,in=90] (-1.2,1.5) -- (-1.2,-0.5) to[out=270,in=270] (-0.5,-0.5);
				\end{scope}
			\end{tikzpicture}
		\end{array}-q\begin{array}{c}
			\begin{tikzpicture}[scale=0.55]
				\draw[thick] (0.5,-0.5) to[out=90,in=270] (-0.5,0.5);
				\draw[thick] (-0.5,-0.5) to[out=90,in=270] (0.5,0.5);
				\begin{scope}[shift={(0,1)}]
					\draw[thick] (0.5,-0.5) to[out=90,in=270] (0.3,0) to[out=90,in=270] (0.5,0.5);
					\draw[thick] (-0.5,-0.5) to[out=90,in=270] (-0.3,0) to[out=90,in=270] (-0.5,0.5);
				\end{scope}
				\draw[thick, postaction={decorate},decoration={markings, mark= at position 0.6 with {\arrow{stealth}}}] (-0.5,1.5) to[out=90,in=90] (-1.2,1.5) -- (-1.2,-0.5) to[out=270,in=270] (-0.5,-0.5);
				\begin{scope}[xscale=-1]
					\draw[thick, postaction={decorate},decoration={markings, mark= at position 0.6 with {\arrow{stealth}}}] (-0.5,1.5) to[out=90,in=90] (-1.2,1.5) -- (-1.2,-0.5) to[out=270,in=270] (-0.5,-0.5);
				\end{scope}
				\draw[fill=\myred] (0,0) circle (0.13);
			\end{tikzpicture}
		\end{array}-q\begin{array}{c}
			\begin{tikzpicture}[scale=0.55]
				\draw[thick] (0.5,-0.5) to[out=90,in=270] (0.3,0) to[out=90,in=270] (0.5,0.5);
				\draw[thick] (-0.5,-0.5) to[out=90,in=270] (-0.3,0) to[out=90,in=270] (-0.5,0.5);
				\begin{scope}[shift={(0,1)}]
					\draw[thick] (0.5,-0.5) to[out=90,in=270] (-0.5,0.5);
					\draw[thick] (-0.5,-0.5) to[out=90,in=270] (0.5,0.5);
				\end{scope}
				\draw[thick, postaction={decorate},decoration={markings, mark= at position 0.6 with {\arrow{stealth}}}] (-0.5,1.5) to[out=90,in=90] (-1.2,1.5) -- (-1.2,-0.5) to[out=270,in=270] (-0.5,-0.5);
				\begin{scope}[xscale=-1]
					\draw[thick, postaction={decorate},decoration={markings, mark= at position 0.6 with {\arrow{stealth}}}] (-0.5,1.5) to[out=90,in=90] (-1.2,1.5) -- (-1.2,-0.5) to[out=270,in=270] (-0.5,-0.5);
				\end{scope}
				\draw[fill=\myred] (0,1) circle (0.13);
			\end{tikzpicture}
		\end{array}+\begin{array}{c}
			\begin{tikzpicture}[scale=0.55]
				\draw[thick,-stealth] (0.5,-0.5) to[out=90,in=270] (-0.5,0.5) -- (-0.5,0.6);
				\draw[thick,-stealth] (-0.5,-0.5) to[out=90,in=270] (0.5,0.5) -- (0.5,0.6);
				\begin{scope}[shift={(0,1)}]
					\draw[thick] (0.5,-0.5) to[out=90,in=270] (-0.5,0.5);
					\draw[thick] (-0.5,-0.5) to[out=90,in=270] (0.5,0.5);
				\end{scope}
				\draw[thick, postaction={decorate},decoration={markings, mark= at position 0.6 with {\arrow{stealth}}}] (-0.5,1.5) to[out=90,in=90] (-1.2,1.5) -- (-1.2,-0.5) to[out=270,in=270] (-0.5,-0.5);
				\begin{scope}[xscale=-1]
					\draw[thick, postaction={decorate},decoration={markings, mark= at position 0.6 with {\arrow{stealth}}}] (-0.5,1.5) to[out=90,in=90] (-1.2,1.5) -- (-1.2,-0.5) to[out=270,in=270] (-0.5,-0.5);
				\end{scope}
				\draw[fill=\myred] (0,0) circle (0.13) (0,1) circle (0.13);
			\end{tikzpicture}
		\end{array}\\
		H^{\rm Hopf} &= A^{-2}\left( q^2\cdot[N]^2 \ \ \ - \ \ \ 2 q \cdot [N][N-1] \ \ \ + \ \ \ [2][N][N-1] \right)
	\end{aligned}
\end{equation}
Note that we have 2 types of elementary resolutions and can organize link resolutions in a hypercube, see~\eqref{Hopf-ex-HOMFLY} as an example.

\paragraph{Examples.}

Here, we provide a correspondences between MOY graphs and their contributions into the HOMFLY polynomials which in Section~\ref{sec:examples} are generalized to the case of the Khovanov--Rozansky polynomials.

\begin{equation}
	 \begin{array}{c}
		\begin{tikzpicture}[scale=0.8]
		\draw[thick,-stealth] (0.5,-0.5) to[out=90,in=270] (-0.5,0.5);
		\draw[thick,-stealth] (-0.5,-0.5) to[out=90,in=270] (0.5,0.5);
		\draw[thick] (-0.5,0.5) to[out=90,in=90] (-0.8,0.5) -- (-0.8,-0.5) to[out=270,in=270] (-0.5,-0.5);
		\begin{scope}[xscale=-1]
			\draw[thick] (-0.5,0.5) to[out=90,in=90] (-0.8,0.5) -- (-0.8,-0.5) to[out=270,in=270] (-0.5,-0.5);
		\end{scope}
		\draw[fill=\myred] (0,0) circle (0.1);
	\end{tikzpicture}
\end{array}\overset{\eqref{MOYloc_II},\,\eqref{MOYloc_III}}{=}[N-1][N]\,,
\end{equation}
\begin{equation}
	\begin{array}{c}
		\begin{tikzpicture}[scale=0.7]
			\draw[thick,-stealth] (0.5,-0.5) to[out=90,in=270] (-0.5,0.5) -- (-0.5,0.6);
			\draw[thick,-stealth] (-0.5,-0.5) to[out=90,in=270] (0.5,0.5) -- (0.5,0.6);
			\begin{scope}[shift={(0,1)}]
				\draw[thick] (0.5,-0.5) to[out=90,in=270] (-0.5,0.5);
				\draw[thick] (-0.5,-0.5) to[out=90,in=270] (0.5,0.5);
			\end{scope}
			\draw[thick, postaction={decorate},decoration={markings, mark= at position 0.6 with {\arrow{stealth}}}] (-0.5,1.5) to[out=90,in=90] (-1.2,1.5) -- (-1.2,-0.5) to[out=270,in=270] (-0.5,-0.5);
			\begin{scope}[xscale=-1]
				\draw[thick, postaction={decorate},decoration={markings, mark= at position 0.6 with {\arrow{stealth}}}] (-0.5,1.5) to[out=90,in=90] (-1.2,1.5) -- (-1.2,-0.5) to[out=270,in=270] (-0.5,-0.5);
			\end{scope}
			\draw[fill=\myred] (0,0) circle (0.1) (0,1) circle (0.1);
		\end{tikzpicture}
	\end{array}\overset{\eqref{MOYloc_II},\,\eqref{MOYloc_III},\,\eqref{MOYloc_IV}}{=}[2][N-1][N]\,.
\end{equation}
For the Khovanov--Rozansky case, quantum numbers $[x]$ are substituted to vector spaces $V_x$ with $\dim_q V_x=[x]$.

\subsection{Categorification}\label{sec:categorification}

Our procedure of the categorifiction of the HOMFLY construction is as follows. 

\medskip

\noindent $\bullet$ We again deal with MOY diagrams but instead of indices, we put on edges even variables $x_b$, and in vertices we put odd (Grassman) variables $\theta_a$. 

\medskip

\noindent $\bullet$ With each (MOY) resolution of a diagram we associate an operator  $\hat D$, linear 
in $\theta$-variables, which is just a sum of $D$’s, associated with particular vertices.

\medskip

\noindent $\bullet$ For a closed link diagram $\hat D^2=0$.

\medskip	
	
\noindent $\bullet$ This allows to associate the cohomology of $\hat D$ with each resolution of a {\it closed} link diagram,
	we call it ``vertical’’ cohomology.
	
\medskip
	
\noindent $\bullet$ Morphisms between different resolutions induce ``horizontal’’ morphisms between 
	vertical cohomologies, and their {\it horizontal} cohomology is isomorphic to the KhR one
	based on matrix factorization approach~\cite{KRI}.
	
\medskip
	
\noindent $\bullet$ The generating Poincar\'e functional is topological invariant and defines the KhR invariant, which can be further promoted to a superpolynomial.

\subsubsection{Vertical morphisms}\label{sec:vert-morph}

\noindent $\bullet$ In the HOMFLY case, we assign tensors to MOY graphs, and in the KhR case, we assign morphisms being differential operators acting on a ring of polynomials in the even and odd variables with rational coefficients $\IQ[\theta_a,x_b]$. 

\medskip

\noindent $\bullet$ For elementary constituent MOY graphs, the categorifiction and the assigned objects are as follows. First, the categorified Kronecker operator is
\begin{equation}\label{CatKron}
\begin{aligned}
		\begin{array}{c}
		\begin{tikzpicture}
			\draw[thick, stealth-] (0,0) -- (1,0);
			\node[above] at (0,0) {$\scriptstyle i$};
			\node[above] at (1,0) {$\scriptstyle j$};
		\end{tikzpicture}
	\end{array} \quad &\longrightarrow \quad \begin{array}{c}
		\begin{tikzpicture}
			\draw[thick, stealth-] (0,0) -- (1,0);
			\node[above] at (0,0) {$\scriptstyle x$};
			\node[above] at (0.5,0) {$\scriptstyle \theta$};
			\node[above] at (1,0) {$\scriptstyle y$};
			\draw[fill=\mygreen] (0.5,0) circle (0.07);
		\end{tikzpicture}
	\end{array} \\
	\delta^i_j \quad \quad &\longrightarrow \quad D_{x,y} = \pi_{xy}\theta + (x-y) \theta^\dagger 
\end{aligned}
\end{equation}
where we recall that $\theta^\dag = \frac{\p}{\p \theta}$. For the $\,\resp\,$ resolution, we have:
\begin{equation}
\begin{aligned}
	\begin{array}{c}
		\begin{tikzpicture}[scale=0.65]
			\draw[thick,-stealth] (-0.5,-0.5) to[out=45,in=270] (-0.15,0) to[out=90,in=315] (-0.5,0.5);
			\draw[thick,-stealth] (0.5,-0.5) to[out=135,in=270] (0.15,0) to[out=90,in=225] (0.5,0.5);
			\node[left] at (-0.5,-0.5) {$\scriptstyle k$};
			\node[right] at (0.5,-0.5) {$\scriptstyle l$};
			\node[left] at (-0.5,0.5) {$\scriptstyle i$};
			\node[right] at (0.5,0.5) {$\scriptstyle j$};
		\end{tikzpicture}
	\end{array} \quad &\longrightarrow \quad \begin{array}{c}
	\begin{tikzpicture}[scale=0.65]
		\draw[thick, -stealth] (-0.5,-0.5) to[out=45,in=270] (-0.2,0) to[out=90,in=315] (-0.5,0.5);
		\draw[thick, -stealth] (0.5,-0.5) to[out=135,in=270] (0.2,0) to[out=90,in=225] (0.5,0.5);
		\draw[fill=\mygreen] (-0.2,0) circle (0.08) (0.2,0) circle (0.08);
		\node[above left] at (-0.5,0.5) {$\scriptstyle x_1$};
		\node[above right] at (0.5,0.5) {$\scriptstyle x_2$};
		\node[below right] at (0.5,-0.5) {$\scriptstyle x_4$};
		\node[below left] at (-0.5,-0.5) {$\scriptstyle x_3$};
		\node[left] at (-0.2,0) {$\scriptstyle \theta$};
		\node[right] at (0.2,0) {$\scriptstyle \eta$};
	\end{tikzpicture}
	\end{array} \\
	\delta^i_k \delta^j_l \quad \quad &\longrightarrow \quad D^{\smallresp} = \pi_{x_1 x_3}\theta + (x_1 - x_3) \theta^\dagger + \pi_{x_2 x_4}\eta + (x_2 - x_4) \eta^\dagger
\end{aligned}
\end{equation}
Operator $D^{\smallresp}$ is a sum of two independent pieces~\eqref{CatKron},
and we depict it by two green dots on the two propagators.
In fact, the resolution $\resp\,$, when descended to a cohomology, can be treated as an identical map,
and no dots are needed.
We will see in Section~\ref{sec:lin_red} that with our definitions they do not affect cohomologies ---
thus, can be either omitted or inserted, depending on the personal taste.
We insert them because elimination theorem in Section~\ref{sec:lin_red}
is a nice prototype of more complicated reductions,
and we use the chance to go first through this simpler example.

Second, for MOY vertex we have:
{\small \begin{equation}\label{CatMOY}
\begin{aligned}
	\begin{array}{c}
		\begin{tikzpicture}[scale=0.65]
			\draw[thick,-stealth] (-0.5,-0.5) -- (0.5,0.5);
			\draw[thick,-stealth] (0.5,-0.5) -- (-0.5,0.5);
			\draw[fill=\myred] (0,0) circle (0.1);
			\node[left] at (-0.5,-0.5) {$\scriptstyle k$};
			\node[right] at (0.5,-0.5) {$\scriptstyle l$};
			\node[left] at (-0.5,0.5) {$\scriptstyle i$};
			\node[right] at (0.5,0.5) {$\scriptstyle j$};
		\end{tikzpicture}
	\end{array} \quad &\longrightarrow \quad \begin{array}{c}
	\begin{tikzpicture}[scale=0.65]
		\draw[thick, -stealth] (-0.5,-0.5) -- (0.5,0.5);
		\draw[thick, -stealth] (0.5,-0.5) -- (-0.5,0.5);
		\draw[fill=\myred] (0,0) circle (0.1);
		\node[above left] at (-0.5,0.5) {$\scriptstyle x_1$};
		\node[above right] at (0.5,0.5) {$\scriptstyle x_2$};
		\node[below right] at (0.5,-0.5) {$\scriptstyle x_4$};
		\node[below left] at (-0.5,-0.5) {$\scriptstyle x_3$};
		\node[left] at (-0.1,0) {$\scriptstyle \theta$};
		\node[right] at (0.1,0) {$\scriptstyle \eta$};
	\end{tikzpicture}
	\end{array} \\
	\CM^{ij}_{kl} \quad \quad &\longrightarrow \quad D^{\smallresm}\left[\begin{array}{cc} x_1 & x_2 \\ x_3 & x_4\end{array}\right] = u_1 \left[\begin{array}{cc} x_1 & x_2 \\ x_3 & x_4\end{array}\right]\theta + (x_1 + x_2 - x_3 - x_4)\theta^\dag + u_2 \left[\begin{array}{cc} x_1 & x_2 \\ x_3 & x_4\end{array}\right]\eta + (x_1 x_2 - x_3 x_4) \eta^\dag
\end{aligned}
\end{equation}}

\noindent $\bullet$ For the categorification of the $\CR$-matrix, we should introduce other variables that carry $t$-degree. They should be odd in order to consume signs. The categorified $\CR$-matrices are 
\begin{equation}\label{categRvar}
	\begin{aligned}
		\begin{array}{c}
			\begin{tikzpicture}[scale=0.7]
				\draw[thick, -stealth] (0.5,-0.5) -- (-0.5,0.5);
				\draw[white, line width = 1.4mm] (-0.5,-0.5) -- (0.5,0.5);
				\draw[thick, -stealth] (-0.5,-0.5) -- (0.5,0.5);
				\node[above left] at (-0.5,0.5) {$\scriptstyle i$};
				\node[above right] at (0.5,0.5) {$\scriptstyle j$};
				\node[below right] at (0.5,-0.5) {$\scriptstyle l$};
				\node[below left] at (-0.5,-0.5) {$\scriptstyle k$};
			\end{tikzpicture}
		\end{array} = \ q\begin{array}{c}
			\begin{tikzpicture}[scale=0.7]
				\draw[thick,-stealth] (-0.5,-0.5) to[out=45,in=270] (-0.15,0) to[out=90,in=315] (-0.5,0.5);
				\draw[thick,-stealth] (0.5,-0.5) to[out=135,in=270] (0.15,0) to[out=90,in=225] (0.5,0.5);
				\node[above left] at (-0.5,0.5) {$\scriptstyle i$};
				\node[above right] at (0.5,0.5) {$\scriptstyle j$};
				\node[below right] at (0.5,-0.5) {$\scriptstyle l$};
				\node[below left] at (-0.5,-0.5) {$\scriptstyle k$};
			\end{tikzpicture}
		\end{array}-\begin{array}{c}
			\begin{tikzpicture}[scale=0.7]
				\draw[thick,-stealth] (-0.5,-0.5) -- (0.5,0.5);
				\draw[thick,-stealth] (0.5,-0.5) -- (-0.5,0.5);
				\draw[fill=\myred] (0,0) circle (0.1);
				\node[above left] at (-0.5,0.5) {$\scriptstyle i$};
				\node[above right] at (0.5,0.5) {$\scriptstyle j$};
				\node[below right] at (0.5,-0.5) {$\scriptstyle l$};
				\node[below left] at (-0.5,-0.5) {$\scriptstyle k$};
			\end{tikzpicture}
		\end{array} \quad \longrightarrow \quad & \begin{array}{c}
			\begin{tikzpicture}[scale=0.7]
				\draw[thick, -stealth] (0.5,-0.5) -- (-0.5,0.5);
				\draw[white, line width = 1.4mm] (-0.5,-0.5) -- (0.5,0.5);
				\draw[thick, -stealth] (-0.5,-0.5) -- (0.5,0.5);
				\node[above left] at (-0.5,0.5) {$\scriptstyle x_1$};
				\node[above right] at (0.5,0.5) {$\scriptstyle x_2$};
				\node[below right] at (0.5,-0.5) {$\scriptstyle x_3$};
				\node[below left] at (-0.5,-0.5) {$\scriptstyle x_4$};
			\end{tikzpicture}
		\end{array}=\begin{array}{c}
			\begin{tikzpicture}[scale=0.7]
				\draw[thick, -stealth] (-0.5,-0.5) to[out=45,in=270] (-0.2,0) to[out=90,in=315] (-0.5,0.5);
				\draw[thick, -stealth] (0.5,-0.5) to[out=135,in=270] (0.2,0) to[out=90,in=225] (0.5,0.5);
				\draw[fill=\mygreen] (-0.2,0) circle (0.08) (0.2,0) circle (0.08);
				\node[above left] at (-0.5,0.5) {$\scriptstyle x_1$};
				\node[above right] at (0.5,0.5) {$\scriptstyle x_2$};
				\node[below right] at (0.5,-0.5) {$\scriptstyle x_3$};
				\node[below left] at (-0.5,-0.5) {$\scriptstyle x_4$};
				\node[left] at (-0.2,0) {$\scriptstyle \theta_1$};
				\node[right] at (0.2,0) {$\scriptstyle \theta_2$};
			\end{tikzpicture}
		\end{array}+ \ \epsilon^{(+)}\begin{array}{c}
			\begin{tikzpicture}[scale=0.7]
				\draw[thick, -stealth] (-0.5,-0.5) -- (0.5,0.5);
				\draw[thick, -stealth] (0.5,-0.5) -- (-0.5,0.5);
				\draw[fill=\myred] (0,0) circle (0.1);
				\node[above left] at (-0.5,0.5) {$\scriptstyle x_1$};
				\node[above right] at (0.5,0.5) {$\scriptstyle x_2$};
				\node[below right] at (0.5,-0.5) {$\scriptstyle x_3$};
				\node[below left] at (-0.5,-0.5) {$\scriptstyle x_4$};
				\node[left] at (-0.1,0) {$\scriptstyle \theta_1$};
				\node[right] at (0.1,0) {$\scriptstyle \theta_2$};
			\end{tikzpicture}
		\end{array}\,;\\
		\begin{array}{c}
			\begin{tikzpicture}[scale=0.7]
				\draw[thick, -stealth] (-0.5,-0.5) -- (0.5,0.5);
				\draw[white, line width = 1.4mm] (0.5,-0.5) -- (-0.5,0.5);
				\draw[thick, -stealth] (0.5,-0.5) -- (-0.5,0.5);
				\node[above left] at (-0.5,0.5) {$\scriptstyle i$};
				\node[above right] at (0.5,0.5) {$\scriptstyle j$};
				\node[below right] at (0.5,-0.5) {$\scriptstyle l$};
				\node[below left] at (-0.5,-0.5) {$\scriptstyle k$};
			\end{tikzpicture}
		\end{array}= \ q^{-1}\begin{array}{c}
		\begin{tikzpicture}[scale=0.7]
			\draw[thick,-stealth] (-0.5,-0.5) to[out=45,in=270] (-0.15,0) to[out=90,in=315] (-0.5,0.5);
			\draw[thick,-stealth] (0.5,-0.5) to[out=135,in=270] (0.15,0) to[out=90,in=225] (0.5,0.5);
			\node[above left] at (-0.5,0.5) {$\scriptstyle i$};
			\node[above right] at (0.5,0.5) {$\scriptstyle j$};
			\node[below right] at (0.5,-0.5) {$\scriptstyle l$};
			\node[below left] at (-0.5,-0.5) {$\scriptstyle k$};
		\end{tikzpicture}
		\end{array}-\begin{array}{c}
		\begin{tikzpicture}[scale=0.7]
			\draw[thick,-stealth] (-0.5,-0.5) -- (0.5,0.5);
			\draw[thick,-stealth] (0.5,-0.5) -- (-0.5,0.5);
			\draw[fill=\myred] (0,0) circle (0.1);
			\node[above left] at (-0.5,0.5) {$\scriptstyle i$};
			\node[above right] at (0.5,0.5) {$\scriptstyle j$};
			\node[below right] at (0.5,-0.5) {$\scriptstyle l$};
			\node[below left] at (-0.5,-0.5) {$\scriptstyle k$};
		\end{tikzpicture}
		\end{array}\quad \longrightarrow \quad & \begin{array}{c}
			\begin{tikzpicture}[scale=0.7]
				\draw[thick, -stealth] (-0.5,-0.5) -- (0.5,0.5);
				\draw[white, line width = 1.4mm] (0.5,-0.5) -- (-0.5,0.5);
				\draw[thick, -stealth] (0.5,-0.5) -- (-0.5,0.5);
				\node[above left] at (-0.5,0.5) {$\scriptstyle x_1$};
				\node[above right] at (0.5,0.5) {$\scriptstyle x_2$};
				\node[below right] at (0.5,-0.5) {$\scriptstyle x_3$};
				\node[below left] at (-0.5,-0.5) {$\scriptstyle x_4$};
			\end{tikzpicture}
		\end{array}=\begin{array}{c}
			\begin{tikzpicture}[scale=0.7]
				\draw[thick, -stealth] (-0.5,-0.5) -- (0.5,0.5);
				\draw[thick, -stealth] (0.5,-0.5) -- (-0.5,0.5);
				\draw[fill=\myred] (0,0) circle (0.1);
				\node[above left] at (-0.5,0.5) {$\scriptstyle x_1$};
				\node[above right] at (0.5,0.5) {$\scriptstyle x_2$};
				\node[below right] at (0.5,-0.5) {$\scriptstyle x_3$};
				\node[below left] at (-0.5,-0.5) {$\scriptstyle x_4$};
				\node[left] at (-0.1,0) {$\scriptstyle \theta_1$};
				\node[right] at (0.1,0) {$\scriptstyle \theta_2$};
			\end{tikzpicture}
		\end{array}+ \ \epsilon^{(-)}\begin{array}{c}
			\begin{tikzpicture}[scale=0.7]
				\draw[thick, -stealth] (-0.5,-0.5) to[out=45,in=270] (-0.2,0) to[out=90,in=315] (-0.5,0.5);
				\draw[thick, -stealth] (0.5,-0.5) to[out=135,in=270] (0.2,0) to[out=90,in=225] (0.5,0.5);
				\draw[fill=\mygreen] (-0.2,0) circle (0.08) (0.2,0) circle (0.08);
				\node[above left] at (-0.5,0.5) {$\scriptstyle x_1$};
				\node[above right] at (0.5,0.5) {$\scriptstyle x_2$};
				\node[below right] at (0.5,-0.5) {$\scriptstyle x_3$};
				\node[below left] at (-0.5,-0.5) {$\scriptstyle x_4$};
				\node[left] at (-0.2,0) {$\scriptstyle \theta_1$};
				\node[right] at (0.2,0) {$\scriptstyle \theta_2$};
			\end{tikzpicture}
		\end{array}\,;\\
		&{\rm tdeg}\,\epsilon^{(\pm)}=+1\,.
	\end{aligned}
\end{equation}

\noindent $\bullet$ A vertical morphism for an arbitrary MOY graph $\Gamma$ (resolution of a tangle $L$ or link $\cal L$) is the sum of the constituent operators put at each vertex $v$:
\be\label{VD-lin-comb}
\hat D_\Gamma=\sum_{{\smallresp}\, \in \Gamma} \hat D^{\smallresp}_v + \sum_{{\smallresm}\, \in \Gamma} \hat D^{\smallresm}_v\,,\quad\mbox{so that}\quad \hat D_\Gamma^2 =: \omega(x_b)\,.
\ee
It is an operator only w.r.t. odd parameters $\theta_a$ --- contains $\theta_a$-derivatives $\theta_a^\dagger :=\frac{\p}{\p\theta_a}$.
Even variables $x_b$ enter without derivatives.
This operator is linear in $\theta_a$ and $\theta_a^\dagger$ but can be non-linear in $x_b$.
Polynomial $\omega(x_b)$ is often called  {\it superpotential}.

\medskip

\noindent $\bullet$ For a closed link $\cal L$, we would like to deal with differentials, i.e. $\hat D_\Gamma^2 = 0$ for all resolutions $\Gamma$ of $\cal L$. For this reason, we choose the superpotential to be
\begin{equation}
	\omega(x_b) = w(\text{outgoing edges}) - w(\text{incoming edges})\,. 
\end{equation}
The choice of Khovanov and Rozansky~\cite{KRI} is $w(x)=x^{N+1}$.

\medskip

\noindent $\bullet$ Thus, we get the relations on still undetermined functions $\pi_{xy}$, $u_{1,2}$. For~\eqref{CatKron}, we have: 
\begin{equation}
	\pi_{xy}(x-y) = x^{N+1} - y^{N+1} \quad \Lra \quad \pi_{xy} = \frac{x^{N+1} - y^{N+1}}{x-y}\,.
\end{equation}
For~\eqref{CatMOY}, we obtain:
\begin{equation}
	u_1 \left[\begin{array}{cc} x_1 & x_2 \\ x_3 & x_4\end{array}\right](x_1 + x_2 - x_3 - x_4) + u_2 \left[\begin{array}{cc} x_1 & x_2 \\ x_3 & x_4\end{array}\right](x_1 x_2 - x_3 x_4) = x_1^{N+1} + x_2^{N+1} - x_3^{N+1} - x_4^{N+1}\,.
\end{equation}
Again, the solution of~\cite{KRI} is
\begin{equation}
	\begin{aligned}
		& u_1\left[\begin{array}{cc} x_1 & x_2 \\ x_3 & x_4\end{array}\right]=\frac{g(x_1+x_2,x_1x_2)-g(x_3+x_4,x_1x_2)}{x_1+x_2-x_3-x_4}\,,\\
		& u_2\left[\begin{array}{cc} x_1 & x_2 \\ x_3 & x_4\end{array}\right]=\frac{g(x_3+x_4,x_1x_2)-g(x_3+x_4,x_3x_4)}{x_1x_2-x_3x_4}
	\end{aligned}
\end{equation}
where 
\begin{equation}
	g(p,q)=\left(\frac{1}{2}\left(p-\sqrt{p^2-4q}\right)\right)^{N+1}+\left(\frac{1}{2}\left(p+\sqrt{p^2-4q}\right)\right)^{N+1}\,.
\end{equation}

\noindent $\bullet$ All variables carry $q$-gradings so that quantum dimensions of cohomologies give values coming from relations~\eqref{MOYloc_II}--\eqref{MOYloc_VI} and coefficients in decomposition~\eqref{R}. This guarantees that the Euler characteristic of the Khovanov--Rozansky complex coincides with the HOMFLY polynomial. First, consider the simplest diagram: 
\begin{equation}\label{1-ferm-unknot}
	\begin{array}{c}
		\begin{tikzpicture}
			\draw[thick,postaction={decorate},decoration={markings, mark= at position 0.5 with {\arrow{stealth}}}] (0,0) circle (0.5);
			\node[right] at (0.5,0) {$\scriptstyle x$};
			\node[left] at (-0.51,0) {$\scriptstyle x$};
			\node[above] at (0,0.5) {$\scriptstyle \theta$};
			\draw[fill=\mygreen] (0,0.5) circle (0.06);
		\end{tikzpicture}
	\end{array}\,,\qquad D_{x,x}= \pi_{xx} \theta\,,\quad \pi_{xx} = (N+1)\,x^{N}\quad \Lra \quad H^*(D_{x,x}) = \IQ[x]/(x^N\IQ[x])\theta\,.
\end{equation}
The quantum dimension of this cohomology must be
\begin{equation}
	\dim_q H^*(D_{x,x}) = [N]
\end{equation}
what implies 
\begin{equation}\label{degs-x-theta-1-val}
	{\rm qdeg }\, x = 2\,,\quad {\rm qdeg}\,\theta = 1 - N\,.
\end{equation}
Second, from~\eqref{1-ferm-unknot},~\eqref{degs-x-theta-1-val}, we get ${\rm qdeg }\, D_{x,x} = N+1$. A generic vertical differential is a linear combination~\eqref{VD-lin-comb} of elementary morphisms, and we expect that it has well-defined $q$-grading. Thus,
\begin{equation}
	{\rm qdeg }\, D_{\Gamma} = N+1\,,
\end{equation}
and in particular, for~\eqref{CatMOY}:
\begin{equation}
	{\rm qdeg }\, D^{\smallresm} = N+1 \quad \Lra \quad {\rm qdeg}\,\theta = 1 - N\,, \quad {\rm qdeg}\,\eta = 3 - N\,.
\end{equation}
Third, consider 
{\small \begin{equation}
	\begin{array}{c}
		\begin{tikzpicture}[scale=0.8]
			\draw[thick,-stealth] (0.5,-0.5) to[out=90,in=270] (-0.5,0.5);
			\draw[thick,-stealth] (-0.5,-0.5) to[out=90,in=270] (0.5,0.5);
			\draw[thick] (-0.5,0.5) to[out=90,in=90] (-0.8,0.5) -- (-0.8,-0.5) to[out=270,in=270] (-0.5,-0.5);
			\begin{scope}[xscale=-1]
				\draw[thick] (-0.5,0.5) to[out=90,in=90] (-0.8,0.5) -- (-0.8,-0.5) to[out=270,in=270] (-0.5,-0.5);
			\end{scope}
			\node[right] at (-0.5,-0.5) {$\scriptstyle x$};
			\node[right] at (-0.5,0.5) {$\scriptstyle x$};
			\node[left] at (0.5,-0.5) {$\scriptstyle z$};
			\node[left] at (0.5,0.5) {$\scriptstyle z$};
			\node[left] at (-0.15,0) {$\scriptstyle \theta_1$};
			\node[right] at (0.25,0) {$\scriptstyle \theta_2$};
			\draw[fill=\myred] (0,0) circle (0.1);
		\end{tikzpicture}
	\end{array}\,,\; D(x,z\,|\,\theta_1,\theta_2) = u_1\left[\begin{array}{cc} x & z \\ x & z \end{array}\right]\theta_1 + u_2\left[\begin{array}{cc} x & z \\ x & z \end{array}\right]\theta_2\; \Lra \; H^*(D(x,z\,|\,\theta_1,\theta_2)) = \IQ[x,z]/(x^N\IQ[x],\,z^{N-1}\IQ[z])\theta_1 \theta_2\,,
\end{equation}}
see details in Section~\ref{sec:MOY-I-ex}. Then, we take
\begin{equation}
	{\rm qdeg }\, x = {\rm qdeg }\, z = 2\,,\quad {\rm qdeg}\,\theta_1 = 1 - N\,, \quad {\rm qdeg}\,\theta_2 = 3 - N
\end{equation}
and obtain
\begin{equation}
	\dim_q H^*(D(x,z\,|\,\theta_1,\theta_2)) = q^{N-1}[N] \cdot q^{N-2}[N-1] \cdot q^{1-N} \cdot q^{3-N} = q[N][N-1]\,.
\end{equation}
Thus, we must make 
\begin{equation}\label{qdeg-epsilons}
	{\rm qdeg }\, \epsilon^{(+)} = -2\,, \quad {\rm qdeg }\, \epsilon^{(-)} = 0\,.
\end{equation}
as it arises in the HOMFLY computations. Thus, we have computed gradings of all variables --- even ones, odd ones, coming from 2-valent and 4-valent vertices, and odd variables $\epsilon^{(\pm)}$.

%
%
%
%
%

\subsubsection{Horizontal morphisms}\label{sec:hor-morphisms}

In this subsection, we repeat the parts and the algorithm from~\cite{2508.05191} for another homomorphism. Horizontal morphisms are fixed \emph{uniquely} up to homotopy by locality and homomorphism requirement \eqref{MFhomo}.
See also \cite{Carqueville:2011zea}.

To describe horizontal morphisms as ring morphisms, it would be more natural to assign variables to MOY diagrams appearing in decomposition \eqref{categRvar} in a uniform way.
To do so, we assign even variables to edges of a very link diagram in question, and assign a couple of odd variables to both intersection resolutions via a single 4-valent vertex or a pair of 2-valent vertices.
Then, we would like to search for horizontal morphisms in terms of a uniform variable choice, there are two types of them:
\begin{equation}\label{categRvar}
	\begin{aligned}
		& \begin{array}{c}
			\begin{tikzpicture}[scale=0.7]
				\draw[thick, -stealth] (0.5,-0.5) -- (-0.5,0.5);
				\draw[white, line width = 1.4mm] (-0.5,-0.5) -- (0.5,0.5);
				\draw[thick, -stealth] (-0.5,-0.5) -- (0.5,0.5);
				\node[above left] at (-0.5,0.5) {$\scriptstyle x_1$};
				\node[above right] at (0.5,0.5) {$\scriptstyle x_2$};
				\node[below right] at (0.5,-0.5) {$\scriptstyle x_3$};
				\node[below left] at (-0.5,-0.5) {$\scriptstyle x_4$};
			\end{tikzpicture}
		\end{array}=\begin{array}{c}
			\begin{tikzpicture}[scale=0.7]
				\draw[thick, -stealth] (-0.5,-0.5) to[out=45,in=270] (-0.2,0) to[out=90,in=315] (-0.5,0.5);
				\draw[thick, -stealth] (0.5,-0.5) to[out=135,in=270] (0.2,0) to[out=90,in=225] (0.5,0.5);
				\draw[fill=\mygreen] (-0.2,0) circle (0.08) (0.2,0) circle (0.08);
				\node[above left] at (-0.5,0.5) {$\scriptstyle x_1$};
				\node[above right] at (0.5,0.5) {$\scriptstyle x_2$};
				\node[below right] at (0.5,-0.5) {$\scriptstyle x_3$};
				\node[below left] at (-0.5,-0.5) {$\scriptstyle x_4$};
				\node[left] at (-0.2,0) {$\scriptstyle \theta_1$};
				\node[right] at (0.2,0) {$\scriptstyle \theta_2$};
			\end{tikzpicture}
		\end{array}+\epsilon^{(+)}\begin{array}{c}
			\begin{tikzpicture}[scale=0.7]
				\draw[thick, -stealth] (-0.5,-0.5) -- (0.5,0.5);
				\draw[thick, -stealth] (0.5,-0.5) -- (-0.5,0.5);
				\draw[fill=\myred] (0,0) circle (0.1);
				\node[above left] at (-0.5,0.5) {$\scriptstyle x_1$};
				\node[above right] at (0.5,0.5) {$\scriptstyle x_2$};
				\node[below right] at (0.5,-0.5) {$\scriptstyle x_3$};
				\node[below left] at (-0.5,-0.5) {$\scriptstyle x_4$};
				\node[left] at (-0.1,0) {$\scriptstyle \theta_1$};
				\node[right] at (0.1,0) {$\scriptstyle \theta_2$};
			\end{tikzpicture}
		\end{array},\quad \chi^{(+)}\left[\begin{array}{cc|c}
			x_1 & x_2 & \theta_1\\
			x_4 & x_3 & \theta_2
		\end{array}\right]:\begin{array}{c}
			\begin{tikzpicture}[scale=0.7]
				\draw[thick, -stealth] (-0.5,-0.5) to[out=45,in=270] (-0.2,0) to[out=90,in=315] (-0.5,0.5);
				\draw[thick, -stealth] (0.5,-0.5) to[out=135,in=270] (0.2,0) to[out=90,in=225] (0.5,0.5);
				\draw[fill=\mygreen] (-0.2,0) circle (0.08) (0.2,0) circle (0.08);
				\node[above left] at (-0.5,0.5) {$\scriptstyle x_1$};
				\node[above right] at (0.5,0.5) {$\scriptstyle x_2$};
				\node[below right] at (0.5,-0.5) {$\scriptstyle x_3$};
				\node[below left] at (-0.5,-0.5) {$\scriptstyle x_4$};
				\node[left] at (-0.2,0) {$\scriptstyle \theta_1$};
				\node[right] at (0.2,0) {$\scriptstyle \theta_2$};
			\end{tikzpicture}
		\end{array}\;\longrightarrow\;\begin{array}{c}
			\begin{tikzpicture}[scale=0.7]
				\draw[thick, -stealth] (-0.5,-0.5) -- (0.5,0.5);
				\draw[thick, -stealth] (0.5,-0.5) -- (-0.5,0.5);
				\draw[fill=\myred] (0,0) circle (0.1);
				\node[above left] at (-0.5,0.5) {$\scriptstyle x_1$};
				\node[above right] at (0.5,0.5) {$\scriptstyle x_2$};
				\node[below right] at (0.5,-0.5) {$\scriptstyle x_3$};
				\node[below left] at (-0.5,-0.5) {$\scriptstyle x_4$};
				\node[left] at (-0.1,0) {$\scriptstyle \theta_1$};
				\node[right] at (0.1,0) {$\scriptstyle \theta_2$};
			\end{tikzpicture}
		\end{array}\,;\\
		& \begin{array}{c}
			\begin{tikzpicture}[scale=0.7]
				\draw[thick, -stealth] (-0.5,-0.5) -- (0.5,0.5);
				\draw[white, line width = 1.4mm] (0.5,-0.5) -- (-0.5,0.5);
				\draw[thick, -stealth] (0.5,-0.5) -- (-0.5,0.5);
				\node[above left] at (-0.5,0.5) {$\scriptstyle x_1$};
				\node[above right] at (0.5,0.5) {$\scriptstyle x_2$};
				\node[below right] at (0.5,-0.5) {$\scriptstyle x_3$};
				\node[below left] at (-0.5,-0.5) {$\scriptstyle x_4$};
			\end{tikzpicture}
		\end{array}=\begin{array}{c}
			\begin{tikzpicture}[scale=0.7]
				\draw[thick, -stealth] (-0.5,-0.5) -- (0.5,0.5);
				\draw[thick, -stealth] (0.5,-0.5) -- (-0.5,0.5);
				\draw[fill=\myred] (0,0) circle (0.1);
				\node[above left] at (-0.5,0.5) {$\scriptstyle x_1$};
				\node[above right] at (0.5,0.5) {$\scriptstyle x_2$};
				\node[below right] at (0.5,-0.5) {$\scriptstyle x_3$};
				\node[below left] at (-0.5,-0.5) {$\scriptstyle x_4$};
				\node[left] at (-0.1,0) {$\scriptstyle \theta_1$};
				\node[right] at (0.1,0) {$\scriptstyle \theta_2$};
			\end{tikzpicture}
		\end{array}+\epsilon^{(-)}\begin{array}{c}
			\begin{tikzpicture}[scale=0.7]
				\draw[thick, -stealth] (-0.5,-0.5) to[out=45,in=270] (-0.2,0) to[out=90,in=315] (-0.5,0.5);
				\draw[thick, -stealth] (0.5,-0.5) to[out=135,in=270] (0.2,0) to[out=90,in=225] (0.5,0.5);
				\draw[fill=\mygreen] (-0.2,0) circle (0.08) (0.2,0) circle (0.08);
				\node[above left] at (-0.5,0.5) {$\scriptstyle x_1$};
				\node[above right] at (0.5,0.5) {$\scriptstyle x_2$};
				\node[below right] at (0.5,-0.5) {$\scriptstyle x_3$};
				\node[below left] at (-0.5,-0.5) {$\scriptstyle x_4$};
				\node[left] at (-0.2,0) {$\scriptstyle \theta_1$};
				\node[right] at (0.2,0) {$\scriptstyle \theta_2$};
			\end{tikzpicture}
		\end{array},\quad \chi^{(-)}\left[\begin{array}{cc|c}
			x_1 & x_2 & \theta_1\\
			x_4 & x_3 & \theta_2
		\end{array}\right]:\begin{array}{c}
			\begin{tikzpicture}[scale=0.7]
				\draw[thick, -stealth] (-0.5,-0.5) -- (0.5,0.5);
				\draw[thick, -stealth] (0.5,-0.5) -- (-0.5,0.5);
				\draw[fill=\myred] (0,0) circle (0.1);
				\node[above left] at (-0.5,0.5) {$\scriptstyle x_1$};
				\node[above right] at (0.5,0.5) {$\scriptstyle x_2$};
				\node[below right] at (0.5,-0.5) {$\scriptstyle x_3$};
				\node[below left] at (-0.5,-0.5) {$\scriptstyle x_4$};
				\node[left] at (-0.1,0) {$\scriptstyle \theta_1$};
				\node[right] at (0.1,0) {$\scriptstyle \theta_2$};
			\end{tikzpicture}
		\end{array}\;\longrightarrow\;\begin{array}{c}
			\begin{tikzpicture}[scale=0.7]
				\draw[thick, -stealth] (-0.5,-0.5) to[out=45,in=270] (-0.2,0) to[out=90,in=315] (-0.5,0.5);
				\draw[thick, -stealth] (0.5,-0.5) to[out=135,in=270] (0.2,0) to[out=90,in=225] (0.5,0.5);
				\draw[fill=\mygreen] (-0.2,0) circle (0.08) (0.2,0) circle (0.08);
				\node[above left] at (-0.5,0.5) {$\scriptstyle x_1$};
				\node[above right] at (0.5,0.5) {$\scriptstyle x_2$};
				\node[below right] at (0.5,-0.5) {$\scriptstyle x_3$};
				\node[below left] at (-0.5,-0.5) {$\scriptstyle x_4$};
				\node[left] at (-0.2,0) {$\scriptstyle \theta_1$};
				\node[right] at (0.2,0) {$\scriptstyle \theta_2$};
			\end{tikzpicture}
		\end{array}\,;\\
		&{\rm qdeg}\,\epsilon^{(+)}=-2,\quad {\rm tdeg}\,\epsilon^{(+)}=+1,\quad {\rm qdeg}\,\epsilon^{(-)}=0,\quad {\rm tdeg}\,\epsilon^{(-)}=+1\,.
	\end{aligned}
\end{equation}
If the uniform choice of variables is made, all rings for various MOY diagrams $\Gamma$ are mutually isomorphic naturally.

One is able to derive expressions for morphisms $\chi^{(\pm)}$ explicitly based on the three properties.
\begin{enumerate}
	\item {\bf Parity.} Horizontal differential \eqref{hor-diff} is expected to be an odd, fermionic operator, so that its nilpotency is a natural property.
	Since $\epsilon^{(\pm)}$ are fermions, it implies that morphisms $\chi^{(\pm)}$ are even linear operators.
	
	\item {\bf Degree.} We expect that morphisms $\chi^{(\pm)}$ have a fixed $q$-degree. The horizontal differential acting between some adjacent ``slices'' of a complex is
	\begin{equation}\label{hor-diff}
		\fD = \sum_v \epsilon^{\alpha(v)}_v \chi^{\alpha(v)}_v
	\end{equation}
	where the sum goes over vertices $v$ of the corresponding ``slice'' of a complex and $\alpha(v)=(\pm)$. Assume that an element $\psi\not\in{\rm ker}\,\fD$ and consider a simple complex:
	\begin{equation}\label{zero-compl}
		\left[\begin{array}{c}
			\begin{tikzpicture}
				\node (A) at (0,0) {$0$};
				\node (B) at (2,0) {$\psi$};
				\node (C) at (4,0) {$\fD\psi$};
				\node (D) at (6,0) {$0$};
				\path (A) edge[->] (B) (B) edge[->] node[above] {$\scriptstyle \fD$} (C) (C) edge[->] (D);
			\end{tikzpicture}
		\end{array}\right]\,.
	\end{equation}
	Its Euler characteristic reads $q^{{\rm qdeg}\,\psi}(1-q^{{\rm qdeg}\,\fD})$. The complex~\eqref{zero-compl} is homotopic to zero. We expect naturally that the Euler characteristic is an invariant and equals to 0 for a zero complex. This reasoning imposes a constraint ${\rm qdeg }\,\fD=0$. Taking into account $q$-gradings of $\epsilon^{(\pm)}$~\eqref{qdeg-epsilons}, we obtain 
	\begin{equation}
		{\rm qdeg }\,\chi^{(+)} = +2\,,\quad {\rm qdeg }\,\chi^{(-)} = 0\,.
	\end{equation}
	\item {\bf Locality.} This property seems to be the most important singling out the KhR construction among the others.
	We assume that for the uniform choice of variables, morphisms $\chi^{(\pm)}$ are local: they depend only on variables $x_1$, $x_2$, $x_3$, $x_4$, $\theta_1$ and $\theta_2$ assigned to a particular intersection \eqref{categRvar}.
\end{enumerate}

Let us calculate morphism $\chi^{(-)}$ explicitly.
First we separate dependence on $\theta_{1,2}$ in the ring basis:
\begin{equation}
	\overset{V_0 \circlearrowleft}{\underset{{\rm parity}=+1}{\left(R\oplus R_{12}\theta_1\theta_2\right)}}\oplus\overset{V_1 \circlearrowleft}{\underset{{\rm parity}=-1}{\left(R_1\theta_1\oplus R_2\theta_2\right)}}\,,
\end{equation}
where $R$, $R_1$, $R_2$, $R_{12}$ are rings of variables $x_i$ and odd variables coming from other intersections.
We combine these subrings in pairs according to their parity.
Since $\chi^{(-)}$ preserves parity, a generic $4\times 4$ linear transform $\chi^{(-)}$ on this space is split in two $2\times 2$ matrix blocks $V_0$ and $V_1$.
We can naturally represent such a map in the form of a differential operator:
\begin{equation}
	\begin{aligned}
		\chi^{(-)}&=\left(V_0\right)_{11}\hat\theta_1^{\dagger}\hat\theta_1\hat\theta_2^{\dagger}\hat\theta_2+\left(V_0\right)_{22}\hat\theta_1\hat\theta_1^{\dagger}\hat\theta_2\hat\theta_2^{\dagger}+\left(V_0\right)_{21}\hat\theta_1\hat\theta_2+\left(V_0\right)_{12}\hat\theta_2^{\dagger}\hat\theta_1^{\dagger}+\\
		&+\left(V_1\right)_{11}\hat\theta_1\hat\theta_1^{\dagger}\hat\theta_2^{\dagger}\hat\theta_2 + \left(V_1\right)_{12}\hat\theta_1\hat\theta_2^{\dagger}+\left(V_1\right)_{21}\hat\theta_2\hat\theta_1^{\dagger} +\left(V_1\right)_{22} \hat\theta_1^{\dagger}\hat\theta_1\hat\theta_2\hat\theta_2^{\dagger}\,.
	\end{aligned}
\end{equation}
Matrix elements of these matrices are polynomials in variables $x_{1,2,3,4}$ of fixed degree due to the $q$-degree constraint:
\begin{equation}
	{\rm qdeg}\,V_0=\left(\begin{array}{cc}
		0 & 2(2-N)\\
		2(N-1)  & 2\\
	\end{array}\right),\quad {\rm qdeg}\,V_1=\left(\begin{array}{cc}
		0 & 2\\
		0  & 2\\
	\end{array}\right)\,.
\end{equation}
Let us assign respective variable homogeneous polynomials to these matrix elements:
\begin{equation}
	V_0=\left(\begin{array}{cc}
		r_3 & 0\\
		g_1  & p_1 \\
	\end{array}\right),\quad V_1=\left(\begin{array}{cc}
		r_1 & p_2 \\
		r_2  & p_3 \\
	\end{array}\right)\,,
\end{equation}
where polynomials $p_{1,2,3}$ are of degree 1, polynomial $g_1$ is of degree $N-1$, and $r_{1,2,3}$ are of degree 0, in other words, the latter are just numbers.
The matrix element $(V_0)_{12}$ has a negative degree for generic $N>2$, therefore it must be 0.

We describe local vertical morphisms for the two resolutions in the first line of \eqref{categRvar} explicitly as:
\begin{equation}\label{MFs}
	\begin{aligned}
		&\MF^{\smallresp}=\pi_{14}\theta_1+\underbrace{\left(x_1-x_4\right)}_{x_{14}}\theta_1^{\dagger}+\pi_{23}\theta_2+\underbrace{\left(x_2-x_3\right)}_{x_{23}}\theta_2^{\dagger}\,,\\
		&\MF^{\smallresm}=u_1\theta_1+\underbrace{\left(x_1+x_2-x_3-x_4\right)}_{s_1}\theta_1^{\dagger}+u_2\theta_2+\underbrace{\left(x_1x_2-x_3x_4\right)}_{s_2}\theta_2^{\dagger}\,,
	\end{aligned}
\end{equation}
where
\begin{equation}
	\pi_{ij} = \frac{x_i^{N+1} - x_j^{N+1}}{x_i - x_j}\,.
\end{equation}
For these operators, homomorphism property descends to the following equation:
\begin{equation}
	\chi^{(-)}\MF^{\smallresm}=\MF^{\smallresp}\,\chi^{(-)}\,.
\end{equation}
This relation produces a collection of subconstraints:
\begin{subequations}
	\begin{align}
		& r_3 s_1 - r_1 x_{14} - r_2 x_{23} = 0\,, \label{sub_a} \\
		& r_3 s_2 - p_2 x_{14} - p_3 x_{23} = 0\,, \label{sub_b} \\
		& r_1 u_1 + p_2 u_2 + g_1 x_{23} - r_3 \pi _{14} = 0\,, \label{sub_c} \\
		& p_2 s_{1} - r_1 s_{2} + p_1 x_{23} = 0\,, \label{sub_d} \\
		& r_2 u_{1} + p_3 u_{2} - g_1 x_{14} - r_3 \pi_{23} = 0\,, \label{sub_e} \\
		& p_3 s_{1} - r_2 s_2 - p_1 x_{14} = 0\,, \label{sub_f} \\
		& g_1 s_{1} - p_1 u_2 - r_2 \pi_{14} + r_1 \pi _{23} = 0\,, \label{sub_g} \\
		& g_1 s_2 + p_1 u_1 - p_3 \pi_{14} + p_2 \pi_{23} = 0\,. \label{sub_h}
	\end{align}
\end{subequations}
Let us start with \eqref{sub_a}. This equation has a solution only if $r_3 \neq 0$. A multiplicative constant of a differential does not affect the quantum dimension of the corresponding cohomology. Thus, for convenience, we set $r_3 = 1$. Since $s_1=x_{14}+x_{23}$ this constraint fixes uniquely $r_1=1$, $r_2=1$.
After substituting this solution, from \eqref{sub_d}, \eqref{sub_f} we re-express $p_2$ and $p_3$ in terms of $p_1$:
\begin{equation}\label{p2,p3}
	p_2=\frac{s_2 - p_1 x_{23}}{s_1}\,,\quad p_3=\frac{s_2 + p_1 x_{14}}{s_1}\,.
\end{equation}
However, for these relations to make sense, the pole in the denominators should be canceled.
This imposes constraints on $p_1$:
\begin{equation}
	\left(s_2 - p_1 x_{23}\right)\big|_{s_1=0}=\left(s_2 + p_1 x_{14}\right)\big|_{s_1=0}=0\,.
\end{equation}
Then, we find:
\begin{equation}
	p_1\big|_{x_2=x_3+x_4-x_1}=\frac{s_2}{x_{23}}\bigg|_{x_2=x_3+x_4-x_1} = -\frac{s_2}{x_{14}}\bigg|_{x_2=x_3+x_4-x_1} = x_1 - x_3\,.
\end{equation}
The most generic solution to this constraint for $p_1$ is the following:
\begin{equation}
	p_1 = (x_1-x_3)-\lambda\, s_1\,.
\end{equation}
And we have found a generic solution indeed since different values of $\lambda$ correspond to homotopies of the double complex.

Now, we substitute this solution into~\eqref{p2,p3} and get
\begin{equation}
	p_2 = x_3 + \lambda x_{23}\,, \quad p_3 = x_1 - \lambda x_{14}\,.
\end{equation}
Then, for example, from~\eqref{sub_e}, we can express the $g_1$ function:
\begin{equation}
	g_1 = \frac{u_1 + p_3 u_2 - \pi_{23}}{x_{14}} = \frac{u_1 + x_1 u_2 - \pi_{23}}{x_{14}} - \lambda u_2\,.
\end{equation} 
The remaining equations are satisfied identically.

Dealing in the same way with $\chi^{(+)}$ (see the detailed derivation in~\cite{2508.05191}), eventually we would arrive to the following morphisms as in \cite{KRI} with free parameters $\mu$ and $\lambda$:
\begin{equation}\label{KRmorphisms}
	\begin{aligned}
		&\chi^{(+)} = \left(\mu  \left(x_1+x_2-x_3-x_4\right)-x_2+x_4\right)\theta_1^{\dagger}\theta_1\theta_2^{\dagger}\theta_2+\theta_1\theta_1^{\dagger}\theta_2\theta_2^{\dagger}-\left((\mu-1)u_2+\frac{u_1+x_1u_2-\pi_{23}}{x_1-x_4} \right)\theta_1\theta_2+\\
		& +\left(x_4+\mu(x_1-x_4) \right)\theta_1\theta_1^{\dagger}\theta_2^{\dagger}\theta_2 + \left( \mu(x_2-x_3)-x_2\right)\theta_1\theta_2^{\dagger}-\theta_2\theta_1^{\dagger} + \theta_1^{\dagger}\theta_1\theta_2\theta_2^{\dagger}\,,\\
		&\chi^{(-)} = \theta_1^{\dagger}\theta_1\theta_2^{\dagger}\theta_2 +\left(\lambda\left(x_3+x_4-x_1-x_2\right)+x_1-x_3\right)\theta_1\theta_1^{\dagger}\theta_2\theta_2^{\dagger} -\left(\lambda u_2+\frac{u_1+x_1u_2-\pi_{23}}{x_4-x_1}\right)\theta_1\theta_2+\\
		& +\theta_1\theta_1^{\dagger}\theta_2^{\dagger}\theta_2  +\left( x_3+\lambda\left(x_2-x_3\right)\right)\theta_1\theta_2^{\dagger}+
		\theta_2\theta_1^{\dagger} +\left( x_1+\lambda\left(x_4-x_1\right)\right)\theta_1^{\dagger}\theta_1\theta_2\theta_2^{\dagger}\,.
	\end{aligned}
\end{equation}

\subsubsection{Khovanov--Rozansky complex
\label{choiceofdiffs}}

\noindent $\bullet$ Now, we have defined vertical and horizontal morphisms and obtain the following bicomplex:
\begin{equation}\label{KRbicompl}
	\begin{array}{c}
		\begin{tikzpicture}
			\node(A) at (0,0) {$R_\epsilon^{*,*}(\MOY_j)$};
			\node(B) at (4,0) {$R_\epsilon^{*,*+1}(\MOY_{j+1})$};
			\node(C) at (8,0) {$R_\epsilon^{*,*+2}(\MOY_{j+2})$};
			\node(E) at (-3,0) {$\ldots$};
			\node(F) at (11,0) {$\ldots$};
			\node(A1) at (0,-1.5) {$R_\epsilon^{*+d,*}(\MOY_j)$};
			\node(B1) at (4,-1.5) {$R_\epsilon^{*+d,*+1}(\MOY_{j+1})$};
			\node(C1) at (8,-1.5) {$R_\epsilon^{*+d,*+2}(\MOY_{j+2})$};
			\node(E1) at (-3,-1.5) {$\ldots$};
			\node(F1) at (11,-1.5) {$\ldots$};
			\path (E) edge[->] (A) (A) edge[->] node[above] {$\scriptstyle \fD_j$} (B) (B) edge[->]  node[above] {$\scriptstyle \fD_{j+1}$}  (C) (C) edge[->] (F) (E1) edge[->] (A1) (A1) edge[->] node[above] {$\scriptstyle \fD_j$} (B1) (B1) edge[->]  node[above] {$\scriptstyle \fD_{j+1}$}  (C1) (C1) edge[->] (F1) (A) edge[->] node[right] {$\scriptstyle \MF_{\MOY_j}$} (A1) (B) edge[->] node[right] {$\scriptstyle \MF_{\MOY_{j+1}}$} (B1) (C) edge[->] node[right] {$\scriptstyle \MF_{\MOY_{j+2}}$} (C1);
		\end{tikzpicture}
	\end{array}
\end{equation}
Here $R_\epsilon^{*,*}(\MOY)$ is a ring of variables $x_k$, $\theta_i$, $\epsilon_j^{\alpha}$ assigned to a MOY diagram $\Gamma$. The first $*$ in the superscript indicates the $q$-grading while the second one indicates the $t$-grading, and $d = N+1$.

\medskip

\noindent $\bullet$ To calculate the Poincar\'e polynomial of a double complex, one descends first on vertical cohomologies
and then considers a new induced mono-complex:
\begin{equation}\label{horizcompl}
	\begin{array}{c}
		\begin{tikzpicture}
			\node(A) at (0,0) {$H^{j}\left(\MF_{ \MOY_j}\right)$};
			\node(B) at (4,0) {$H^{j+1}\left(\MF_{\MOY_{j+1}}\right)$};
			\node(C) at (8,0) {$H^{j+2}\left(\MF_{\MOY_{j+2}}\right)$};
			\node(E) at (-2.5,0) {$\ldots$};
			\node(F) at (10.5,0) {$\ldots$};
			\draw[->] (E) -- (A);
			\draw[->] (A) -- (B) node[pos=0.5,above] {$\scriptstyle \tilde\fD_j$};
			\draw[->] (B) -- (C) node[pos=0.5,above] {$\scriptstyle \tilde\fD_{j+1}$};
			\draw[->] (C) -- (F);
		\end{tikzpicture}
	\end{array}
\end{equation}

\noindent $\bullet$ Khovanov-Rozansky categorified HOMFLY polynomial is the Poincar\'e polynomial of this resulting complex:
\begin{equation}
	\KhR(A,q,T)=\sum\lm_{f} T^f\;{\rm dim}_q\,H^{f}(\tilde \fD_f)\,.
\end{equation}
The HOMFLY polynomial corresponds to $\KhR(A,q,-1)$.

\subsection{Summary}

Let us briefly outline the main points of the constructions from Section~\ref{sec:HOMFLY} and~\ref{sec:categorification}. The HOMFLY polynomial is a sum of vertices contributions over a hypercube. In the KhR case being the categorification of the HOMFLY one, these contributions are substituted to vector spaces being vertical cohomologies, and summation of these contributions is substituted to taking cohomologies of the resulting horizontal complex. 

The question is: How to compute these vertical and horizontal cohomologies? It turns out that this computation can be simplified using vertical (Section~\ref{sec:vert-red}) and horizontal (see the example in Section~\ref{sec:IR}) reductions. Afterwards, in the $N=2$ and the bipartite cases, one can also perform the reduction to the Khovanov(-like) monocomplex, see Section~\ref{sec:Kh-like}.

There are two elementary vertical reductions, see Section~\ref{sec:elem_red} --- these are linear and quadratic reductions. They, in turn, produce some of the categorified MOY moves in Section~\ref{sec:cat-MOY}. Using these vertical reductions, one calculates vertical cohomologies. 

Horizontal reductions always work the same way. Namely, one just breaks the horizontal complex of vertical cohomologies into a direct sum that includes exact subcomplexes that do not contribute to horizontal cohomologies.

\section{Vertical reductions}\label{sec:vert-red}

In this section, we deal with vertical reductions. Their types are described in Sections~\ref{sec:elem_red} and~\ref{sec:cat-MOY}. In Section~\ref{sec:examples}, we provide examples of these reductions and prove some of them in the subsequent Sections~\ref{sec:lin_red}--\ref{sec:box-2}.

\subsection{Two elementary reductions}\label{sec:elem_red}

After the operators $\hat D_\Gamma$ are defined by summing their single-vertex prototypes over all vertices~\eqref{VD-lin-comb},
the next step is to calculate their cohomologies.
This is done by a kind of inverse procedure --- eliminating vertices from a MOY graph $\Gamma$ and reducing cohomologies
to those for simpler diagrams.
Elimination of vertices is named {\it reduction}, and there are two {\it elementary} reductions,
corresponding to the elimination of $\,\respexp\,$ and $\,\resm\,$.
Actually in the case of $\,\respexp\,$, this is further split in two identical steps,
and the term {\it elementary} will refer to each of them.
With a reference to the power of functions $B$ and $E$~\eqref{Dresp-Dresm-gen} in $x$, the two elementary reductions are named the linear and quadratic reductions.
While the {\bf linear} one is a kind of trivial --- cohomologies are not changed, the {\bf quadratic} one is not ---
original cohomologies are not quite expressed through cohomologies for simpler diagrams,
the relation is a little more involved.

We denote the rest of a diagram by $G$. This remaining graph $G$ also contains a set of variables but we do not indicate it explicitly.
Then, the two cases to be analyzed are depicted as

\begin{equation}\label{linear_reduction}
	\begin{array}{c}
		\begin{tikzpicture}[scale=1.2]
			\draw[thick, postaction={decorate},decoration={markings, mark= at position 0.75 with {\arrow{stealth}}, mark= at position 0.25 with {\arrow{stealth}}}] (-0.5,-0.5) to[out=45,in=315] (-0.5,0.5);
			\draw[thick, postaction={decorate},decoration={markings, mark= at position 0.75 with {\arrow{stealth}}, mark= at position 0.25 with {\arrow{stealth}}}] (0.5,-0.5) to[out=135,in=225] (0.5,0.5);
			\node at (-0.2,0.45) {$\scriptstyle x_1$};
			\node at (-0.2,-0.45) {$\scriptstyle x_3$};
			\node at (0.2,0.45) {$\scriptstyle x_2$};
			\node at (0.2,-0.45) {$\scriptstyle x_4$};
			\draw[fill=\mygreen] (-0.3,0) circle (0.05);
			\draw[fill=\mygreen] (0.3,0) circle (0.05);
			\node at (-0.5,0) {$\scriptstyle \theta_1$};
			\node at (0.5,0) {$\scriptstyle \theta_2$};
			\begin{scope}[rotate=45]
				\node at (1.1,0) {$\scriptstyle G$};
			\end{scope}
			\draw[fill=\myblue, even odd rule] (0,0) circle (0.7)  (0,0) circle (0.9);
		\end{tikzpicture}
	\end{array}\quad\quad
	\begin{aligned}
		&\hat D^{\smallresp}\left[\begin{array}{cc} x_1 & x_2 \\ x_3 & x_4\end{array}\right]
=\hat d_G\left[\begin{array}{cc} x_1 & x_2 \\ x_3 & x_4\end{array}\right] + \pi_{13} \theta_1 + \underline{(x_1 - x_3)} \theta_1^\dag + \pi_{24} \theta_2 + \underline{(x_2 - x_4)} \theta_2^\dag\,, 
\\ 
&\pi_{ij} = \frac{x_i^{N+1} - x_j^{N+1}}{x_i - x_j}\,,
\\
		&\hat d_G^{\,2}= - \pi_{13}(x_1 - x_3) - \pi_{24}(x_2 - x_4)\,, 
	\end{aligned}
\end{equation}

and

\begin{equation}\label{quad_reduction}
	\begin{array}{c}
		\begin{tikzpicture}[scale=1.2]
			\node[left] at (-1,0) {$\scriptstyle G$};
			\draw[fill=\myblue] (0,0) circle (1);
			\draw[fill=white] (0,0) circle (0.7);
			\draw[thick, postaction={decorate},decoration={markings, mark= at position 0.75 with {\arrow{stealth}}, mark= at position 0.25 with {\arrow{stealth}}}] (-0.5,-0.5) -- (0.5,0.5) node[left,pos=0.3] {$\scriptstyle x_3$} node[right,pos=0.7] {$\scriptstyle x_2$};
			\draw[thick, postaction={decorate},decoration={markings, mark= at position 0.75 with {\arrow{stealth}}, mark= at position 0.25 with {\arrow{stealth}}}] (0.5,-0.5) -- (-0.5,0.5) node[right,pos=0.3] {$\scriptstyle x_4$} node[left,pos=0.7] {$\scriptstyle x_1$};
			\draw[fill=burgundy] (0,0) circle (0.07);
		\end{tikzpicture}
	\end{array}\quad\quad
	\begin{aligned}
		&\hat D^{\smallresm}\left[\begin{array}{cc} x_1 & x_2 \\ x_3 & x_4\end{array}\right]
=\hat d_G\left[\begin{array}{cc} x_1 & x_2 \\ x_3 & x_4\end{array}\right]
        +u_1\left[\begin{array}{cc} x_1 & x_2 \\ x_3 & x_4\end{array}\right]\theta_1+s_1\theta_1^{\dagger}+u_2\left[\begin{array}{cc} x_1 & x_2 \\ x_3 & x_4\end{array}\right]\theta_2+s_2\theta_2^{\dagger}\,,\\
		&s_1=x_1+x_2-x_3-x_4\,,\\
		&s_2=\underline{x_1x_2-x_3x_4}\,,\\
		&d_G^2=-u_1\left[\begin{array}{cc} x_1 & x_2 \\ x_3 & x_4\end{array}\right]s_1-u_2\left[\begin{array}{cc} x_1 & x_2 \\ x_3 & x_4\end{array}\right]s_2\,.
	\end{aligned}
\end{equation}

\bigskip

Linear and quadratic refers to the powers of $x$ in the underlined expressions.
Note that for particular choice of $N=2$ linearity occurs also in another structure function ---
and this leads to significant simplifications see Section~\ref{sec:N=2}.

In fact, the first of these pictures can be handled in two steps, see details in Section~\ref{sec:lin_red}.


\subsection{Categorified MOY moves}\label{sec:cat-MOY}

There are also vertical reductions categorifying MOY relations~\eqref{MOYloc_II}--\eqref{MOYloc_VI}:

\begin{subequations}
	\begin{align}
		&\label{MOYcat-loc_II}\begin{array}{c}
			\begin{tikzpicture}
				\draw[thick,postaction={decorate},decoration={markings, mark= at position 0.5 with {\arrow{stealth}}}] (0,0) circle (0.35);
				\node[right] at (0.35,0) {$\scriptstyle x$};
			\end{tikzpicture}
		\end{array}\cong V_{N},\quad{\rm dim}_q\,V_N=[N]\,, \\
		&\label{MOYcat-loc_III}\begin{array}{c}
			\begin{tikzpicture}[scale=0.7]
				\draw[thick,stealth-,postaction={decorate},decoration={markings, mark= at position 0.7 with {\arrow{stealth}}}] (0,1) -- (0,0.5) to[out=180,in=90] (-0.5,0) to[out=270,in=180] (0,-0.5) -- (0,-1);
				\draw[fill=\myred] (-0.1,0.5) to[out=90,in=180] (0,0.6) to[out=0,in=90] (0.6,0) to[out=270,in=0] (0,-0.6) to[out=180,in=270] (-0.1,-0.5) to[out=90,in=180] (0,-0.4) to[out=0,in=270] (0.4,0) to[out=90,in=0] (0,0.4) to[out=180,in=270] (-0.1,0.5);
				\node[right] at (0,1) {$\scriptstyle x$};
				\node[right] at (0,-1) {$\scriptstyle z$};
				\node[left] at (-0.5,0) {$\scriptstyle y$};
			\end{tikzpicture}
		\end{array}\cong\begin{array}{c}
			\begin{tikzpicture}
				\draw[thick, -stealth] (0,-0.5) -- (0,0.5);
				\node[right] at (0,0.5) {$\scriptstyle x$};
				\node[right] at (0,-0.5) {$\scriptstyle z$};
			\end{tikzpicture}
		\end{array}\otimes V_{N-1}, \quad {\rm dim}_q\,V_{N-1}=[N-1]\,,\\
		&\label{MOYcat-loc_IV}\begin{array}{c}
			\begin{tikzpicture}[scale=0.7]
				\begin{scope}[shift={(0,1)}]
					\draw[thick,-stealth] (-0.5,-0.5) to[out=90,in=270] (0.5,0.5);
					\draw[thick,-stealth] (0.5,-0.5) to[out=90,in=270] (-0.5,0.5);
					\draw[fill=\myred] (0,0) circle (0.1);
				\end{scope}
				\draw[thick,-stealth] (-0.5,-0.5) to[out=90,in=270] (0.5,0.5) -- (0.5,0.6);
				\draw[thick,-stealth] (0.5,-0.5) to[out=90,in=270] (-0.5,0.5) -- (-0.5,0.6);
				\draw[fill=\myred] (0,0) circle (0.1);
				\node[left] at (-0.5,-0.5) {$\scriptstyle x_4$};
				\node[right] at (0.5,-0.5) {$\scriptstyle x_3$};
				\node[left] at (-0.5,0.5) {$\scriptstyle y$};
				\node[right] at (0.5,0.5) {$\scriptstyle z$};
				\node[left] at (-0.5,1.5) {$\scriptstyle x_1$};
				\node[right] at (0.5,1.5) {$\scriptstyle x_2$};
			\end{tikzpicture}
		\end{array}\cong \begin{array}{c}
			\begin{tikzpicture}[scale=0.7]
				\draw[thick,-stealth] (-0.5,-0.5) to[out=90,in=270] (0.5,0.5);
				\draw[thick,-stealth] (0.5,-0.5) to[out=90,in=270] (-0.5,0.5);
				\draw[fill=\myred] (0,0) circle (0.1);
				\node[left] at (-0.5,-0.5) {$\scriptstyle x_4$};
				\node[right] at (0.5,-0.5) {$\scriptstyle x_3$};
				\node[left] at (-0.5,0.5) {$\scriptstyle x_1$};
				\node[right] at (0.5,0.5) {$\scriptstyle x_2$};
			\end{tikzpicture}
		\end{array}\otimes V_2,\quad {\rm dim}_q\,V_2=[2]\,,\\
		&\label{MOYcat-loc_V}\begin{array}{c}
			\begin{tikzpicture}[scale=0.7]
				\draw[thick,postaction={decorate},decoration={markings, mark= at position 0.7 with {\arrow{stealth}}}] (-1,-1) -- (-0.5,-0.5);
				\draw[thick,postaction={decorate},decoration={markings, mark= at position 0.7 with {\arrow{stealth}}}] (1,1) -- (0.5,0.5);
				\draw[thick,postaction={decorate},decoration={markings, mark= at position 0.7 with {\arrow{stealth}}}] (-0.5,0.5) -- (-1,1);
				\draw[thick,postaction={decorate},decoration={markings, mark= at position 0.7 with {\arrow{stealth}}}] (0.5,-0.5) -- (1,-1);
				\draw[thick,postaction={decorate},decoration={markings, mark= at position 0.7 with {\arrow{stealth}}}] (-0.5,0.5) -- (0.5,0.5);
				\draw[thick,postaction={decorate},decoration={markings, mark= at position 0.7 with {\arrow{stealth}}}] (0.5,-0.5) -- (-0.5,-0.5);
				\begin{scope}[shift={(-0.5,0)}]
					\draw[fill=\myred] (0.1,0.5) to[out=90,in=0] (0,0.6) to[out=180,in=90] (-0.1,0.5) -- (-0.1,-0.5) to[out=270,in=180] (0,-0.6) to[out=0,in=270] (0.1,-0.5) -- (0.1,0.5);
				\end{scope}
				\begin{scope}[shift={(0.5,0)}]
					\draw[fill=\myred] (0.1,0.5) to[out=90,in=0] (0,0.6) to[out=180,in=90] (-0.1,0.5) -- (-0.1,-0.5) to[out=270,in=180] (0,-0.6) to[out=0,in=270] (0.1,-0.5) -- (0.1,0.5);
				\end{scope}
				\node[left] at (-1,-1) {$\scriptstyle x_1$};
				\node[left] at (-1,1) {$\scriptstyle x_2$};
				\node[right] at (1,1) {$\scriptstyle x_3$};
				\node[right] at (1,-1) {$\scriptstyle x_4$};
				\node[above] at (0,0.5) {$\scriptstyle x_5$};
				\node[below] at (0,-0.5) {$\scriptstyle x_6$};
			\end{tikzpicture}
		\end{array}\;\cong\;\begin{array}{c}
			\begin{tikzpicture}
				\draw[thick, stealth-] (-0.5,0.5) to[out=315,in=225] (0.5,0.5);
				\draw[thick, stealth-] (0.5,-0.5) to[out=135,in=45] (-0.5,-0.5);
				\node[left] at (-0.5,-0.5) {$\scriptstyle x_1$};
				\node[left] at (-0.5,0.5) {$\scriptstyle x_2$};
				\node[right] at (0.5,0.5) {$\scriptstyle x_3$};
				\node[right] at (0.5,-0.5) {$\scriptstyle x_4$};
			\end{tikzpicture}
		\end{array} \oplus \left(\begin{array}{c}
			\begin{tikzpicture}
				\draw[thick, stealth-] (-0.5,0.5) to[out=315,in=45] (-0.5,-0.5);
				\draw[thick, stealth-] (0.5,-0.5) to[out=135,in=225] (0.5,0.5);
				\node[left] at (-0.5,-0.5) {$\scriptstyle x_1$};
				\node[left] at (-0.5,0.5) {$\scriptstyle x_2$};
				\node[right] at (0.5,0.5) {$\scriptstyle x_3$};
				\node[right] at (0.5,-0.5) {$\scriptstyle x_4$};
			\end{tikzpicture}
		\end{array} \otimes V_{N-2}\right),\quad {\rm dim}_q\,V_{N-2}=[N-2]\,,
\end{align}
\end{subequations}
\begin{equation}
		\label{MOYcat-loc_VI}\begin{array}{c}
			\begin{tikzpicture}[scale=0.6]
				\draw[thick,-stealth] (0,0) to[out=90,in=225] (1.5,1.5) to[out=45,in=270] (2,3);
				\draw[thick,-stealth] (2,0) to[out=90,in=315] (1.5,1.5) to[out=135,in=270] (0,3);
				\draw[thick,-stealth] (1,0) to[out=90,in=270] (0,1.5) to[out=90,in=270] (1,3);
				\node[left] at (0,0) {$\scriptstyle x_1$};
				\node[left] at (1,0) {$\scriptstyle x_2$};
				\node[right] at (2,0) {$\scriptstyle x_3$};
				\node[left] at (0,3) {$\scriptstyle x_4$};
				\node[left] at (1,3) {$\scriptstyle x_5$};
				\node[right] at (2,3) {$\scriptstyle x_6$};
				\draw[fill=\myred] (1.5,1.5) circle (0.13) (0.45,0.78) circle (0.13) (0.45,2.22) circle (0.13);
			\end{tikzpicture}
		\end{array}\oplus\begin{array}{c}
			\begin{tikzpicture}[scale=0.6]
				\draw[thick,-stealth] (0,0) -- (0,3);
				\draw[thick,-stealth] (1,0) to[out=90,in=270] (2,3);
				\draw[thick,-stealth] (2,0) to[out=90,in=270] (1,3);
				\node[left] at (0,0) {$\scriptstyle x_1$};
				\node[left] at (1,0) {$\scriptstyle x_2$};
				\node[right] at (2,0) {$\scriptstyle x_3$};
				\node[left] at (0,3) {$\scriptstyle x_4$};
				\node[left] at (1,3) {$\scriptstyle x_5$};
				\node[right] at (2,3) {$\scriptstyle x_6$};
				\draw[fill=\myred] (1.5,1.5) circle (0.13);
			\end{tikzpicture}
		\end{array}\cong \begin{array}{c}
			\begin{tikzpicture}[scale=0.6,xscale=-1]
				\draw[thick,-stealth] (0,0) to[out=90,in=225] (1.5,1.5) to[out=45,in=270] (2,3);
				\draw[thick,-stealth] (2,0) to[out=90,in=315] (1.5,1.5) to[out=135,in=270] (0,3);
				\draw[thick,-stealth] (1,0) to[out=90,in=270] (0,1.5) to[out=90,in=270] (1,3);
				\node[right] at (0,0) {$\scriptstyle x_3$};
				\node[left] at (1,0) {$\scriptstyle x_2$};
				\node[left] at (2,0) {$\scriptstyle x_1$};
				\node[right] at (0,3) {$\scriptstyle x_6$};
				\node[left] at (1,3) {$\scriptstyle x_5$};
				\node[left] at (2,3) {$\scriptstyle x_4$};
				\draw[fill=\myred] (1.5,1.5) circle (0.13) (0.45,0.78) circle (0.13) (0.45,2.22) circle (0.13);
			\end{tikzpicture}
		\end{array}\oplus\begin{array}{c}
			\begin{tikzpicture}[scale=0.6,xscale=-1]
				\draw[thick,-stealth] (0,0) -- (0,3);
				\draw[thick,-stealth] (1,0) to[out=90,in=270] (2,3);
				\draw[thick,-stealth] (2,0) to[out=90,in=270] (1,3);
				\node[right] at (0,0) {$\scriptstyle x_3$};
				\node[left] at (1,0) {$\scriptstyle x_2$};
				\node[left] at (2,0) {$\scriptstyle x_1$};
				\node[right] at (0,3) {$\scriptstyle x_6$};
				\node[left] at (1,3) {$\scriptstyle x_5$};
				\node[left] at (2,3) {$\scriptstyle x_4$};
				\draw[fill=\myred] (1.5,1.5) circle (0.13);
			\end{tikzpicture}
		\end{array}\,.
\end{equation}
We have already validated relation~\eqref{MOYcat-loc_II} in~\eqref{1-ferm-unknot}. We consider the example of the first MOY move~\eqref{MOYcat-loc_III} in Section~\ref{sec:MOY-I-ex} and prove it in Section~\ref{sec:MOY-I}. To prove the remaining relations, one uses quadratic reductions. We illustrate the validity of relation~\eqref{MOYcat-loc_IV} in Section~\ref{sec:quad-red-II-MOY-ex} and provide proofs of relations~\eqref{MOYcat-loc_IV} and~\eqref{MOYcat-loc_V} in Sections~\ref{sec:MOY-II} and \ref{sec:box} correspondingly.

\subsection{List of examples}\label{sec:examples}

In this subsection, 
we provide examples of {\it vertical} reductions (see Table~\ref{tab:examples-1}) which then generalize into formulas~\eqref{psisepar},~\eqref{MOY-I-WF},~\eqref{4v-fPsi}, and the respective isomorphisms of the cohomologies from Table~\ref{tab:isom}.


\begin{itemize}
\item{}
First, we provide the example of the linear reduction for the unknot with two fermions~\eqref{2-ferm-unknot}. 

\item{}
Second, we show that the 4-valent vertex connected to form the figure-eight graph is equivalent to the unknot with one fermion times $(N-1)$-dimensional $q$-graded vector space --- the example of the MOY reduction~\eqref{MOYcat-loc_III}.

\item{}
Third, we again consider the figure-eight graph and also the parallel Hopf link to show the quadratic reduction. The parallel Hopf link also illustrates the second MOY relation~\eqref{MOYcat-loc_IV}.

%
%
%
%

\end{itemize}

\subsubsection{Linear reduction}\label{sec:lin-red-ex} 

Let us first consider the following example:
\begin{equation}\label{2-ferm-unknot}
    \begin{array}{c}
				\begin{tikzpicture}
					\draw[thick,postaction={decorate},decoration={markings, mark= at position 0.5 with {\arrow{stealth}}}] (0,0) circle (0.5);
					\node[right] at (0.5,0) {$\scriptstyle x$};
                    \node[left] at (-0.51,0) {$\scriptstyle y$};
                    \node[above] at (0,0.5) {$\scriptstyle \theta_1$};
                    \node[below] at (0,-0.5) {$\scriptstyle \theta_2$};
                    \draw[fill=\mygreen] (0,0.5) circle (0.06);
                    \draw[fill=\mygreen] (0,-0.5) circle (0.06);
				\end{tikzpicture}
			\end{array}\,,\qquad D_{x,y}= \underbrace{\pi_{xy} \theta_1 + (y-x) \theta_1^\dag}_{d_G(x,y\,|\,\theta_1)} + \pi_{xy} \theta_2 + (x-y) \theta_2^\dag\,,\quad \pi_{xy} = \frac{x^{N+1} - y^{N+1}}{x-y}\,.
\end{equation}
We split the $\theta_2$-dependence of the wave function:
\begin{equation}
    \Psi(x,y\,|\,\theta_1,\theta_2) = \psi_0(x,y\,|\,\theta_1) + \theta_2\, \psi_2(x,y\,|\,\theta_1)\,.
\end{equation}
From the zero mode condition, it follows that
\ba
    0 = D_{x,y}\Psi &= \left( \pi_{xy} \theta_1 \psi_0(x,y\,|\,\theta_1) - (x-y) \theta_1^\dag \psi_0(x,y\,|\,\theta_1) + (x-y) \psi_2 (x,y\,|\,\theta_1) \right) + \\
    &+ \theta_2 \left( \pi_{xy}\psi_0(x,y\,|\,\theta_1) -\pi_{xy} \theta_1 \psi_2(x,y\,|\,\theta_1) + (x-y)\theta_1^\dag \psi_2 (x,y\,|\,\theta_1) \right)\,.
\ea
Each bracket vanishes separately. Thus, we get the following system of equations:
\begin{equation}\label{cyc-2-sys}
\begin{cases}
    \pi_{xy} \theta_1 \psi_0(x,y\,|\,\theta_1) - (x-y) \theta_1^\dag \psi_0(x,y\,|\,\theta_1) + (x-y) \psi_2 (x,y\,|\,\theta_1) = 0\,, \\
    \pi_{xy}\psi_0(x,y\,|\,\theta_1) -\pi_{xy} \theta_1 \psi_2(x,y\,|\,\theta_1) + (x-y)\theta_1^\dag \psi_2 (x,y\,|\,\theta_1) = 0\,.
\end{cases}
\end{equation}
At $x=y$ we have
\begin{equation}\label{cyc-2-x=y}
\begin{cases}
    \pi_{xx} \theta_1 \psi_0(x,x\,|\,\theta_1) = d_G(x,x\,|\,\theta_1) = 0 \quad \Longrightarrow \quad \boxed{\psi_0(x,x\,|\,\theta_1) = \theta_1 f(x)} \\
    \pi_{xx} \psi_0 (x,x\,|\,\theta_1) - \pi_{xx} \theta_1 \psi_2(x,x\,|\,\theta_1) = 0 \quad \Lra \quad \psi_2(x,x\,|\,\theta_1) = f(x) + \theta_1 g(x)
\end{cases}
\end{equation}
At $x\neq y$, the first equation in~\eqref{cyc-2-sys} reads:
\begin{equation}\label{cyc-2-psi2}
    \psi_2 (x,y\,|\,\theta_1) = \theta_1^\dag \psi_0(x,y\,|\,\theta_1) - \frac{\pi_{xy}\theta_1 \psi_0 (x,y\,|\,\theta_1) - \pi_{xx}\theta_1 \psi_0(x,x\,|\,\theta_1)}{x-y}\,,
\end{equation}
and the second relation in~\eqref{cyc-2-x=y} actually follows from~\eqref{cyc-2-psi2} at the limit $y\rightarrow x$. One can also check that the second equation in~\eqref{cyc-2-sys} at $x\neq y$ does not give any additional restrictions.

Now, we substitute this solution for $\psi_2(x,y\,|\,\theta_1)$ into the whole cohomology $\Psi$ and rewrite it in the form
\ba
    \Psi(x,y\,|\,\theta_1,\theta_2) &= \overbrace{\psi_0(x,x\,|\,\theta_1) + \underline{(x-y) \frac{\psi_0(x,y\,|\,\theta_1) - \psi_0(x,x\,|\,\theta_1)}{x-y}}}^{\psi_0(x,y\,|\,\theta_1)} - \theta_2 \left( \frac{\pi_{xy}-\pi_{xx}}{x-y}\theta_1 - \theta_1^\dag \right)\psi_0(x,x\,|\,\theta_1) - \\
    &\underline{- \theta_2 \left( \pi_{xy}\theta_1 \frac{\psi_0(x,y\,|\,\theta_1) - \psi_0(x,x\,|\,\theta_1)}{x-y} - \theta_1^\dag \psi_0(x,y\,|\,\theta_1) + \theta_1^\dag \psi_0(x,x\,|\,\theta_1) \right)} = \\
    &= \bigg( 1 - \theta_2 \underbrace{\left( \frac{\pi_{xy} - \pi_{xx}}{x-y}\theta_1 - \theta_1^\dag \right)}_{\frac{d_G(x,y\,|\,\theta_1) - d_G(x,x\,|\,\theta_1)}{x-y}} \bigg) \psi_0(x,x\,|\,\theta_1) + D_{x,y} \left( \theta_2 \frac{\psi_0(x,y\,|\,\theta_1) - \psi_0(x,x\,|\,\theta_1)}{x-y} \right)\,.
\ea
The underlined terms in the first equality gather into the image of $D_{x,y}$. The second relation is the particular case of the generic formula~\eqref{psisepar}.

Let us now check that a cohomology of $d_G(x,x\,|\,\theta_1)$, see the first identity in~\eqref{cyc-2-x=y}, indeed gives a cohomology of $D_{x,y}$:
\begin{equation}
    \Psi(x,y\,|\,\theta_1,\theta_2) = \left( 1 - \theta_2 \left( \frac{\pi_{xy} - \pi_{xx}}{x-y}\theta_1 - \theta_1^\dag \right) \right) \theta_1 f(x) = (\theta_1 + \theta_2) f(x)\,.
\end{equation}
Then, obviously, $D_{x,y} \Psi(x,y\,|\,\theta_1,\theta_2) \equiv 0$.

\subsubsection{First MOY move}\label{sec:MOY-I-ex}
 
We consider the following example:
\begin{equation}\label{I-MOY-ex}
    \begin{array}{c}
	\begin{tikzpicture}[scale=0.8]
	    \draw[thick,-stealth] (0.5,-0.5) to[out=90,in=270] (-0.5,0.5);
	    \draw[thick,-stealth] (-0.5,-0.5) to[out=90,in=270] (0.5,0.5);
	    \draw[thick] (-0.5,0.5) to[out=90,in=90] (-0.8,0.5) -- (-0.8,-0.5) to[out=270,in=270] (-0.5,-0.5);
	    \begin{scope}[xscale=-1]
	    \draw[thick] (-0.5,0.5) to[out=90,in=90] (-0.8,0.5) -- (-0.8,-0.5) to[out=270,in=270] (-0.5,-0.5);
	    \end{scope}
	    \node[right] at (-0.5,-0.5) {$\scriptstyle x$};
	    \node[right] at (-0.5,0.5) {$\scriptstyle x$};
	    \node[left] at (0.5,-0.5) {$\scriptstyle z$};
	    \node[left] at (0.5,0.5) {$\scriptstyle z$};
	     \node[left] at (-0.15,0) {$\scriptstyle \theta_1$};
	     \node[right] at (0.25,0) {$\scriptstyle \theta_2$};
	    \draw[fill=\myred] (0,0) circle (0.1);
	\end{tikzpicture}
    \end{array}\,,\qquad D(x,z\,|\,\theta_1,\theta_2) = u_1\left[\begin{array}{cc} x & z \\ x & z \end{array}\right]\theta_1 + u_2\left[\begin{array}{cc} x & z \\ x & z \end{array}\right]\theta_2\,.
\end{equation}
We rewrite this differential in the form
\begin{equation}
\begin{aligned}
    D(x,z\,|\,\theta_1,\theta_2) &= \pi_{xx}\theta_1 - z u_2\left[\begin{array}{cc} x & z \\ x & z \end{array}\right]\theta_1 + u_2\left[\begin{array}{cc} x & z \\ x & z \end{array}\right]\theta_2 = \\
    &= D'(x,z\,|\,\theta_1,\theta_2) - [z\theta_1 \theta_2^\dag,D'(x,z\,|\,\theta_1,\theta_2)] = e^{-z\theta_1 \theta_2^\dag}D'(x,z\,|\,\theta_1,\theta_2)e^{z\theta_1 \theta_2^\dag} 
\end{aligned}
\end{equation}
where
\begin{equation}\label{D'ex}
    D'(x,z\,|\,\theta_1,\theta_2) = \pi_{xx} \theta_1 + u_2\left[\begin{array}{cc} x & z \\ x & z \end{array}\right]\theta_2\,.
\end{equation}
Now, we separate the dependence on $\theta_2$ of the cohomology $\Psi'(x,z\,|\,\theta_1,\theta_2) \in H^*(D'(x,z\,|\,\theta_1,\theta_2))\,$:
\begin{equation}\label{Psi'}
    \Psi'(x,z\,|\,\theta_1,\theta_2) = \psi'_0(x,z\,|\,\theta_1) + \theta_2 \psi'_2(x,z\,|\,\theta_1)
\end{equation}
and write the zero mode condition:
\begin{equation}
    0 = D' \Psi' = \pi_{xx} \theta_1 \psi'_0 + \theta_2 \left( u_2\left[\begin{array}{cc} x & z \\ x & z \end{array}\right] \psi'_0 - \pi_{xx}\theta_1 \psi'_2 \right)
\end{equation}
from which we have
\begin{equation}\label{psi0=thf0}
\begin{cases}
    \pi_{xx} \theta_1 \psi'_0 = 0 \quad \Lra \quad \psi'_0(x,z\,|\,\theta_1) = \theta_1 f_0(x,z)\,, \\
    u_2\left[\begin{array}{cc} x & z \\ x & z \end{array}\right] \psi'_0 = \pi_{xx}\theta_1 \psi'_2\,.
\end{cases}           
\end{equation}
We substitute the first relation into the second one and get
\begin{equation}
	u_2\left[\begin{array}{cc} x & z \\ x & z \end{array}\right] \theta_1 f_0(x,z) = \pi_{xx}\theta_1 \psi'_2\,.
\end{equation}
So, the solutions are
\begin{equation}\label{sol}
	\begin{cases}
		\ \ \ \ \ \ f_0 = \pi_{xx}\,p(x,z) + \theta_1 g(x,z)\,, \vspace{0.1cm} \\
		\psi'_2 = u_2\left[\begin{array}{cc} x & z \\ x & z \end{array}\right] p(x,z) + \theta_1 \chi(x,z)\,,
	\end{cases} \\ 
\end{equation}
where $p(x,z)$, $g(x,z)$ and $\chi(x,z)$ are some polynomials. And plugging~\eqref{sol} and the first relation in~\eqref{psi0=thf0} into~\eqref{Psi'}, we get
\begin{equation}
    \Psi' = \theta_2 \theta_1 \chi(x,z) + D'(p(x,z))\,.
\end{equation}
Note that the derivation starting from~\eqref{D'ex} is independent on the concrete view of the function $u_2$. If now we imply that $u_2\left[\begin{array}{cc} x & z \\ x & z \end{array}\right]$ is a polynomial of degree $N-2$ in $x$, $z$ variables, the cohomology $\Psi(x,z\,|\,\theta_1,\theta_2)$ of $D(x,z\,|\,\theta_1,\theta_2)$ is expressed through the cohomology $\Psi'(x,z\,|\,\theta_1,\theta_2)$ of $D'(x,z\,|\,\theta_1,\theta_2)$ as follows:
\begin{equation}\label{Psi-I-MOY-ex}
    \Psi(x,z\,|\,\theta_1,\theta_2) = e^{-z\theta_1 \theta_2^\dag} \Psi'(x,z\,|\,\theta_1,\theta_2) = (1 - z\theta_1 \theta_2^\dag)\Psi'(x,z\,|\,\theta_1,\theta_2) = (1 - z\theta_1 \theta_2^\dag)x^k z^j \theta_1 \theta_2 = x^k z^j \theta_1 \theta_2
\end{equation}
where $0\leq k \leq N-1$ and $0\leq j \leq N-2$, and the answer~\eqref{Psi-I-MOY-ex} is indeed the correct answer for~\eqref{I-MOY-ex}. 

This example is the simplest one that illustrates the isomorphisms~\eqref{isom-2}.

\subsubsection{The simplest example of quadratic reduction}\label{sec:quad-red-ex}

Consider the following example:
\begin{equation}\label{QR-ex}
	\begin{array}{c}
		\begin{tikzpicture}[scale=1.0]
			\draw[thick,-stealth] (0.5,-0.5) to[out=90,in=270] (-0.5,0.5);
			\draw[thick,-stealth] (-0.5,-0.5) to[out=90,in=270] (0.5,0.5);
			\draw[thick] (-0.5,0.5) to[out=90,in=90] (-0.8,0.5) -- (-0.8,-0.5) to[out=270,in=270] (-0.5,-0.5);
			\begin{scope}[xscale=-1]
				\draw[thick] (-0.5,0.5) to[out=90,in=90] (-0.8,0.5) -- (-0.8,-0.5) to[out=270,in=270] (-0.5,-0.5);
			\end{scope}
			\node[right] at (-0.5,-0.5) {$\scriptstyle x_4$};
			\node[right] at (-0.5,0.5) {$\scriptstyle x_1$};
			\node[left] at (0.5,-0.5) {$\scriptstyle x_3$};
			\node[left] at (0.5,0.5) {$\scriptstyle x_2$};
			\node[left] at (-0.15,0) {$\scriptstyle \theta_1$};
			\node[left] at (-0.85,0) {$\scriptstyle \theta_3$};
			\node[right] at (0.25,0) {$\scriptstyle \theta_2$};
			\node[right] at (0.85,0) {$\scriptstyle \theta_4$};
			\draw[fill=\myred] (0,0) circle (0.07);
			\draw[fill=\mygreen] (-0.8,0) circle (0.06);
			\draw[fill=\mygreen] (0.8,0) circle (0.06);
		\end{tikzpicture}
	\end{array}\,,\quad D\left[\begin{array}{cc|c} x_1 & x_2 & \theta_1 \\ x_4 & x_3 & \theta_2 \end{array}\right] = u_1\theta_1 + u_2\theta_2 + s_1 \theta_1^\dag + s_2 \theta_2^\dag + \underbrace{\pi_{14}\theta_3 + (x_4 - x_1)\theta_3^\dag + \pi_{23}\theta_4 + (x_3 - x_2)\theta_4^\dag}_{d_G}\,.
\end{equation}
From the one hand, in order to find the cohomology, we can get use of the quadratic reduction. Namely, we split it
\begin{equation}
	\Psi\left[\begin{array}{cc|c} x_1 & x_2 & \theta_1 \\ x_4 & x_3 & \theta_2 \end{array}\right]=\psi_0\left[\begin{array}{cc} x_1 & x_2 \\ x_4 & x_3 \end{array}\right]+\theta_1\psi_1\left[\begin{array}{cc} x_1 & x_2 \\ x_4 & x_3 \end{array}\right]+\theta_2\psi_2\left[\begin{array}{cc} x_1 & x_2 \\ x_4 & x_3 \end{array}\right]+\theta_1\theta_2\psi_{12}\left[\begin{array}{cc} x_1 & x_2 \\ x_4 & x_3 \end{array}\right]\,,
\end{equation}
where we also decompose $\psi_0$ according to Appendix~\ref{sec:4f-deco}:
\begin{equation}
	\psi_0\left[\begin{array}{cc} x_1 & x_2 \\ x_4 & x_3 \end{array}\right] = \frac{1}{2}(1+h)\underbrace{\psi_0\left[\begin{array}{cc} x_1 & x_2 \\ x_2 & x_1 \end{array}\right]}_{\psi_-(x_1,x_2)} + \frac{1}{2}(1-h)\underbrace{\psi_0\left[\begin{array}{cc} x_1 & x_2 \\ x_1 & x_2 \end{array}\right]}_{\psi_+(x_1,x_2)}+s_1 v_1(\psi_0) + s_2 v_2(\psi_0)
\end{equation}
and write the zero mode condition:
{\small \begin{equation}\label{4v-DPsi=0}
		D\Psi = \left(d_G\psi_0+s_1\psi_1+s_2\psi_2\right)+\theta_1\left(-d_G\psi_1+u_1\psi_0-s_2\psi_{12}\right)+\theta_2\left(-d_G\psi_2+u_2\psi_0+s_1\psi_{12}\right)+\theta_1\theta_2\left(d_G\psi_{12}+u_1\psi_2-u_2\psi_1\right) = 0\,.
\end{equation}}
Each round bracket vanishes separately. At $x_1 = x_3$, $x_2 = x_4$ and $x_1=x_4$, $x_2=x_3$ from the first equation
\begin{equation}\label{4v-eq1}
	d_G\psi_0+s_1\psi_1+s_2\psi_2 = 0\,,
\end{equation}
we get
\begin{equation}
	D_\pm \psi_\pm=0
\end{equation}
respectively, with

\ba
	D_- &= \pi_{12}\theta_3 + (x_2 - x_1) \theta_3^\dag + \pi_{12}\theta_4 + (x_1 - x_2)\theta_4^\dag\,, \\
	D_+ &= \pi_{11}\theta_3 + \pi_{22}\theta_4\,.
\ea
The zero modes must satisfy the condition $\psi_-(x,x)=\psi_+(x,x)$. After factorization by the images of $D_\pm$, we get the solution
\begin{equation}
\begin{aligned}
	\psi_+ &= (x_1 - x_2) x_1^k x_2^{k'} \theta_3 \theta_4\,,\quad 0\leq k \leq N-1,\; k' \leq N-2\,, \\
	\psi_- &\equiv 0\,.
\end{aligned}
\end{equation}
Then, we find
\begin{equation}
\begin{aligned}
	d_G \psi_0 &= x_1^k x_2^{k'}\left(\frac{1}{2}(x_1-x_2-x_3+x_4)\left( (x_4-x_1)\theta_4 + (x_2 - x_3)\theta_3 \right)\right) + s_1 d_G v_1(\psi_0) + s_2 d_G v_2(\psi_0) = \\
	&= - \frac{1}{2}s_1 x_1^k x_2^{k'} \left( (x_2+x_3)\theta_3 + (x_1 + x_4)\theta_4 \right) + s_2 x_1^k x_2^{k'}(\theta_3 + \theta_4) + s_1 d_G v_1(\psi_0) + s_2 d_G v_2(\psi_0)
\end{aligned}
\end{equation}
so that equation~\eqref{4v-eq1} dictates that 
\begin{equation}
\begin{aligned}
	\psi_1 &= x_1^k x_2^{k'}\left(\frac{1}{2}(x_2 + x_3)\theta_3 + \frac{1}{2}(x_1 + x_4)\theta_4\right) - d_G v_1(\psi_0)\,, \\
	\psi_2 &= x_1^k x_2^{k'}\left(- \theta_3 - \theta_4\right) - d_G v_2(\psi_0)\,.
\end{aligned}
\end{equation}
The function $\psi_{12}$ can then be determined from the second or the third bracket in~\eqref{4v-DPsi=0}. We use the third one:
\begin{equation}
	\psi_{12} = - \frac{u_2 \psi_0 - d_G \psi_2}{s_1} = -x_1^k x_2^{k'}\left(\frac{\frac{1}{2}u_2 (x_1-x_2-x_3+x_4) + \pi_{14} - \pi_{23}}{s_1}\theta_3 \theta_4+1\right) - u_2 v_1(\psi_0) + u_1 v_2(\psi_0)\,,
\end{equation}
and the cohomology is
\begin{equation}
\begin{aligned}
	\Psi &= x_1^k x_2^{k'} \bigg(\theta_1 \theta_2 + \frac{1}{2}(x_1-x_2-x_3+x_4)\theta_3\theta_4 + \frac{1}{2}(x_2 + x_3)\theta_1\theta_3 + \frac{1}{2}(x_1 + x_4)\theta_1\theta_4 - \theta_2\theta_3 - \theta_2\theta_4 - \\ 
	&-\frac{\frac{1}{2}u_2 (x_1-x_2-x_3+x_4) + \pi_{14} - \pi_{23}}{s_1}\theta_1 \theta_2 \theta_3 \theta_4 \bigg) + D(\theta_1 v_1(\psi_0) + \theta_2 v_2 (\psi_0))\,.
\end{aligned}
\end{equation}
From the other hand, one can apply linear reduction twice or make 2-step quadratic reduction from Section~\ref{sec:quad-red-2} and get the answer for the full cohomology. One can check that the three answers in these two methods for the full wave function differ by an image of $D$. Thus, we have shown that in this example, general formula~\eqref{4v-fPsi} works. 

\subsubsection{Example of quadratic reduction and second MOY move}\label{sec:quad-red-II-MOY-ex} 

Consider another example:

{\small \begin{equation}\label{QR-ex-2}
	\begin{array}{c}
		\begin{tikzpicture}[scale=0.7]
			\draw[thick,-stealth] (0.5,-0.5) to[out=90,in=270] (-0.5,0.5) -- (-0.5,0.6);
			\draw[thick,-stealth] (-0.5,-0.5) to[out=90,in=270] (0.5,0.5) -- (0.5,0.6);
			\begin{scope}[shift={(0,1)}]
				\draw[thick] (0.5,-0.5) to[out=90,in=270] (-0.5,0.5);
				\draw[thick] (-0.5,-0.5) to[out=90,in=270] (0.5,0.5);
			\end{scope}
			\draw[thick, postaction={decorate},decoration={markings, mark= at position 0.6 with {\arrow{stealth}}}] (-0.5,1.5) to[out=90,in=90] (-1.2,1.5) -- (-1.2,-0.5) to[out=270,in=270] (-0.5,-0.5);
			\begin{scope}[xscale=-1]
				\draw[thick, postaction={decorate},decoration={markings, mark= at position 0.6 with {\arrow{stealth}}}] (-0.5,1.5) to[out=90,in=90] (-1.2,1.5) -- (-1.2,-0.5) to[out=270,in=270] (-0.5,-0.5);
			\end{scope}
			\draw[fill=\myred] (0,0) circle (0.1) (0,1) circle (0.1);
			\node[left] at (-0.5,-0.5) {$\scriptstyle x_1$};
			\node[right] at (0.5,-0.5) {$\scriptstyle x_2$};
			\node[left] at (-0.5,0.5) {$\scriptstyle x_3$};
			\node[right] at (0.5,0.5) {$\scriptstyle x_4$};
			\node[left] at (-0.5,1.5) {$\scriptstyle x_1$};
			\node[right] at (0.5,1.5) {$\scriptstyle x_2$};
			\node[left] at (-0.3,0) {$\scriptstyle \theta_1$};
			\node[right] at (0.3,0) {$\scriptstyle \theta_2$};
			\node[left] at (-0.3,1) {$\scriptstyle \theta_3$};
			\node[right] at (0.3,1) {$\scriptstyle \theta_4$};
		\end{tikzpicture}
	\end{array}\,,\quad D= u_1\left[\begin{array}{cc} x_1 & x_2 \\ x_3 & x_4 \end{array}\right]\theta_3 + u_2\left[\begin{array}{cc} x_1 & x_2 \\ x_3 & x_4 \end{array}\right]\theta_4 + s_1 \theta_3^\dag + s_2 \theta_4^\dag + \underbrace{u_1\left[\begin{array}{cc} x_3 & x_4 \\ x_1 & x_2 \end{array}\right]\theta_1 + u_2\left[\begin{array}{cc} x_3 & x_4 \\ x_1 & x_2 \end{array}\right]\theta_2 - s_1 \theta_1^\dag - s_2 \theta_2^\dag}_{d_G}\,.
\end{equation}}
We split the cohomology as follows:
\begin{equation}
	\Psi = \psi_0 + \theta_3 \psi_3 + \theta_4 \psi_4 + \theta_3 \theta_4 \psi_{34}\,.
\end{equation}
The zero mode condition reads:
{\small \begin{equation}\label{QR-ZM}
	0 = D\Psi = \left(d_G\psi_0+s_1\psi_3+s_2\psi_4\right)+\theta_3\left(-d_G\psi_3+u_1\psi_0-s_2\psi_{34}\right)+\theta_4\left(-d_G\psi_4+u_2\psi_0+s_1\psi_{34}\right)+\theta_3\theta_4\left(d_G\psi_{34}+u_1\psi_4-u_2\psi_3\right)\,.
\end{equation}}

\noindent We again split $\psi_0$ being the function of 4 even variables according to Appendix~\ref{sec:4f-deco}:
\begin{equation}
	\psi_0 = \underbrace{\frac{1}{2}(1+h)\psi_- + \frac{1}{2}(1-h)\psi_+}_{\tilde \psi_0} + s_1 v_1(\psi_0) + s_2 v_2(\psi_0) 
\end{equation}
where $\psi_\pm$ satisfy the equations
\begin{equation}\label{D+-psi+-=0-2}
	D_\pm \psi_\pm = 0
\end{equation}
with
\begin{equation}
	D_- = u_1\left[\begin{array}{cc} x_2 & x_1 \\ x_1 & x_2 \end{array}\right]\theta_1 + u_2\left[\begin{array}{cc} x_2 & x_1 \\ x_1 & x_2 \end{array}\right]\theta_2\,, \quad D_+ = u_1\left[\begin{array}{cc} x_1 & x_2 \\ x_1 & x_2 \end{array}\right]\theta_1 + u_2\left[\begin{array}{cc} x_1 & x_2 \\ x_1 & x_2 \end{array}\right]\theta_2\,,
\end{equation}
and $\hat{d} = D_- = D_+$. In this case, it is convenient to perform the variables change so that
\begin{equation}
	\psi_{\pm}(x_1,x_2) = f(x_1,x_2) \pm (x_1 - x_2) g(x_1,x_2)\,.
\end{equation}
Thus, we get from~\eqref{D+-psi+-=0-2}:
\begin{equation}
	\hat{d}f = \hat{d}g = 0\,,
\end{equation}
and
\begin{equation}\label{degen-psi0}
	\tilde \psi_0 = f(x_1,x_2) + (x_1 - x_2) h \, g(x_1,x_2) = f(x_1,x_2) + (x_4 - x_3) g(x_1,x_2) = \sum\lm_{i=0}^{N-1}\sum\lm_{j=0}^{N-2}\left(d_{ij}x_1^ix_2^j+(x_3-x_4)e_{ij}x_1^ix_2^j\right)\theta_1 \theta_2\,.
\end{equation}
Then,
\begin{equation}
	d_G \psi_0 = \sum\lm_{i=0}^{N-1}\sum\lm_{j=0}^{N-2}\left(d_{ij}x_1^ix_2^j+(x_3-x_4)e_{ij}x_1^ix_2^j\right)\left( -s_1\theta_2 + s_2 \theta_1 \right) + s_1 d_G v_1(\psi_0) + s_2 d_G v_2(\psi_0)\,. 
\end{equation}
So that the equation coming from the first bracket in~\eqref{QR-ZM}
\begin{equation}
	d_G\psi_0+s_1\psi_3+s_2\psi_4 = 0
\end{equation}
has the solution
\begin{equation}
\begin{aligned}
	\psi_3 &= \sum\lm_{i=0}^{N-1}\sum\lm_{j=0}^{N-2}\left(d_{ij}x_1^ix_2^j+(x_3-x_4)e_{ij}x_1^ix_2^j\right) \theta_2 - d_G v_1(\psi_0)\,, \\
	\psi_4 &= -\sum\lm_{i=0}^{N-1}\sum\lm_{j=0}^{N-2}\left(d_{ij}x_1^ix_2^j+(x_3-x_4)e_{ij}x_1^ix_2^j\right) \theta_1 - d_G v_2(\psi_0)\,.
\end{aligned}
\end{equation}
The function $\psi_{34}$ can then be expressed from the second or the third bracket in~\eqref{QR-ZM}. We use the third one:
\begin{equation}
\begin{aligned}
	\psi_{34} &= - \frac{u_2\left[\begin{array}{cc} x_1 & x_2 \\ x_3 & x_4 \end{array}\right] \psi_0 - d_G \psi_4}{s_1} = \\
	&= \sum\lm_{i=0}^{N-1}\sum\lm_{j=0}^{N-2}\left(d_{ij}x_1^ix_2^j+(x_3-x_4)e_{ij}x_1^ix_2^j\right) \left(\frac{u_1\left[\begin{array}{cc}
			x_1 & x_2 \\
			x_3 & x_4 \\
		\end{array}\right]-u_1\left[\begin{array}{cc}
			x_3 & x_4 \\
			x_1 & x_2 \\
		\end{array}\right]}{s_2}\theta_1 \theta_2 + 1\right) - u_2 v_1(\psi_0) + u_1 v_2(\psi_0)\,.
\end{aligned}
\end{equation}
The full cohomology is
\begin{equation}
\begin{aligned}
	\Psi &= \sum\lm_{i=0}^{N-1}\sum\lm_{j=0}^{N-2}\left(d_{ij}x_1^ix_2^j+(x_3-x_4)e_{ij}x_1^ix_2^j\right) \left( \frac{u_1\left[\begin{array}{cc}
			x_1 & x_2 \\
			x_3 & x_4 \\
		\end{array}\right]-u_1\left[\begin{array}{cc}
			x_3 & x_4 \\
			x_1 & x_2 \\
		\end{array}\right]}{s_2}\theta_1 \theta_2 \theta_3 \theta_4 + \theta_3 \theta_4 - \theta_2 \theta_3 + \theta_1 \theta_4 + \theta_1 \theta_2 \right) + \\
	&+ D\left( \theta_3 v_1(\psi_0) + \theta_4 v_2(\psi_0) \right)\,.
\end{aligned}
\end{equation}
This answer is in full accordance with the general procedure in Section~\ref{sec:quad_red} and with the corresponding formula~\eqref{Psi11-Hopf}. 

\subsection{Elementary linear reduction}\label{sec:lin_red}


The first diagram in Section~\ref{sec:elem_red} can be made from a simpler one.
Namely, consider just a single arc, containing two variables $x$ and $y$ connecting two ends of a graph $G\,$:
\begin{equation}\label{2Gclosure}
	\begin{array}{c}
		\begin{tikzpicture}
			\draw[thick, postaction={decorate},decoration={markings, mark = at position 0.25 with {\arrowreversed{stealth}}, mark = at position 0.75 with {\arrowreversed{stealth}}}] (0,0) to[out=45,in=180] (0.8,0.5) to[out=0,in=90] (1.2,0) to[out=270,in=0] (0.8,-0.5) to[out=180,in=315] (0,0);
			\draw[fill=\myblue] (0,0) circle (0.5);
			\node[white] at (0,0) {$\scriptstyle G$};
			\node[right] at (1,0.5) {$\scriptstyle x$};
			\node[right] at (1,-0.5) {$\scriptstyle y$};
            \node[right] at (1.25,0) {$\scriptstyle \theta$};
            \draw[fill=\mygreen] (1.2,0) circle (0.06);
		\end{tikzpicture}
	\end{array},\qquad D_{x,y}=d_G(x,y)+\pi_{xy}\theta+(x-y)\theta^{\dagger}\,.
\end{equation}
The example of such a graph and the derivation of the isomorphism of the cohomologies are present is Section~\ref{sec:lin-red-ex}. The operator $d_G(x,y)$ depends on many other variables coming from a graph $G$, even and odd, which do not matter in this section.
According to (\ref{D^2-gen}) for $w = x^{N+1}$, in this case of one incoming and one outgoing vertex,
\be
d_G^2(x,y)  = -x^{N+1}+y^{N+1} = -(x-y)\pi_{xy}\,,
\label{dgsquare}
\ee
and
\be
D_{x,y}^2 = 0\,.
\label{Dxysquare0}
\ee
This is an illustration of consistency --- this time of the condition (\ref{D^2-gen}) for the double-end tangle $G$
and its end-free closure depicted in (\ref{2Gclosure}).
Note also that at coincident points
\be\label{dxx^2=0}
d_G^2(x,x) = 0\,.
\ee

\bigskip

\noindent Let us now prove that the two nilpotent operators $D_{x,y}$ and $d_G(x,x)$ have the same cohomologies.

For this, we begin by looking at the zero modes of $D_{x,y}$.
They have the form
\begin{equation}\label{Psi(x,y)}
	\Psi(x,y\,|\,\theta) =\psi_0(x,y)+\theta\,\psi_1(x,y)
\end{equation}
where $\psi_0$ and $\psi_1$ have different parities,
and can also depend on the other suppressed variables within $G$.

The reasoning is split in several steps.

\medskip

\noindent $\bullet$
The zero mode satisfies
\begin{equation}\label{zm_lin}
	\begin{aligned}
		D_{x,y}\Psi= 
			\Big(\underline{d_G(x,y)\psi_0(x,y)+ (x-y)\psi_1(x,y)}\Big)
		+\theta \Big(\underline{-d_G(x,y)\psi_1(x,y)+\pi_{xy}\psi_0(x,y)}\Big) + \underbrace{\theta^\dagger  (x-y)\psi_0(x,y)}_{=0} = 0\,.
	\end{aligned}
\end{equation}
The third term vanishes because $\theta^\dag = \frac{\p}{\p \theta}$, and $\psi_0$ does not depend on $\theta$. The two underlined terms should vanish separately.
From the first one we obtain two restrictions:
\be
d_G(x,x)\psi_0(x,x) = 0
\label{dGxxpsi}
\ee
at $y=x$ \footnote{It is important that in this case $(d_G(x,y)\psi_0(x,y))|_{x=y} = d_G(x,x)\psi_0(x,x)$. This fact is actually the reason why the linear reduction works. In the non-linear case, this will be no longer true, and we cannot describe a reduction in the same way.}, and
\begin{equation}
	\psi_1(x,y)=-\frac{d_G(x,y)}{x-y}\,\psi_0(x,y)
\label{psi1vspsi0}
\end{equation}
for $y\neq x$.
The condition, coming from the second term,  states that
\be
d_G(x,y)\psi_1(x,y)=\pi_{xy}\psi_0(x,y)\,.
\ee
It does not give new restrictions because it follows automatically from (\ref{psi1vspsi0}) due to (\ref{dgsquare}).

\medskip

\noindent $\bullet$
Equation (\ref{psi1vspsi0}), however, is not nice enough, because the r.h.s. is not obligatory a polynomial.
However, it actually is --- provided $\psi_0(x,y)$ is not arbitrary, but satisfies (\ref{dGxxpsi}).

This is expressed by the following chain of identities:
$$
\Psi(x,y) =\psi_0(x,y) -\theta\, \frac{d_G(x,y) }{x-y}\,\psi_0(x,y) =
$$
$$
= \overbrace{  {\psi_0(x,x)} +  \underline{(x-y) \frac{\psi_0(x,y)-\psi_0(x,x)}{x-y}  }}^{\psi_0(x,y)}
-\theta
\overbrace{\left( {\frac{d_G(x,y)-d_G(x,x)}{x-y}\,\psi_0(x,x)}
 +  \underline{d_G(x,y)  \frac{\psi_0(x,y)-\psi_0(x,x)}{x-y}}\right)}
^{\frac{d_G(x,y) }{x-y}\,\psi_0(x,y)- \overline{{\frac{d_G(x,x) }{x-y}\,\psi_0(x,x)}}}
 =
$$
\be\label{psisepar}
={ \left(1 -\theta\cdot \frac{d_G(x,y)-d_G(x,x)}{x-y}\right)\psi_0(x,x)} +
\underline{D_{x,y}\left(  \theta \cdot\frac{\psi_0(x,y)-\psi_0(x,x)}{x-y}\right)}\,.
\ee
The first equality holds because the overlined term vanishes due to (\ref{dGxxpsi}).
In the second equality, the underlined terms combine into a $D_{x,y}$-exact expression ---
what can be seen with the help of  (\ref{zm_lin}).
Note that all the fractions in the last two lines are actually polynomials:
their numerators vanish when $x=y$.

\medskip

\noindent $\bullet$ Second, a cohomology is defined up to an arbitrary function from the image of a differential. Thus, we can consider 
\begin{equation}
\begin{aligned}
    \tilde \Psi(x,y) &= \left(1 -\theta\cdot \frac{d_G(x,y)-d_G(x,x)}{x-y}\right)\psi_0(x,x) + D_{x,y} \left(1 -\theta\cdot \frac{d_G(x,y)-d_G(x,x)}{x-y}\right)\Xi_0(x,x) = \\
    &= \left(1 -\theta\cdot \frac{d_G(x,y)-d_G(x,x)}{x-y}\right)(\psi_0(x,x) + d_G(x,x)\,\Xi_0(x,x)) \in \ker D_{x,y}\,.
\end{aligned}
\end{equation}
Thus, any wave function vanished by $d_G(x,x)$ produces a zero mode of $D_{x,y}$.

\medskip

\noindent $\bullet$  What is the outcome?

A zero mode of $D_{x,y}$ is made from every zero mode of $d_G(x,x)$ ---
and it is unique modulo $D_{x,y}$. The vice versa is obvious.

Thus, the conclusion is that the cohomologies of $D_{x,y}$ and $d_G(x,x)$ coincide:
\begin{tcolorbox}
	\begin{equation}\label{isom-1}
		H^*(D_{x,y})\cong H^*\left(d_G(x,x)\right)\,.
	\end{equation}
\end{tcolorbox}

Coming back to the diagram (\ref{linear_reduction}),
we need to apply (\ref{isom-1}) twice to obtain
\begin{tcolorbox}
	\begin{equation}\label{isom-plus}
		H^*\left(\hat D^{\smallresp}\left[\begin{array}{cc} x_1 & x_2 \\ x_3 & x_4\end{array}\right]\right)
\cong H^*\left(\hat d_G\left[\begin{array}{cc} x_1 & x_2 \\ x_1 & x_2 \end{array}\right] \right)
	\end{equation}
\end{tcolorbox}
In this sense, the resolution $\,\respexp \,$ is indeed equivalent to the unity.

\bigskip

\noindent A few {\bf comments} are now in order.

\begin{itemize}
\item

Instead of (\ref{psi1vspsi0}) one could simply {\it suggest} (guess) another solution,
which is obviously polynomial\,:
\begin{equation}\label{zerom}
	\boxed{\Psi(x,y) = \left(1 -  \theta\,\frac{d_G(x,y)-d_G(x,x)}{x-y} \right) \psi_0(x,x)\,,}
\end{equation}
and {\it note} that it is indeed a zero mode, when $d_G(x,x)\psi(x,x) = 0$:
\be
D_{x,y}\Psi(x,y) = \overbrace{d_G(x,y)\psi_0(x,x)-(x-y)\frac{d_G(x,y)-d_G(x,x)}{x-y}\psi_0(x,x)}^{
d_G(x,x) \psi_0(x,x)\ \stackrel{(\ref{dGxxpsi})}{=} \ 0} +
\nn \\
+ \theta\underbrace{\left(d_G(x,y)\frac{d_G(x,y)-d_G(x,x)}{x-y}\psi_0(x,x) +\pi(x,y)\psi_0(x,x)\right)}_{
\stackrel{(\ref{dgsquare})}{=} -\frac{d_G(x,y)}{x-y} d_G(x,x)\psi_0(x,x)\ \stackrel{(\ref{dGxxpsi})}{=} \ 0       }
= 0\,.
\ee

\item But is it the {\it only} zero modes or there are more?

\item The answer is obviously no, since there are gauge transformations
\be
	\tilde\Psi = \Psi+D_{x,y}\,\Xi\,,\quad \Xi = \xi_0 + \theta \xi_1 \quad \Longrightarrow \quad
\left\{\begin{array}{c} \tilde\psi_0 = \psi_0+d_G\xi_0+(x-y)\xi_1\,, \\
\tilde\psi_1 = \psi_1- d_G\xi_1 + \pi_{xy}\xi_0\,.
\end{array} \right.
\ee
which due to (\ref{Dxysquare0}) leave  the zero-mode condition $D_{x,y}\Psi=0$  intact.

\item But do we obtain all the polynomial zero modes by gauge transformations of (\ref{zerom})?

We have described above a procedure which avoids guesses and leads to unambiguous statement (\ref{isom-1}).
In other words, in this case, the answer is --- yes.

However, in a more complicated situation the rigorous derivation can be problematic,
and this type of heuristic reasoning can be useful.
One should only be careful with drawing precise conclusions --- keeping in mind the series of questions in this comment.

\end{itemize}

\paragraph{On the isomorphism of cohomologies.} In this remark, we explain when we can state that $H^*(\hat{A})\cong H^*(\hat{B})$ with differentials $\hat{A}$ and $\hat{B}$ being nilpotent: $\hat{A}^2 = \hat{B}^2 = 0$. So that if we find particular solutions $\hat{A}\, \Psi_A = 0$, $\hat{B}\, \Psi_B = 0$, then general solutions are $\Psi_A + A \, \Xi_A \in H^*(\hat{A})$ and $\Psi_B + B \, \Xi_B \in H^*(\hat{B})$ for arbitrary functions $\Xi_A$, $\Xi_B$. In other words, wave functions are defined up to gauge transformations $\Psi_A \longrightarrow \Psi_A + A \, \Xi_A$, $\Psi_B \longrightarrow \Psi_B + B \, \Xi_B$.

First, if the isomorphism of cohomologies holds, there must exist a linear map $\hat{\Phi}:\; H^*(\hat{B}) \longrightarrow H^*(\hat{A})$:
\begin{equation}\label{Psi_A=PhiPsi_B}
    \Psi_A + A \, \Xi_A = \hat{\Phi}(\Psi_B + B\, \Xi_B) \in H^*(\hat{A}) \quad \Longrightarrow \quad \boxed{\hat{A} \, \hat{\Phi} = \hat{\Phi}' \hat{B}\,.}
\end{equation}
Second, the isomorphism implies that there exists $\hat{\Phi}^{-1}:\; H^*(\hat{A}) \longrightarrow H^*(\hat{B})$, so
\begin{equation}
    \Psi_B + B\, \Xi_B = \hat{\Phi}^{-1}(\Psi_A + A \, \Xi_A)\,,
\end{equation}
and this expression indeed lies in $H^*(\hat{B})$ because it follows from the boxed equation in~\eqref{Psi_A=PhiPsi_B} that $\hat{B}\, \hat{\Phi}^{-1} = (\hat{\Phi}')^{-1} \hat{A}$ if $(\hat{\Phi}')^{-1}$ exists.

Thus, the isomorphism of the cohomologies assumes that invertible linear maps $\hat{\Phi}$, $\hat{\Phi}'$ exist.

\subsection{First MOY move}\label{sec:MOY-I}

Consider a diagram with a looped 4-valent vertex:

\begin{equation}\label{MOY-1}
	\begin{array}{c}
		\begin{tikzpicture}[scale=0.6]
			\draw[thick] (0.5,0.5) to[out=0,in=0] (0.5,-0.5);
			\draw[thick,postaction={decorate},decoration={markings, mark= at position 0.9 with {\arrow{stealth}}}] (0.5,-0.5) to[out=180,in=300] (-0.7,0.7);
			\draw[thick, postaction={decorate},decoration={markings, mark= at position 0.8 with {\arrow{stealth}}}] (-0.7,-0.7) to[out=60,in=180] (0.5,0.5);
			\draw[fill=burgundy] (-0.3,0) circle (0.1);
			\draw[fill=\myblue, even odd rule] (0.4,0) circle (1.3) (0.4,0) circle (1.8);
			\node[right] at (-0.7,0.7) {$\scriptstyle x$};
			\node[left] at (-1.4,0) {$\scriptstyle G$};
			\node[right] at (-0.7,-0.7) {$\scriptstyle y$};
			\node[above] at (0.5,0.5) {$\scriptstyle z$};
			\node[below] at (0.5,-0.5) {$\scriptstyle z$};
		\end{tikzpicture}
	\end{array}\,,\quad D(x,y,z\,|\,\theta_1,\theta_2) = d_G(x,y) + u_1\left[\begin{array}{cc} x & z \\ y & z \end{array}\right]\theta_1 + u_2\left[\begin{array}{cc} x & z \\ y & z \end{array}\right] \theta_2 +(x-y)\theta_1^{\dagger}+z(x-y)\theta_2^{\dagger}\,.
\end{equation}
The example of such a graph and the derivation of the isomorphism of the cohomologies are present is Section~\ref{sec:MOY-I-ex}. First, we rewrite the differential equivalently:
\begin{equation}
	D(x,y,z\,|\,\theta_1,\theta_2) = d_G(x,y) + \pi_{xy} \theta_1 + (x-y) \theta_1^\dag - z\, u_2 \left[\begin{array}{cc} x & z \\ y & z \end{array}\right] \theta_1 + u_2 \left[\begin{array}{cc} x & z \\ y & z \end{array}\right] \theta_2 + z(x-y)\theta_2^\dag 
\end{equation}
as $u_1\left[\begin{array}{cc} x & z \\ y & z \end{array}\right] = \pi_{xy} - z u_2 \left[\begin{array}{cc} x & z \\ y & z \end{array}\right]$. Note that the function $u_2 \left[\begin{array}{cc} x & z \\ y & z \end{array}\right]$ is the polynomial of degree $N-1$. Now, it is easy to see that
\begin{equation}\label{D-D'}
	D(x,y,z\,|\,\theta_1,\theta_2) = D'(x,y,z\,|\,\theta_1,\theta_2) - [z\theta_1 \theta_2^\dag,D'(x,y,z\,|\,\theta_1,\theta_2)] = e^{-z\theta_1 \theta_2^\dag}D'(x,y,z\,|\,\theta_1,\theta_2)e^{z\theta_1 \theta_2^\dag} 
\end{equation}
where
\begin{equation}\label{D'-IMOY}
	D'(x,y,z\,|\,\theta_1,\theta_2) = \underbrace{d_G(x,y) + \pi_{xy} \theta_1 + (x-y)\theta_1^\dag}_{D_{x,y}~\eqref{2Gclosure}} + u_2 \left[\begin{array}{cc} x & z \\ y & z \end{array}\right]\theta_2\,,
\end{equation}
and the cohomology $\Psi(x,y,z\,|\,\theta_1,\theta_2) \in H^*(D(x,y,z\,|\,\theta_1,\theta_2))$ is expressed through the wave function $\Psi'(x,y,z\,|\,\theta_1,\theta_2) \in H^*(D'(x,y,z\,|\,\theta_1,\theta_2))$ as follows:
\begin{equation}\label{MOY-I-WF}
	\Psi(x,y,z\,|\,\theta_1,\theta_2) = e^{-z\theta_1 \theta_2^\dag} \Psi'(x,y,z\,|\,\theta_1,\theta_2) = (1 -z\theta_1 \theta_2^\dag) \Psi'(x,y,z\,|\,\theta_1,\theta_2)\,.
\end{equation}
To establish the isomorphism of cohomologies, we also need to prove that for {\it any} cohomology of $D'(x,y,z\,|\,\theta_1,\theta_2)$, we can find a cohomology of $D(x,y,z\,|\,\theta_1,\theta_2)$, and vice versa. So, we consider
\begin{equation}
	\begin{aligned}
		\tilde\Psi(x,y,z\,|\,\theta_1,\theta_2) &= e^{-z\theta_1 \theta_2^\dag} (\Psi'(x,y,z\,|\,\theta_1,\theta_2) + D'(x,y,z\,|\,\theta_1,\theta_2)\,\Xi(x,y,z\,|\,\theta_1,\theta_2) ) \overset{\eqref{D-D'}}{=} \\ &= e^{-z\theta_1 \theta_2^\dag} \Psi'(x,y,z\,|\,\theta_1,\theta_2) + D(x,y,z\,|\,\theta_1,\theta_2)(e^{-z\theta_1 \theta_2^\dag}\,\Xi(x,y,z\,|\,\theta_1,\theta_2)) \; \in \; \ker D(x,y,z\,|\,\theta_1,\theta_2)\,.
	\end{aligned}
\end{equation}
The converse statement is obvious because the operator $e^{-z\theta_1 \theta_2^\dag}$ is invertible.

Thus, we have derived that
\begin{tcolorbox}
	\begin{equation}\label{isom-2}
		H^*(D(x,y,z\,|\,\theta_1,\theta_2))\cong H^*(D'(x,y,z\,|\,\theta_1,\theta_2))\cong \bigoplus_{k=0}^{n-2} H^*(D_{x,y})\cdot z^k \theta_2 \overset{\eqref{isom-1}}{\cong} H^*\left(d_G(x,x)\right) \otimes V_{N-1}\,.
	\end{equation}
\end{tcolorbox}
Actually, the last two isomorphisms need more comments. We can use linear reduction from Section~\ref{sec:lin_red} in order to reduce the differential~\eqref{D'-IMOY} to the operator
\begin{equation}
    \widetilde D(x,z\,|\,\theta_2) = d_G(x,x) + u_2 \left[\begin{array}{cc} x & z \\ x & z \end{array}\right]\theta_2\,.
\end{equation}
This differential satisfies the condition
\begin{equation}\label{dG^2=0-ex}
	0 = \widetilde{D}^2 = d_G^2(x,x) + u_2 \left[\begin{array}{cc} x & z \\ x & z \end{array}\right](\theta_2 d_G(x,x) + d_G(x,x) \theta_2) = d_G^2(x,x) \quad \Lra \quad \boxed{d_G^2(x,x)=0\,.}
\end{equation}
Let us now separate the $\theta_2$-dependence in the corresponding wave function:
\begin{equation}
    \widetilde \Psi (x,z\,|\,\theta_2) = \tilde \psi_0(x,z) + \theta_2 \tilde \psi_2(x,z)\,.
\end{equation}
Then, we write the zero mode condition:
\begin{equation}
    0 = \widetilde D \widetilde \Psi = d_G(x,x)\tilde \psi_0(x,z) + \theta_2 \left(-d_G(x,x) \tilde \psi_2 (x,z) + u_2 \left[\begin{array}{cc} x & z \\ x & z \end{array}\right] \tilde \psi_0(x,z)\right)\,.
\end{equation}
The $\theta_2^0$-, $\theta_2^1$-terms vanish separately, so we have
\begin{equation}\label{tpsi0-tpsi2}
    \begin{cases}
        d_G(x,x)\tilde \psi_0(x,z) = 0\,, \\
        -d_G(x,x)\tilde \psi_2(x,z) + u_2 \left[\begin{array}{cc} x & z \\ x & z \end{array}\right] \tilde \psi_0(x,z) = 0\,.
    \end{cases}
\end{equation}
We should first solve the first equation in~\eqref{tpsi0-tpsi2}. Due to~\eqref{dG^2=0-ex}, one can separate the image term:
\begin{equation}
	\tilde \psi_0(x,z) = d_G(x,x)f(x,z) + g(x,z)\,,
\end{equation}
and substitute this solution into the second equation:
\begin{equation}
	-d_G(x,x)\tilde \psi_2(x,z) + u_2 \left[\begin{array}{cc} x & z \\ x & z \end{array}\right] d_G(x,x) f(x,z) + u_2 \left[\begin{array}{cc} x & z \\ x & z \end{array}\right] g(x,z) = 0\,.
\end{equation}
From this equation, we get that
\begin{equation}
\begin{cases}
	\tilde \psi_2 = u_2 \left[\begin{array}{cc} x & z \\ x & z \end{array}\right] p_2(x,z) + d_G(x,x) \chi(x,z)\,, \\
	g(x,z) = d_G(x,x)(p_2(x,z) - f(x,z))
\end{cases}
\end{equation}
with $p_2(x,z)$ and $\chi(x,z)$ being some polynomials, so that the full wave function is
\begin{equation}\label{tens-prod-coh}
	\tilde \Psi = d_G(x,x)p_2(x,z) + \theta_2 u_2 \left[\begin{array}{cc} x & z \\ x & z \end{array}\right] p_2(x,z) + \theta_2\, d_G(x,x) \chi(x,z) = \widetilde{D}(p_2(x,z)) + \theta_2\, d_G(x,x) \chi(x,z)\,.
\end{equation}
Note that this explanation is independent on the choice of $d_G(x,x)$ and $u_2$.

\paragraph{Isomorphism of cohomologies having conjugated differentials.} Let us consider a generic case when a differential $\hat B$ is conjugated to a differential $\hat A$, i.e.
\begin{equation}\label{B-A-conj}
    \hat B = \hat \Phi^{-1} \hat A \hat \Phi \quad \text{or} \quad \hat A = \hat \Phi \hat B \hat \Phi^{-1}\,.
\end{equation}
Consider a generic cohomology of $\hat B$
\begin{equation}
    \Psi_B + \hat B \chi_B = \Psi_B + \hat \Phi^{-1} \hat A \hat \Phi \chi_B \; \in \; H^*(\hat B)\,.
\end{equation}
Then, we find the following cohomology of $\hat A$ by multiplying the last expression leftward by $\hat \Phi$:
\begin{equation}
    \hat \Phi \Psi_B + \hat A \hat \Phi \chi_B \; \in \; H^*(\hat A)
\end{equation}
as
\begin{equation}
    \hat A (\hat \Phi \Psi_B + \hat A \hat \Phi \chi_B) = \hat A \hat \Phi \Psi_B \overset{\eqref{B-A-conj}}{=} \hat \Phi \hat B \Psi_B = 0\,.
\end{equation}
Conversely, if we have a cohomology of $\hat A$:
\begin{equation}
    \Psi_A + \hat A \chi_A = \Psi_A + \hat \Phi \hat B \hat \Phi^{-1} \chi_A \; \in \; H^*(\hat A)\,.
\end{equation}
Then, we find the following cohomology of $\hat B$ by multiplying the last expression leftward by $\hat \Phi^{-1}$:
\begin{equation}
    \hat\Phi^{-1} \Psi_A + \hat B \hat \Phi^{-1} \chi_A \; \in \; H^*(\hat B)
\end{equation}
as
\begin{equation}
    \hat B (\hat\Phi^{-1} \Psi_A + \hat B \hat \Phi^{-1} \chi_A) = \hat B \hat \Phi^{-1} \Psi_A \overset{\eqref{B-A-conj}}{=} \hat \Phi^{-1} \hat A \Psi_A = 0\,.
\end{equation}
Thus, we have proved that for conjugated differentials $\hat A$ and $\hat B$
\begin{equation}
    H^*(\hat A) \cong H^*(\hat B)\,.
\end{equation}

\subsection{Quadratic reduction} \label{sec:quad_red}

Now we consider the differential for a diagram containing the dot resolution:
\begin{equation}\label{quad_reduction}
	\begin{array}{c}
		\begin{tikzpicture}
			\node[left] at (-1,0) {$\scriptstyle G$};
			\draw[fill=\myblue] (0,0) circle (1);
			\draw[fill=white] (0,0) circle (0.7);
			\draw[thick, postaction={decorate},decoration={markings, mark= at position 0.75 with {\arrow{stealth}}, mark= at position 0.25 with {\arrow{stealth}}}] (-0.5,-0.5) -- (0.5,0.5) node[left,pos=0.3] {$\scriptstyle x_4$} node[right,pos=0.7] {$\scriptstyle x_2$};
			\draw[thick, postaction={decorate},decoration={markings, mark= at position 0.75 with {\arrow{stealth}}, mark= at position 0.25 with {\arrow{stealth}}}] (0.5,-0.5) -- (-0.5,0.5) node[right,pos=0.3] {$\scriptstyle x_3$} node[left,pos=0.7] {$\scriptstyle x_1$};
			\draw[fill=\myred] (0,0) circle (0.07);
		\end{tikzpicture}
	\end{array}\quad\quad\quad
	\begin{aligned}
		&D\left[\begin{array}{cc|c} x_1 & x_2 & \theta_1 \\ x_4 & x_3 & \theta_2 \end{array}\right]=d_G\left[\begin{array}{cc} x_1 & x_2 \\ x_4 & x_3 \end{array}\right]+u_1\theta_1+s_1\theta_1^{\dagger}+u_2\theta_2+s_2\theta_2^{\dagger}\,,\\
		&s_1=x_1+x_2-x_3-x_4\,,\\
		&s_2=x_1x_2-x_3x_4\,.
	\end{aligned}
\end{equation}
The examples of such a graph and the derivations of the isomorphism of the cohomologies are presented is Sections~\ref{sec:quad-red-ex} and~\ref{sec:quad-red-II-MOY-ex}. Here, we again omit the dependence of the operators $D$ and $d_G$ on the variables of a graph $G$. According to (\ref{D^2-gen}) for $w = x^{N+1}$, in this case of two incoming and two outgoing edges,
\begin{equation}\label{4vert-d^2}
	d_G^2= -x_1^{N+1} - x_2^{N+1} + x_3^{N+1} + x_4^{N+1} = -u_1\left[\begin{array}{cc} x_1 & x_2 \\ x_4 & x_3 \end{array}\right]s_1-u_2\left[\begin{array}{cc} x_1 & x_2 \\ x_4 & x_3 \end{array}\right]s_2
\end{equation}
so that $D^2=0$, as it must be for the closed diagram.

Now, let us prove that the cohomology of the nilpotent operator $D\left[\begin{array}{cc|c} x_1 & x_2 & \theta_1 \\ x_4 & x_3 & \theta_2 \end{array}\right]$ splits into the cohomologies of $D_+$ and $D_-$, corresponding to diagrams in~\eqref{G+-}, subject to a boundary condition. 
\begin{equation}\label{G+-}
	\bar\MOY_+=\begin{array}{c}
		\begin{tikzpicture}
			\draw[thick, postaction={decorate},decoration={markings, mark= at position 0.75 with {\arrow{stealth}}, mark= at position 0.25 with {\arrow{stealth}}}] (-0.5,-0.5) to[out=45,in=315] node[left,pos=0.5,shift={(0.1,0)}] {$\scriptstyle x_1$} (-0.5,0.5);
			\draw[thick, postaction={decorate},decoration={markings, mark= at position 0.75 with {\arrow{stealth}}, mark= at position 0.25 with {\arrow{stealth}}}] (0.5,-0.5) to[out=135,in=225] node[right,pos=0.5,shift={(-0.1,0)}] {$\scriptstyle x_2$} (0.5,0.5);
			\begin{scope}[rotate=45]
				\node at (1.1,0) {$\scriptstyle G$};
			\end{scope}
			\draw[fill=\myblue, even odd rule] (0,0) circle (0.7)  (0,0) circle (0.9);
		\end{tikzpicture}
	\end{array},\quad \bar\MOY_-=\begin{array}{c}
		\begin{tikzpicture}
			\draw[dashed,thick, postaction={decorate},decoration={markings, mark= at position 0.75 with {\arrow{stealth}}, mark= at position 0.25 with {\arrow{stealth}}}] (-0.5,-0.5) -- (0.5,0.5) node[left,pos=0.3] {$\scriptstyle x_2$} node[right,pos=0.7] {$\scriptstyle x_2$};
			\draw[thick, postaction={decorate},decoration={markings, mark= at position 0.75 with {\arrow{stealth}}, mark= at position 0.25 with {\arrow{stealth}}}] (0.5,-0.5) -- (-0.5,0.5) node[right,pos=0.3] {$\scriptstyle x_1$} node[left,pos=0.7] {$\scriptstyle x_1$};
			\begin{scope}[rotate=45]
				\node at (1.1,0) {$\scriptstyle G$};
			\end{scope}
			\draw[fill=\myblue, even odd rule] (0,0) circle (0.7)  (0,0) circle (0.9);
		\end{tikzpicture}
	\end{array}\mbox{(NO VERTEX!!!)}\,.
\end{equation}
Consider a wave function:
\begin{equation}
	\Psi\left[\begin{array}{cc|c} x_1 & x_2 & \theta_1 \\ x_4 & x_3 & \theta_2 \end{array}\right]=\psi_0\left[\begin{array}{cc} x_1 & x_2 \\ x_4 & x_3 \end{array}\right]+\theta_1\psi_1\left[\begin{array}{cc} x_1 & x_2 \\ x_4 & x_3 \end{array}\right]+\theta_2\psi_2\left[\begin{array}{cc} x_1 & x_2 \\ x_4 & x_3 \end{array}\right]+\theta_1\theta_2\psi_{12}\left[\begin{array}{cc} x_1 & x_2 \\ x_4 & x_3 \end{array}\right]\,,
\end{equation}
where we decompose $\psi_0$ according to Appendix~\ref{sec:4f-deco}:
\begin{equation}
	\psi_0\left[\begin{array}{cc} x_1 & x_2 \\ x_4 & x_3 \end{array}\right] = \frac{1}{2}(1+h)\underbrace{\psi_0\left[\begin{array}{cc} x_1 & x_2 \\ x_2 & x_1 \end{array}\right]}_{\psi_-(x_1,x_2)} + \frac{1}{2}(1-h)\underbrace{\psi_0\left[\begin{array}{cc} x_1 & x_2 \\ x_1 & x_2 \end{array}\right]}_{\psi_+(x_1,x_2)}+s_1 v_1(\psi_0) + s_2 v_2(\psi_0)\,.
\end{equation}
It is important that 
\begin{equation}\label{psi+=psi-}
	\psi_+(x,x) = \psi_-(x,x)\,.
\end{equation}
Being a zero mode, the wave function satisfies
{\small \begin{equation}
	D\Psi = \left(d_G\psi_0+s_1\psi_1+s_2\psi_2\right)+\theta_1\left(-d_G\psi_1+u_1\psi_0-s_2\psi_{12}\right)+\theta_2\left(-d_G\psi_2+u_2\psi_0+s_1\psi_{12}\right)+\theta_1\theta_2\left(d_G\psi_{12}+u_1\psi_2-u_2\psi_1\right) = 0\,.
\end{equation}}

\noindent The terms in round brackets must vanish separately. From the first one, we obtain the following restrictions:
\begin{equation}\label{D+-psi+-=0}
\begin{aligned}
	D_+ \psi_+ &= D_- \psi_- = 0\,, \\
	\psi_1 &= -v_1(d_G \psi_0)\,,\quad \psi_2 = -v_2(d_G \psi_0)\,.
\end{aligned}
\end{equation}
where
\begin{equation}
	D_+ := d_G\left[\begin{array}{cc} x_1 & x_2 \\ x_1 & x_2 \end{array}\right]\,,\quad D_- := d_G\left[\begin{array}{cc} x_1 & x_2 \\ x_2 & x_1 \end{array}\right]\,.
\end{equation}
The second and the third brackets give
\begin{equation}\label{psi12}
	\psi_{12} = \frac{u_1 \psi_0 - d_G \psi_1}{s_2} = - \frac{u_2 \psi_0 - d_G \psi_2}{s_1}
\end{equation}
while the last equation does not give new restrictions due to~\eqref{psi12} and~\eqref{4vert-d^2}. Thus, the full wave function is
\ba\label{4v-fPsi} 
	\Psi &= \left(1-\theta_1v_1 d_G -\theta_2 v_2 d_G - \theta_1 \theta_2 \frac{u_2+d_G v_2 d_G}{s_1}\right)\psi_0 = \\
	&= \left(1-\theta_1v_1 d_G -\theta_2 v_2 d_G - \theta_1 \theta_2 \frac{u_2+d_G v_2 d_G}{s_1}\right)\underbrace{\left( \frac{1}{2}(1+h)\psi_- + \frac{1}{2}(1-h)\psi_+ \right)}_{\tilde \psi_0} + D(\theta_1 v_1(\psi_0) + \theta_2 v_2(\psi_0))\,.
\ea
It can also be proved that {\it any} wave functions $\psi_+(x_1,x_2)$ and $\psi_-(x_1,x_2)$ such that $D_+\psi_+ = D_- \psi_- = 0$ and $\psi_+(x,x)=\psi_-(x,x)$ produce a zero mode of $D$. Thus, we conclude that\footnote{At the r.h.s. inside the cohomology, we indicate the space at the first place and the differential acting on this space at the second place.}
\begin{tcolorbox}
	\begin{equation}
		H^*(D)\cong H^*\left(\Xi(x_1,x_2),{\rm diag}\left(D_+,D_-\right)\right)
	\end{equation}
\end{tcolorbox}
\noindent where
\begin{equation}\label{xi_const}
	\Xi(x_1,x_2):=\left\{\left(\begin{array}{c}
		\psi_+(x_1,x_2)\\
		\psi_-(x_1,x_2)
	\end{array}\right)\;\bigg|\;\psi_+(x,x)=\psi_-(x,x), \;\forall x\right\}\,.
\end{equation}
Note that the general solution of~\eqref{psi+=psi-} is given by
\begin{equation}\label{+=- sol}
	\psi_+(x_1,x_2) = \psi_-(x_1,x_2) + (x_1 - x_2) g(x_1,x_2)\,.
\end{equation}
If we take into account equation~\eqref{D+-psi+-=0}, we get the following restrictions:
\begin{equation}\label{+- cond}
	\begin{aligned}
		D_+\psi_- + (x_1 - x_2) D_+ g = 0\,,  \\
		D_- \psi_+ - (x_1 - x_2) D_- g = 0\,,
	\end{aligned}
\end{equation}
that can make $\psi_{\pm} \sim (x_1 - x_2)$, see Section~\ref{sec:box} for an example.
\paragraph{Degenerate case.} Let us consider the degenerate case of $\hat{d}:=d_G\big|_{\substack{x_3=x_1\\ x_4=x_2}}=d_G\big|_{\substack{x_3=x_2\\ x_4=x_1}}$ explicitly. If we make the variables change
\begin{equation}
	\psi_{\pm}(x_1,x_2) = f(x_1,x_2) \pm (x_1 - x_2) g(x_1,x_2)\,,
\end{equation}
we get from~\eqref{D+-psi+-=0}:
\begin{equation}
	\hat{d}f = \hat{d}g = 0\,,
\end{equation}
and
\begin{equation}\label{degen-psi0}
	\tilde \psi_0 = f(x_1,x_2) + (x_1 - x_2) h \, g(x_1,x_2) = f(x_1,x_2) + (x_3 - x_4) g(x_1,x_2)\,.
\end{equation}
So that
\begin{tcolorbox}
\begin{equation}\label{quad-red-degen}
	H^*(D) \cong H^*\left( D_+ = D_- \right) \oplus (x_3 - x_4)\, H^*\left( D_+ = D_- \right) \cong V_2 \otimes H^*\left( D_+ = D_- \right)\,.
\end{equation} 
\end{tcolorbox}

\subsection{2-step quadratic reduction}\label{sec:quad-red-2}

We again consider the differential operator for a diagram with a 4-valent vertex~\eqref{quad_reduction}. But now we would like to split the quadratic reduction from Section~\ref{sec:quad_red} into two steps. First, we do the linear reduction with respect to $s_1 = x_1 + x_2 - x_3 - x_4$. Second, we reduce by $s_2|_{x_4 = x_1 + x_2 - x_3} = (x_1 - x_3)(x_2 - x_3)$ where $s_2 = x_1 x_2 - x_3 x_4$. Thus, we first split the differential as follows:
\begin{equation}
	D\left[\begin{array}{cc|c} x_1 & x_2 & \theta_1 \\ x_4 & x_3 & \theta_2 \end{array}\right] = D_G \left[\begin{array}{cc|c} x_1 & x_2 &  \\ x_4 & x_3 & \theta_2 \end{array}\right] +u_1\theta_1+s_1\theta_1^{\dagger}\,,
\end{equation}
where
\begin{equation}
	D_G \left[\begin{array}{cc|c} x_1 & x_2 &  \\ x_4 & x_3 & \theta_2 \end{array}\right] = d_G \left[\begin{array}{cc} x_1 & x_2   \\ x_4 & x_3 \end{array}\right] + u_2 \theta_2 + s_2 \theta_2^\dag\,.
\end{equation}
We do the same splitting for the wave function
\begin{equation}
	\Psi \left[\begin{array}{cc|c} x_1 & x_2 & \theta_1 \\ x_4 & x_3 & \theta_2 \end{array}\right] = \Psi_0 \left[\begin{array}{cc|c} x_1 & x_2 &  \\ x_4 & x_3 & \theta_2 \end{array}\right] + \theta_1 \Psi_1 \left[\begin{array}{cc|c} x_1 & x_2 &  \\ x_4 & x_3 & \theta_2\,. \end{array}\right]
\end{equation}
Then, we write the zero mode condition:
\begin{equation}
	0 = D \Psi = \left( D_G \Psi_0 + s_1 \Psi_1 \right) + \theta_1 \left( u_1 \Psi_0 - D_G \Psi_1 \right)\,.
\end{equation}
Each bracket vanishes separately, and we get the following system:
\begin{equation}
\begin{cases}
	D_G \Psi_0 = -s_1 \Psi_1\,, \\
	u_1 \Psi_0 = D_G \Psi_1\,.
\end{cases}
\end{equation}
The first equation at the point $s_1 = 0$ leads to the condition
\begin{equation}
	D_G|_{s_1 = 0} \Psi_0|_{s_1 = 0} = 0\,.
\end{equation}
At $s_1 \neq 0$, we get
\begin{equation}
	\Psi_1 = - \frac{D_G}{s_1}\Psi_0 \quad \Lra \quad \Psi = \left( 1 - \theta_1 \frac{D_G}{s_1} \right)\Psi_0\,.
\end{equation}
One can now rewrite this cohomology in the explicit polynomial form:
\begin{equation}
	\Psi = \left( 1 - \theta_1 \frac{D_G - D_G|_{s_1 = 0}}{s_1} \right)\Psi_0|_{s_1 = 0} + D\left( \theta_1 \cdot \frac{\Psi_0 - \Psi_0|_{s_1 = 0}}{s_1} \right)\,.
\end{equation}
So that a full cohomology $\Psi$ is expressed through a cohomology $\Psi_0$ at the particular point $s_1 = 0$, and the last cohomology satisfies the simpler equation $D_G|_{s_1 = 0} \Psi_0|_{s_1 = 0} = 0$. Now, we split the differential $D_G|_{s_1 = 0}$:
\begin{equation}
	D_G|_{s_1 = 0} = u_2|_{s_1 = 0}\, \theta_2 + \underbrace{(x_1 - x_3)(x_2 - x_3)}_{\tilde s_2} \theta_2^\dag + d_G|_{s_1 = 0}\,,
\end{equation}
and the wave function:
\begin{equation}
	\Psi_0|_{s_1 = 0} = \psi_0(x_1,x_2,x_3) + \theta_2 \psi_2(x_1,x_2,x_3) \,.
\end{equation}
Let us now make the 3-variables function decomposition. It can be simply obtained\footnote{For the straightforward derivation see Appendix~\ref{sec:3f-deco}.} as the limit $s_1 = 0$ from 4-variables function decomposition, see Appendix~\ref{sec:4f-deco}, so that 
\begin{equation}
	\psi_0(x_1,x_2,x_3) = \frac{x_3 - x_2}{x_1 - x_2} \underbrace{\psi_0(x_1,x_2,x_1)}_{\psi_-(x_1,x_2)} + \frac{x_1 - x_3}{x_1 - x_2} \underbrace{\psi_0(x_1,x_2,x_2)}_{\psi_+(x_1,x_2)} + (x_1 - x_3)(x_2 - x_3)\tilde v_2(\psi_0)\,.
\end{equation}
It is important that there is the condition
\begin{equation}
	\psi_+(x,x) = \psi_-(x,x)\,.
\end{equation}
One can then solve the zero mode condition
\begin{equation}
	0 = D_G|_{s_1 = 0} \Psi_0|_{s_1 = 0} = \left( d_G|_{s_1 = 0}\,\psi_0 + (x_1 - x_3)(x_2 - x_3)\psi_2 \right) + \theta_2 \left( u_2|_{s_1 = 0}\, \psi_0 - d_G|_{s_1 = 0}\, \psi_2 \right)\,.
\end{equation}
Again, each bracket vanishes separately. At the points $x_1 = x_3$ and $x_2 = x_3$ we get the equations
\begin{equation}
	\underbrace{d_G \left[\begin{array}{cc} x_1 & x_2   \\ x_2 & x_1 \end{array}\right]}_{D_-} \psi_-(x_1,x_2) = 0\,,\quad \underbrace{d_G \left[\begin{array}{cc} x_1 & x_2   \\ x_1 & x_2 \end{array}\right]}_{D_+} \psi_+(x_1,x_2) = 0\,,
\end{equation}
and outside of these points
\begin{equation}
	\psi_2 = -\tilde v_2(d_G|_{s_1 = 0}\psi_0) \quad \Lra \quad \Psi_0|_{s_1 = 0} = \left( 1 - \theta_2 \tilde v_2 d_G|_{s_1 = 0} \right)\psi_0\,. 
\end{equation}
We rewrite the last cohomology:
\begin{equation}\label{res-2-step-QR}
	\Psi_0|_{s_1 = 0} = \left( 1 - \theta_2 \tilde v_2 d_G|_{s_1 = 0} \right)\left( \frac{x_3 - x_2}{x_1 - x_2}\, \psi_-(x_1,x_2) + \frac{x_1 - x_3}{x_1 - x_2}\, \psi_+(x_1,x_2) \right) + D_G|_{s_1 = 0}(\theta_2 \tilde v_2 (\psi_0))\,. 
\end{equation}
Thus, we again obtain that 
\begin{tcolorbox}
	\begin{equation}
		H^*(D)\cong H^*\left(\Xi(x_1,x_2),{\rm diag}\left(D_+,D_-\right)\right)
	\end{equation}
\end{tcolorbox}
\noindent where
\begin{equation}\label{xi_const}
	\Xi(x_1,x_2):=\left\{\left(\begin{array}{c}
		\psi_+(x_1,x_2)\\
		\psi_-(x_1,x_2)
	\end{array}\right)\;\bigg|\;\psi_+(x,x)=\psi_-(x,x), \;\forall x\right\}\,.
\end{equation}
Thus, one can either use the quadratic reduction from Section~\ref{sec:quad_red}, or one from this section.

\subsection{Second MOY move}\label{sec:MOY-II}

Let us consider the following MOY diagram:

\begin{equation}
	\begin{array}{c}
		\begin{tikzpicture}[scale=0.7]
			\node[left] at (-1.3,0) {$\scriptstyle G$};
			\begin{scope}[shift={(0,1)}]
				\draw[thick,-stealth] (-0.5,-0.5) to[out=90,in=270] (0.5,0.5);
				\draw[thick,-stealth] (0.5,-0.5) to[out=90,in=270] (-0.5,0.5);
				\draw[fill=\myred] (0,0) circle (0.1);
			\end{scope}
			\draw[thick,-stealth] (-0.5,-0.5) to[out=90,in=270] (0.5,0.5) -- (0.5,0.6);
			\draw[thick,-stealth] (0.5,-0.5) to[out=90,in=270] (-0.5,0.5) -- (-0.5,0.6);
			\draw[fill=\myred] (0,0) circle (0.1);
			\node[left] at (-0.25,-0.1) {$\scriptstyle x_4$};
			\node[right] at (0.25,-0.1) {$\scriptstyle x_3$};
			\node[left] at (-0.5,0.5) {$\scriptstyle y$};
			\node[right] at (0.5,0.5) {$\scriptstyle z$};
			\node[left] at (-0.25,1.0) {$\scriptstyle x_1$};
			\node[right] at (0.25,1.0) {$\scriptstyle x_2$};
			\draw[fill=\myblue, even odd rule] (0,0.5) circle (1.1)  (0,0.5) circle (1.4);
		\end{tikzpicture}
	\end{array}\quad\quad
	\begin{aligned}
	&D=d_G\left[\begin{array}{cc} x_1 & x_2 \\ x_4 & x_3 \end{array}\right]+u_1\left[\begin{array}{cc} x_1 & x_2 \\ y & z \end{array}\right]\theta_1+\underline{(x_1 + x_2 - y - z)}\theta_1^{\dagger}+u_2\left[\begin{array}{cc} x_1 & x_2 \\ y & z \end{array}\right]\theta_2+\\
	&+ \underline{(x_1 x_2 - y z)}\theta_2^{\dagger}+u_1\left[\begin{array}{cc} y & z \\ x_4 & x_3 \end{array}\right]\theta_3 + (y + z - x_3 - x_4)\theta_3^\dagger + u_2\left[\begin{array}{cc} y & z \\ x_4 & x_3 \end{array}\right]\theta_4 + (y z - x_3 x_4)\theta_4^\dagger\,.
	\end{aligned}
\end{equation}
We first make the quadratic reduction (according to Section~\ref{sec:quad_red}) with respect to $(x_1 + x_2 - y - z)$ and $(x_1 x_2 - y z)$ and end up with
\begin{equation}
\begin{aligned}
	D_+ = d_G\left[\begin{array}{cc} x_1 & x_2 \\ x_4 & x_3 \end{array}\right]+ u_1\left[\begin{array}{cc} x_1 & x_2 \\ x_4 & x_3 \end{array}\right]\theta_3 + (x_1 + x_2 - x_3 - x_4)\theta_3^\dagger + u_2\left[\begin{array}{cc} x_1 & x_2 \\ x_4 & x_3 \end{array}\right]\theta_4 + (x_1 x_2 - x_3 x_4)\theta_4^\dagger\,, \\
	D_- = d_G\left[\begin{array}{cc} x_1 & x_2 \\ x_4 & x_3 \end{array}\right]+ u_1\left[\begin{array}{cc} x_2 & x_1 \\ x_4 & x_3 \end{array}\right]\theta_3 + (x_1 + x_2 - x_3 - x_4)\theta_3^\dagger + u_2\left[\begin{array}{cc} x_2 & x_1 \\ x_4 & x_3 \end{array}\right]\theta_4 + (x_1 x_2 - x_3 x_4)\theta_4^\dagger\,.
\end{aligned}
\end{equation}
These differentials coincide. Thus, we arrive to the degenerate case of quadratic reduction and get~\eqref{quad-red-degen}: 
\begin{tcolorbox}
\begin{equation}
	H^*(D) \cong V_2 \otimes H^*(D_+ = D_-)\,.
\end{equation}
\end{tcolorbox}
In other words, with our sketchy notations, we get the isomorphism of the corresponding cohomologies:
\begin{equation}
	\begin{array}{c}
		\begin{tikzpicture}[scale=0.7]
			\node[left] at (-1.3,0) {$\scriptstyle G$};
			\begin{scope}[shift={(0,1)}]
				\draw[thick,-stealth] (-0.5,-0.5) to[out=90,in=270] (0.5,0.5);
				\draw[thick,-stealth] (0.5,-0.5) to[out=90,in=270] (-0.5,0.5);
				\draw[fill=\myred] (0,0) circle (0.1);
			\end{scope}
			\draw[thick,-stealth] (-0.5,-0.5) to[out=90,in=270] (0.5,0.5) -- (0.5,0.6);
			\draw[thick,-stealth] (0.5,-0.5) to[out=90,in=270] (-0.5,0.5) -- (-0.5,0.6);
			\draw[fill=\myred] (0,0) circle (0.1);
			\node[left] at (-0.25,-0.1) {$\scriptstyle x_4$};
			\node[right] at (0.25,-0.1) {$\scriptstyle x_3$};
			\node[left] at (-0.5,0.5) {$\scriptstyle y$};
			\node[right] at (0.5,0.5) {$\scriptstyle z$};
			\node[left] at (-0.25,1.0) {$\scriptstyle x_1$};
			\node[right] at (0.25,1.0) {$\scriptstyle x_2$};
			\draw[fill=\myblue, even odd rule] (0,0.5) circle (1.1)  (0,0.5) circle (1.4);
		\end{tikzpicture}
	\end{array}\cong \begin{array}{c}
		\begin{tikzpicture}[scale=0.7]
			\node[left] at (-1,0) {$\scriptstyle G$};
			\draw[thick,-stealth] (-0.5,-0.5) to[out=90,in=270] (0.5,0.5);
			\draw[thick,-stealth] (0.5,-0.5) to[out=90,in=270] (-0.5,0.5);
			\draw[fill=\myred] (0,0) circle (0.1);
			\node[left] at (-0.6,-0.8) {$\scriptstyle x_4$};
			\node[right] at (0.6,-0.8) {$\scriptstyle x_3$};
			\node[left] at (-0.6,0.8) {$\scriptstyle x_1$};
			\node[right] at (0.6,0.8) {$\scriptstyle x_2$};
			\draw[fill=\myblue, even odd rule] (0,0) circle (0.725)  (0,0) circle (1.0);
		\end{tikzpicture}
	\end{array}\otimes V_2\,.
\end{equation}

\subsection{Third MOY move I}\label{sec:box}

Consider an MOY diagram with two antiparallelly connected MOY vertices: 

{\small \begin{equation}\label{box-vert-diff}
	\begin{array}{c}
		\begin{tikzpicture}[scale=0.7]
			\node[left] at (-1.5,0) {$\scriptstyle G$};
			\draw[thick, postaction={decorate},decoration={markings, mark= at position 0.9 with {\arrow{stealth}}}] (0,0.4) to[out=0,in=120] (1,-0.7);
			\draw[thick, postaction={decorate},decoration={markings, mark= at position 0.9 with {\arrow{stealth}}}] (0,-0.4) to[out=180,in=300] (-1,0.7);
			\draw[thick, -stealth] (-1,-0.7) to[out=60,in=180] (0,0.4) -- (0.1,0.4);
			\draw[thick, -stealth] (1,0.7) to[out=240,in=0] (0,-0.4) -- (-0.1,-0.4);
			\draw[fill=burgundy] (-0.63,0) circle (0.1) (0.63,0) circle (0.1);
			\draw[fill=\myblue, even odd rule] (0,0) circle (1.2) (0,0) circle (1.5);
			\node[right] at (-1,0.7) {$\scriptstyle x_1$};
			\node[left] at (1,-0.7) {$\scriptstyle x_2$};
			\node[left] at (1,0.7) {$\scriptstyle x_3$};
			\node[right] at (-1,-0.7) {$\scriptstyle x_4$};
			\node[above] at (0,0.4) {$\scriptstyle x_5$};
			\node[below] at (0,-0.4) {$\scriptstyle x_6$};
		\end{tikzpicture}
	\end{array}\quad\quad \begin{aligned}
	&\MF=d_G\left[\begin{array}{cc}
		x_1 & x_3\\
		x_4 & x_2\\
	\end{array}\right]+u_1\left[\begin{array}{cc}
		x_1 & x_5\\
		x_4 & x_6\\
	\end{array}\right]\theta_1+\underline{(x_1 + x_5 - x_4 - x_6)}\theta_1^{\dagger}+u_2\left[\begin{array}{cc}
		x_1 & x_5\\
		x_4 & x_6\\
	\end{array}\right]\theta_2+\\
	&+(x_1 x_5 - x_4 x_6)\theta_2^{\dagger}+u_1\left[\begin{array}{cc}
		x_2 & x_6\\
		x_3 & x_5\\
	\end{array}\right]\theta_3+\underline{(x_2 + x_6 - x_3 - x_5)}\theta_3^{\dagger}+u_2\left[\begin{array}{cc}
		x_2 & x_6\\
		x_3 & x_5\\
	\end{array}\right]\theta_4+(x_2 x_6 - x_3 x_5)\theta_4^{\dagger}\,.
	\end{aligned}
\end{equation}}
\noindent We would like to introduce the following notations:
\begin{equation}
	d_+:=d_G\left[\begin{array}{cc}
		x_1 & x_2\\
		x_1 & x_2\\
	\end{array}\right] \qquad \begin{array}{c}
		\begin{tikzpicture}[scale=0.7]
			\node[left] at (-1.5,0) {$\scriptstyle G$};
			\draw[thick, postaction={decorate},decoration={markings, mark= at position 0.9 with {\arrow{stealth}}}] (-1,-0.7) to[out=60,in=300] (-1,0.7);
			\draw[thick, postaction={decorate},decoration={markings, mark= at position 0.9 with {\arrow{stealth}}}] (1,0.7) to[out=240,in=120] (1,-0.7);
			\draw[fill=\myblue, even odd rule] (0,0) circle (1.2) (0,0) circle (1.5);
			\node[right] at (-1,0.7) {$\scriptstyle x_1$};
			\node[left] at (1,-0.7) {$\scriptstyle x_2$};
			\node[left] at (1,0.7) {$\scriptstyle x_2$};
			\node[right] at (-1,-0.7) {$\scriptstyle x_1$};
		\end{tikzpicture}
	\end{array},\quad d_-:=d_G\left[\begin{array}{cc}
	x_1 & x_1\\
	x_2 & x_2\\
	\end{array}\right]\qquad \begin{array}{c}
		\begin{tikzpicture}[scale=0.7]
			\node[left] at (-1.5,0) {$\scriptstyle G$};
			\draw[thick, postaction={decorate},decoration={markings, mark= at position 0.9 with {\arrow{stealth}}}] (-1,-0.7) to[out=60,in=120] (1,-0.7);
			\draw[thick, postaction={decorate},decoration={markings, mark= at position 0.9 with {\arrow{stealth}}}] (1,0.7) to[out=240,in=300] (-1,0.7);
			\draw[fill=\myblue, even odd rule] (0,0) circle (1.2) (0,0) circle (1.5);
			\node[right] at (-1,0.7) {$\scriptstyle x_1$};
			\node[left] at (1,-0.7) {$\scriptstyle x_2$};
			\node[left] at (1,0.7) {$\scriptstyle x_1$};
			\node[right] at (-1,-0.7) {$\scriptstyle x_2$};
		\end{tikzpicture}
	\end{array}\,.
\end{equation}
First, we provide the two linear reductions with respect to underlined in~\eqref{box-vert-diff} terms and arrive to
\begin{equation}\label{box-vert-diff-red}
\begin{aligned}
	D' &= d_G\big|_{\substack{x_3 = x_2 - x_5 + x_6 \\
			x_4 = x_1 + x_5 - x_6}} + u_2\left[\begin{array}{cc}
		x_1 & x_5\\
		x_4 & x_6\\
	\end{array}\right]\bigg|_{x_4 = x_1 + x_5 - x_6}\theta_2 + u_2\left[\begin{array}{cc}
		x_2 & x_6\\
		x_3 & x_5\\
	\end{array}\right]\bigg|_{x_3 = x_2 - x_5 + x_6}\theta_4 + \\
	&+(x_1 - x_6)(x_5 - x_6)\theta_2^\dag + \underline{(x_5 - x_2)(x_5 - x_6)}\theta_4^\dag
\end{aligned}
\end{equation}
such that
\begin{equation}
	H^*(D) = H^*(D')\,.
\end{equation}
Second, we make the quadratic reduction with respect to the last term in~\eqref{box-vert-diff-red}, in other words, to the following module: 
\begin{equation}
	\IQ[x_1,x_2,x_5,x_6]/\langle (x_5 - x_2)(x_5 - x_6) \rangle\,. 
\end{equation}
At the point $x_5 = x_2$, $x_3 = x_6$, $x_4 = x_1 + x_2 - x_6$, we have:
\begin{equation}\label{D-Box}
	D_- = d_G\big|_{\substack{x_3 = x_6 \\
	x_4 = x_1 + x_2 - x_6}} + u_2\left[\begin{array}{cc}
	x_1 & x_2\\
	x_4 & x_6\\
\end{array}\right]\bigg|_{x_4 = x_1 + x_3 - x_6}\theta_2 + \underline{(x_1 - x_6)(x_2 - x_6)}\theta_2^\dag\,. 
\end{equation}
To find its cohomology we should further reduce with respect to the last summand. According to~\eqref{res-2-step-QR}, the result is  
\begin{equation}\label{Box-Psi-}
	\Psi_- = \left( 1 - \theta_2 \tilde v_2\, d_G\big|_{\substack{x_3 = x_6 \\
			x_4 = x_1 + x_2 - x_6}} \right)\left( \frac{x_6 - x_2}{x_1 - x_2}\, \psi_- + \frac{x_1 - x_6}{x_1 - x_2}\, \psi_+ \right) + D_- \Theta_-
\end{equation}
with
\begin{equation}
	d_\pm \psi_\pm = 0\,.
\end{equation}
At the point $x_5 = x_6$, $x_3 = x_2$, $x_4 = x_1$, we get:
\begin{equation}\label{D+Box}
	D_+ = d_+ + u_2\left[\begin{array}{cc}
		x_1 & x_6\\
		x_1 & x_6\\
	\end{array}\right]\theta_2\,.
\end{equation}
Its cohomology is spanned by $\theta_2 \psi_+$:
\begin{equation}
	\Psi_+ = \sum_{k = 0}^{N-2} x_6^k \theta_2 \xi_+ + D_+ \Theta_+\,, \quad d_+ \xi_+ = 0\,.
\end{equation}
For cohomologies of these differentials~\eqref{D+Box},~\eqref{D-Box}, we have the following condition:
\begin{equation}\label{Box-pm-cond}
	\Psi_-\big|_{x_2 = x_6} = \Psi_+\big|_{x_2 = x_6}\,. 
\end{equation}
Thus, we have:
\begin{equation}
	\sum_{k = 0}^{N-2} x_6^k \theta_2 \xi_+ - \psi_+ = \tilde D \tilde \Theta\,,
\end{equation}
where
\begin{equation}
	\tilde D = D_+\big|_{x_2 = x_6} = D_-\big|_{x_2 = x_6}\,.
\end{equation}
The only way for $\Psi_-\big|_{x_2 = x_6}$ and $\Psi_+\big|_{x_2 = x_6}$ not to lie in the image of $\tilde D$ is $\Psi_\pm$ to be proportional to the root $x_2 - x_6$. The component $\psi_+$ of the wave function $\Psi_-$ is a projection to $x_6 = x_2$ so it is annihilated Thus, the initial cohomology actually splits into the direct sum:
\begin{equation}
	H^*(D) = H^*(D_-) \oplus H^*(D_+)\,,
\end{equation}
and the wave functions of $D_\pm$ are 
\begin{equation}
\begin{aligned}
	\Psi_+ &= (x_2 - x_6)\sum_{k=0}^{N-3}x_6^k \theta_2 \xi_+ + D_+ \Theta_+\,, \\
	\Psi_- &= \left( 1 - \theta_2 \tilde v_2\, d_G\big|_{\substack{x_3 = x_6 \\
			x_4 = x_1 + x_2 - x_6}} \right)\frac{x_6 - x_2}{x_1 - x_2}\, \psi_- + D_-\Theta_-\,.
\end{aligned}
\end{equation}
The only remaining contribution of $\Psi_-$ lives at the point $x_1 = x_6 = x_3$, $x_5 = x_2 = x_4$, and
\begin{equation}
	H^*(D_-) \cong (x_2 - x_6) H^*(d_-)\,.
\end{equation}
Thus, in total:
\begin{tcolorbox}
\begin{equation}
	H^*(D) \cong (x_2 - x_6)\left[ H^*(d_-) \oplus \left( H^*(d_+) \otimes V_{N-2} \right) \right]\,.
\end{equation}
\end{tcolorbox}

\subsection{Third MOY move II}\label{sec:box-2}

In this section we present more rigorous proof of the third MOY relation~\eqref{MOYcat-loc_V}.

We again first reduce~\eqref{box-vert-diff} with respect to $\theta_1^\dag$ and $\theta_3^\dag$ and variables $x_3$ and $x_4$ (these are linear reductions so the nesting quotients are simply isomorphic) and arrive to the differential~\eqref{box-vert-diff-red}. Let us change the variables in the following way ($x_{ij}:=x_i-x_j$, $\bar x_{ij}:=x_i+x_j$):
\begin{equation}
	\theta_2=\frac{\xi_2+\xi_4}{2},\quad \theta_4=\frac{\xi_2-\hat \xi_4}{2},\quad \theta_2^{\dagger}=\xi_2^{\dagger}+ \xi_4^{\dagger},\quad \theta_4^{\dagger}=\xi_2^{\dagger}- \xi_4^{\dagger},\quad 
	\bar x_{56}=y+\bar x_{12}\,.
\end{equation}
Then we have:
\begin{equation}\label{MF0}
\begin{aligned}
		D' &= d_G\big|_{\substack{x_3 = x_2 - x_5 + x_6 \\
			x_4 = x_1 + x_5 - x_6}} + \underbrace{\frac{1}{2}\left( u_2\left[\begin{array}{cc}
		x_1 & x_5\\
		x_4 & x_6\\
	\end{array}\right]\bigg|_{x_4 = x_1 + x_5 - x_6} + u_2\left[\begin{array}{cc}
		x_2 & x_6\\
		x_3 & x_5\\
	\end{array}\right]\bigg|_{x_3 = x_2 - x_5 + x_6}\right)}_{w_2(x_1,x_2,x_5,x_6)}\xi_2 + \\ 
	&+ \underbrace{\frac{1}{2}\left( u_2\left[\begin{array}{cc}
			x_1 & x_5\\
			x_4 & x_6\\
		\end{array}\right]\bigg|_{x_4 = x_1 + x_5 - x_6} - u_2\left[\begin{array}{cc}
			x_2 & x_6\\
			x_3 & x_5\\
		\end{array}\right]\bigg|_{x_3 = x_2 - x_5 + x_6}\right)}_{w_4(x_1,x_2,x_5,x_6)}\xi_4 + \underbrace{x_{56}(x_{56}+x_{12})}_{S(x_{12},x_{56})}\xi_2^\dag -y x_{56}\xi_4^{\dagger}\,.
\end{aligned}
\end{equation}
Polynomial $w_4$ has the following properties:
\begin{equation}
	w_4\big|_{x_{56}=0}=x_{12}\,p_{N-2}(\underbrace{y|_{x_{56}=0}}_{\tilde y}|x_1,x_2),\quad w_4\big|_{x_{56}=-x_{12}}=0\,,
\end{equation}
where polynomial $p_{N-2}$ is a $\IQ$-monic polynomial of degree $N-2$ in $y$.
Further, we reduce with respect to $\xi_2^{\dagger}$, in other words, to the following module:
\begin{equation}
	\IQ[x_1,x_2,y,x_{56}]/\langle S \rangle\,.
\end{equation}
Consider a generic function of $x_1$, $x_2$, $y$, $x_{56}$ as a series in $x_{56}$:
\begin{equation}
	F(x_1,x_2,y,x_{56})=F_0(x_1,x_2,y)+x_{56}F_1(x_1,x_2,y)+\sum\lm_{k=2}^{\infty}x_{56}^{k-1}(x_{56}+x_{12})F_k(x_1,x_2,y)\,.
\end{equation}
So the module has the following structure:
\begin{equation}
	\begin{aligned}
		&F=F_+(x_1,x_2,y)+\frac{x_{56}}{x_{12}}\left(F_+(x_1,x_2,y)-F_-(x_1,x_2,y)\right)\;{\rm mod}\;S \,,\\
		&F_+(x_1,x_2,y):=F(x_1,x_2,y,0),\quad F_-(x_1,x_2,y):=F(x_1,x_2,y,-x_{12}),\quad F_+(x,x,y)=F_-(x,x,y)\,,\\
		& 1*_S x_{56}=x_{56}*_S 1=x_{56},\quad x_{56}*_Sx_{56}=-x_{12}x_{56}\,,\\
		& \left(F_+ +\frac{x_{56}}{x_{12}}(F_+ - F_-)\right) *_S\left(G_+ +\frac{x_{56}}{x_{12}}(G_+ - G_-)\right) =F_+ G_+ + \frac{x_{56}}{x_{12}}(F_+ G_+ - F_- G_-)\,.
	\end{aligned}
\end{equation}
Thus, for operator \eqref{MF0} we have:
\begin{equation}
	\begin{aligned}
		&D'\big|_{S}=\MF_+ +\frac{x_{56}}{x_{12}}\left(\MF_+-\MF_-\right)\,,\\
		&\MF_+=d_+ + x_{12}\,p_{N-2}(\tilde y|x_1,x_2)\,\xi_4\,,\\
		&\MF_-=d_- + x_{12}\,\underbrace{y|_{x_{56}=-x_{12}}}_{\tilde{\tilde y}}\,\xi_4^{\dagger}\,.
	\end{aligned}
\end{equation}
So, we conclude that the initial problem is isomorphic to the following problem
\begin{equation}
	\begin{aligned}
		&\MF_+\Psi_+(x_1,x_2,\tilde y,\xi_4)=0,\quad \MF_-\Psi_-(x_1,x_2,\tilde{\tilde y},\xi_4)=0,\quad \Psi_1(x,x,\tilde y,\xi_4)=\Psi_2(x,x,\tilde y,\xi_4)\,,\\ &\left(\Psi_+,\Psi_-\right)\sim\left(\Psi_++\MF_+\Theta_+,\Psi_-+\MF_-\Theta_-\right),\quad \Theta_+(x,x,\tilde y,\xi_4)=\Theta_-(x,x,\tilde y,\xi_4)\,.
	\end{aligned}
\end{equation}
Let us stress that:
\begin{equation}
	\MF_+\big|_{x_2=x_1}=\MF_-\big|_{x_2=x_1}=:\MF_0\,.
\end{equation}
We can calculate that a generic unconstrained solution to $\MF_+\Psi_+=0$, $\MF_-\Psi_-=0$ reads:
\begin{equation}
	\Psi_+(x_1,x_2,\tilde y,\xi_4)=\xi_4 \rho_+(x_1,x_2,\tilde y)+\MF_+\Theta_+,\quad \Psi_-(x_1,x_2,\tilde{\tilde y},\xi_4)= \rho_-(x_1,x_2,\tilde{\tilde y})+\MF_-\Theta_-,\quad d_{\pm}\rho_{\pm}=0\,.
\end{equation}
Now, we impose the constraint of coinciding points:
\begin{equation}
	\xi_4\rho_+(x,x,\tilde y)+\MF_0\Theta_+(x,x,\tilde y,\xi_4)=\rho_-(x,x,\tilde y)+\MF_0\Theta_-(x,x,\tilde y,\xi_4)\,.
\end{equation}
Clearly, $\rho_\pm(x,x,y)=\MF_0g_\pm(x,y)$, then shifting by $\Theta_{+}\to \Theta_{+}-g_{+}(x,x,y)\xi_4\,$, $\Theta_{-}\to \Theta_{-}-g_{-}(x,x,y)$ we will force $\rho_\pm$ to be annihilated in the point $x_1=x_2$, so without loss of generality, we choose:
\begin{equation}
	\rho_\pm=x_{12}\varphi_{\pm}\,.
\end{equation}
It is clear that $\varphi_\pm$ are annihilated by $d_\pm$, and conversely, ${\rm Im}\,d_{\pm}$ are mapped to ${\rm Im}\,\MF_{\pm}$.
So, we conclude that $\varphi_{\pm}\sim\psi_{\pm}$, where $\psi_{\pm}\in H^*(\MF_{\pm})$.

Eventually, let us note the following:
\begin{equation}\label{locktwocomponents}
	\begin{aligned}
		&x_{12}\left(\xi_4p_{N-2}(\tilde y|x_1,x_2)\psi_+,\frac{d_--d_+}{x_{12}}\psi_+\right)=\left(x_{12}\xi_4p_{N-2}(\tilde y|x_1,x_2)\psi_+,d_-\psi_+\right)=\left(\MF_+\psi_+,\MF_-\psi_+\right)\,,\\
		&x_{12}\left(\frac{d_+-d_-}{x_{12}}\xi_4\psi_-,\tilde{\tilde y}\psi_-\right)=\left(d_+\xi_4\psi_-,x_{12}\tilde{\tilde y}\psi_-\right)=\left(\MF_+\xi_4\psi_-,\MF_-\xi_4\psi_-\right)\,.
	\end{aligned}
\end{equation}
So, we conclude that cohomologies of $\MF$~\eqref{box-vert-diff} are spanned by the following vectors:
\begin{equation}
	\label{psilock}
	(x_{12}\xi_4y^k\psi_+(x_1,x_2),0),\;0\leq k \leq N-3,\quad (0,x_{12}\psi_-(x_1,x_2)),\quad \psi_{\pm}\in H^*(\MF_{\pm})\,.
\end{equation}

%
%
%


\section{Simplest example of horizontal reduction --- Reidemeister move I}\label{sec:IR}

%

Here, by the simplest example, we show how the horizontal reduction works. Consider a diagram with a loop:

\begin{equation}
	\begin{array}{c}
		\begin{tikzpicture}[scale=0.6]
			\draw[thick] (0.5,0.5) to[out=0,in=0] (0.5,-0.5);
			\draw[thick,postaction={decorate},decoration={markings, mark= at position 0.9 with {\arrow{stealth}}}] (0.5,-0.5) to[out=180,in=300] (-0.7,0.7);
			\draw[white, line width = 1.2mm] (-0.7,-0.7) to[out=60,in=180] (0.5,0.5);
			\draw[thick, postaction={decorate},decoration={markings, mark= at position 0.8 with {\arrow{stealth}}}] (-0.7,-0.7) to[out=60,in=180] (0.5,0.5);
			\draw[fill=\myblue, even odd rule] (0.4,0) circle (1.3) (0.4,0) circle (1.8);
			\node[left] at (-1.2,0.7) {$\scriptstyle G$};
			\node[right] at (-0.7,0.7) {$\scriptstyle x$};
			\node[right] at (-0.7,-0.7) {$\scriptstyle y$};
			\node[above] at (0.5,0.5) {$\scriptstyle z$};
			\node[below] at (0.5,-0.5) {$\scriptstyle z$};
		\end{tikzpicture}
	\end{array}=\;\left[\begin{array}{c}
		\begin{tikzpicture}[scale=0.6]
			\draw[thick,postaction={decorate},decoration={markings, mark= at position 0.9 with {\arrow{stealth}}}] (0.5,0.5) to[out=0,in=0] (0.5,-0.5) to[out=180,in=270] (-0.2,0) to[out=90,in=180] (0.5,0.5);
			\draw[thick,postaction={decorate},decoration={markings, mark= at position 0.9 with {\arrow{stealth}}}] (-0.7,-0.7) to[out=60,in=270] (-0.4,0) to[out=90,in=300] (-0.7,0.7);
			\draw[fill=\myblue, even odd rule] (0.4,0) circle (1.3) (0.4,0) circle (1.8);
			\node[left] at (-1.3,0.7) {$\scriptstyle G$};
			\node[right] at (-0.7,0.7) {$\scriptstyle x$};
			\node[right] at (-0.7,-0.7) {$\scriptstyle y$};
			\node[above] at (0.5,0.5) {$\scriptstyle z$};
			\node[below] at (0.5,-0.5) {$\scriptstyle z$};
		\end{tikzpicture}
	\end{array}\xrightarrow{\scriptstyle \epsilon^{(+)}\chi^{(+)}\left[\begin{array}{cc} x & z \\ y & z \end{array}\right]}\, \epsilon^{(+)}\begin{array}{c}
		\begin{tikzpicture}[scale=0.6]
			\draw[thick] (0.5,0.5) to[out=0,in=0] (0.5,-0.5);
			\draw[thick,postaction={decorate},decoration={markings, mark= at position 0.9 with {\arrow{stealth}}}] (0.5,-0.5) to[out=180,in=300] (-0.7,0.7);
			\draw[thick, postaction={decorate},decoration={markings, mark= at position 0.8 with {\arrow{stealth}}}] (-0.7,-0.7) to[out=60,in=180] (0.5,0.5);
			\draw[fill=burgundy] (-0.3,0) circle (0.1);
			\draw[fill=\myblue, even odd rule] (0.4,0) circle (1.3) (0.4,0) circle (1.8);
			\node[left] at (-1.3,0.7) {$\scriptstyle G$};
			\node[right] at (-0.7,0.7) {$\scriptstyle x$};
			\node[right] at (-0.7,-0.7) {$\scriptstyle y$};
			\node[above] at (0.5,0.5) {$\scriptstyle z$};
			\node[below] at (0.5,-0.5) {$\scriptstyle z$};
		\end{tikzpicture}
	\end{array}\right]\,.
\end{equation}
Here, $G$ denotes an arbitrary graph such that the whole diagram is closed. At first, we should find the vertical cohomologies. The zeroth differential is
\begin{equation}
	D_0 = d_G(x,y) + \pi_{xy}\theta_1 + (x-y)\theta_1^\dag + \pi_{zz} \theta_2\,.
\end{equation}
Linear reduction with respect to $(x-y)\theta_1^\dag$ reduces the differential to
\begin{equation}
	D'_0 = d_G(x,x) + \pi_{zz}\theta_2\,.
\end{equation}
So that according to~\eqref{zerom},~\eqref{tens-prod-coh}, the total wave function is
\begin{equation}
	\Psi_0 = \underbrace{\left( 1 - \theta_1 \frac{d_G(x,y)-d_G(x,x)}{x-y} \right)}_{U_0}\underbrace{\left[ \left( \sum_{j=0}^{N-1} a_j z^j \right)\theta_2 \psi_G(x) \right]}_{\psi_0}
\end{equation}
where the cohomology $\psi_G(x)$ satisfies $d_G(x,x)\psi_G(x) = 0$.

The first differential is
\begin{equation}
	D_1 = d_G(x,y) + u_1\left[\begin{array}{cc} x & z \\ y & z \end{array}\right]\theta_1 + (x-y)\theta_1^\dag + u_2\left[\begin{array}{cc} x & z \\ y & z \end{array}\right]\theta_2 + z(x-y)\theta_2^\dag\,.
\end{equation}
This operator is dealt in Section~\ref{sec:MOY-I}, and its reduction is called the first MOY move. Its wave function is
\begin{equation}
	\Psi_1 = \underbrace{\left( 1 - z \theta_1 \theta_2^\dag \right)\left( 1 - \theta_1 \frac{d_G(x,y) - d_G(x,x) + \left( u_2\left[\begin{array}{cc} x & z \\ y & z \end{array}\right] - u_2\left[\begin{array}{cc} x & z \\ x & z \end{array}\right] \right)\theta_2}{x-y} \right)}_{U_1}\underbrace{\left[ \left( \sum_{j=0}^{N-2} b_j z^j \right)\theta_2 \psi_G(x) \right]}_{\psi_1}\,,
\end{equation} 
and the reduced differential is
\begin{equation}
	D'_1 = d_G(x,x) + u_2\left[\begin{array}{cc} x & z \\ x & z \end{array}\right]\theta_2\,.
\end{equation}
Now, let us descend the action of $\chi^{(+)}\left[\begin{array}{cc} x & z \\ y & z \end{array}\right]$ to the cohomologies $\psi_{0,1}$. The following diagram is commutative: 
\begin{equation}
	\begin{array}{c}
	\begin{tikzpicture}
		\node(A) at (0,0) {$\psi_{0}$};
		\node(B) at (2.5,0) {$\psi_{1}$};
		\node(C) at (0,-1.5) {$\Psi_0$};
		\node(D) at (2.5,-1.5) {$\Psi_1$};
		\draw[-stealth] ([shift={(0,0.1)}]A.east) -- ([shift={(0,0.1)}]B.west) node[pos=0.5,above] {$\scriptstyle \chi'_0$};
		\draw[-stealth] ([shift={(0,0.1)}]C.east) -- ([shift={(0,0.1)}]D.west) node[pos=0.5,above] {$\scriptstyle \chi_{0}$};
		\draw[-stealth] ([shift={(-0.1,0)}]A.south) -- ([shift={(-0.1,0)}]C.north) node[pos=0.5,left] {$\scriptstyle U_{0}$};
		\draw[-stealth] ([shift={(-0.1,0)}]B.south) -- ([shift={(-0.1,0)}]D.north) node[pos=0.5,left] {$\scriptstyle U_{1}$};
	\end{tikzpicture}
\end{array}\,.
\end{equation}
Thus, 
\begin{equation}
	\chi'_0 \psi_0 = U_1^{-1}(\chi_0 \Psi_0)= (\chi_0 \Psi_0)\big|_{\substack{x=y\\
			\theta_1=0}}=\psi_0\,.
\end{equation}
The kernel of $\chi'_0$ is spanned by $u_2\left[\begin{array}{cc} x & z \\ x & z \end{array}\right]\psi_0$ as
\begin{equation}
	\chi'_0 \left(u_2\left[\begin{array}{cc} x & z \\ x & z \end{array}\right]\psi_0\right) = u_2\left[\begin{array}{cc} x & z \\ x & z \end{array}\right]\psi_0 = D'_1\left[ \left( \sum_{j=0}^{N-1} a_j z^j \right)\psi_G(x) \right]\,.
\end{equation}
Therefore, we get the following split of the horizontal complex:
\begin{equation}
\begin{aligned}
	0 \; \xrightarrow{0} \; \left( \sum_{j=0}^{N-2} a_j z^j \right)\theta_2 \psi_G(x) \; &\xrightarrow{\rm id} \; \left( \sum_{j=0}^{N-2} b_j z^j \right)\theta_2 \psi_G(x) \; \xrightarrow{0} \; 0  \\
	&\oplus \\
	0 \; \xrightarrow{0} \; &u_2\left[\begin{array}{cc} x & z \\ x & z \end{array}\right]\theta_2 \psi_G(x) \; \xrightarrow{0} \; 0
\end{aligned}
\end{equation}
The first subcomplex is exact, thus, it does not contribute to the cohomology. Therefore, we have the following isomorphism of the cohomologies:
\begin{equation}
	\begin{array}{c}
		\begin{tikzpicture}[scale=0.6]
			\draw[thick] (0.5,0.5) to[out=0,in=0] (0.5,-0.5);
			\draw[thick,postaction={decorate},decoration={markings, mark= at position 0.9 with {\arrow{stealth}}}] (0.5,-0.5) to[out=180,in=300] (-0.7,0.7);
			\draw[white, line width = 1.2mm] (-0.7,-0.7) to[out=60,in=180] (0.5,0.5);
			\draw[thick, postaction={decorate},decoration={markings, mark= at position 0.8 with {\arrow{stealth}}}] (-0.7,-0.7) to[out=60,in=180] (0.5,0.5);
			\draw[fill=\myblue, even odd rule] (0.4,0) circle (1.3) (0.4,0) circle (1.8);
			\node[right] at (-0.7,0.7) {$\scriptstyle x$};
			\node[right] at (-0.7,-0.7) {$\scriptstyle y$};
			\node[above] at (0.5,0.5) {$\scriptstyle z$};
			\node[below] at (0.5,-0.5) {$\scriptstyle z$};
		\end{tikzpicture}
	\end{array}\cong \begin{array}{c}
		\begin{tikzpicture}[scale=0.6]
			\draw[thick,postaction={decorate},decoration={markings, mark= at position 0.9 with {\arrow{stealth}}}] (-0.7,-0.7) to[out=60,in=270] (-0.4,0) to[out=90,in=300] (-0.7,0.7);
			\draw[fill=\myblue, even odd rule] (0.4,0) circle (1.3) (0.4,0) circle (1.8);
			\node[right] at (-0.7,0.7) {$\scriptstyle x$};
			\node[right] at (-0.7,-0.7) {$\scriptstyle x$};
		\end{tikzpicture}
	\end{array}\otimes u_2\left[\begin{array}{cc} x & z \\ x & z \end{array}\right]\theta_2\,.
\end{equation}


\section{Examples}\label{sec:ex}

In this section, we consider examples of computation of Poincar\'e polynomials. 

\subsection{1-unknot}

For the unknot one-intersection resolutions, we have the following complex:
\begin{equation}
	\begin{array}{c}
		\begin{tikzpicture}[scale=0.7]
			\draw[thick,-stealth] (0.5,-0.5) to[out=90,in=270] (-0.5,0.5);
			\draw[thick, white, line width = 1.5mm] (-0.5,-0.5) to[out=90,in=270] (0.5,0.5);
			\draw[thick,-stealth] (-0.5,-0.5) to[out=90,in=270] (0.5,0.5);
			\draw[thick] (-0.5,0.5) to[out=90,in=90] (-0.8,0.5) -- (-0.8,-0.5) to[out=270,in=270] (-0.5,-0.5);
			\begin{scope}[xscale=-1]
				\draw[thick] (-0.5,0.5) to[out=90,in=90] (-0.8,0.5) -- (-0.8,-0.5) to[out=270,in=270] (-0.5,-0.5);
			\end{scope}
		\end{tikzpicture}
	\end{array}=\left[0\longrightarrow \Gamma_0\overset{\epsilon_0\chi_0}{\longrightarrow}\epsilon_0\,\Gamma_1\longrightarrow 0\right]\,,
\end{equation}
where we use the following vertical morphisms:
\begin{equation}\label{unknot-ver-dif}
	\begin{aligned}
		&\Gamma_0=\begin{array}{c}
			\begin{tikzpicture}[scale=0.75]
				\draw[thick,-stealth] (0.5,-0.5) to[out=90,in=270] (0.3,0) to[out=90,in=270] (0.5,0.5);
				\draw[thick,-stealth] (-0.5,-0.5) to[out=90,in=270] (-0.3,0) to[out=90,in=270] (-0.5,0.5);
				\draw[thick] (-0.5,0.5) to[out=90,in=90] (-0.8,0.5) -- (-0.8,-0.5) to[out=270,in=270] (-0.5,-0.5);
				\begin{scope}[xscale=-1]
					\draw[thick] (-0.5,0.5) to[out=90,in=90] (-0.8,0.5) -- (-0.8,-0.5) to[out=270,in=270] (-0.5,-0.5);
				\end{scope}
				\node[right] at (-0.5,-0.5) {$\scriptstyle x$};
				\node[right] at (-0.5,0.5) {$\scriptstyle x$};
				\node[left] at (0.5,-0.5) {$\scriptstyle y$};
				\node[left] at (0.5,0.5) {$\scriptstyle y$};
				\node[right] at (0.25,0) {$\scriptstyle \theta_2$};
				\node[left] at (-0.2,0) {$\scriptstyle \theta_1$};
				\draw[fill=\mygreen] (0.3,0) circle (0.08);
				\draw[fill=\mygreen] (-0.3,0) circle (0.08);
			\end{tikzpicture}
		\end{array},\quad D_0 = x^N \theta_1 + y^N \theta_2\,,  \\
		&\Gamma_1=\begin{array}{c}
			\begin{tikzpicture}[scale=0.75]
				\draw[thick,-stealth] (0.5,-0.5) to[out=90,in=270] (-0.5,0.5);
				\draw[thick,-stealth] (-0.5,-0.5) to[out=90,in=270] (0.5,0.5);
				\draw[thick] (-0.5,0.5) to[out=90,in=90] (-0.8,0.5) -- (-0.8,-0.5) to[out=270,in=270] (-0.5,-0.5);
				\begin{scope}[xscale=-1]
					\draw[thick] (-0.5,0.5) to[out=90,in=90] (-0.8,0.5) -- (-0.8,-0.5) to[out=270,in=270] (-0.5,-0.5);
				\end{scope}
				\node[right] at (-0.5,-0.5) {$\scriptstyle x$};
				\node[right] at (-0.5,0.5) {$\scriptstyle x$};
				\node[left] at (0.5,-0.5) {$\scriptstyle y$};
				\node[left] at (0.5,0.5) {$\scriptstyle y$};
				\node[right] at (0.23,0) {$\scriptstyle \theta_2$};
				\node[left] at (-0.2,0) {$\scriptstyle \theta_1$};
				\draw[fill=\myred] (0,0) circle (0.1);
			\end{tikzpicture}
		\end{array},\quad D_1 = u_1 \left[\begin{array}{cc}
			x & y \\
			x & y \\
		\end{array}\right]\theta_1 + u_2 \left[\begin{array}{cc}
			x & y \\
			x & y \\
		\end{array}\right]\theta_2\,,
	\end{aligned}
\end{equation}
where
\begin{equation}
	u_1\left[\begin{array}{cc}
		x & y \\
		x & y \\
	\end{array}\right]=(N+1)\sum_{k=0}^N x^ky^{N-k},\quad u_2\left[\begin{array}{cc}
	x & y \\
	x & y \\
	\end{array}\right]=-(N+1)\sum_{k=0}^{N-1}x^k y^{N-1-k}\,.
\end{equation}
The cohomologies of similar operators have been already found in Section~\ref{sec:MOY-I}. However, here from the beginning, let us consider a generic case of the operator
\begin{equation}
	D(x,y|\theta_1,\theta_2) = f_1(x,y)\theta_1 + f_2(x,y)\theta_2
\end{equation}
with some polynomial functions $f_{1,2}$. We split the cohomology:
\begin{equation}
	\Psi(x,y|\theta_1,\theta_2) = \psi_0(x,y|\theta_1) + \theta_2 \psi_2(x,y|\theta_1)\,,
\end{equation}
and write the zero mode condition:
\begin{equation}
	0 = D\Psi = f_1 \theta_1 \psi_0 + \theta_2 (f_2 \psi_0 - u_1 \theta_1 \psi_2)\,.
\end{equation}
Each component vanishes separately:
\begin{equation}
\begin{cases}
	f_1 \theta_1 \psi_0 = 0 \quad \Lra \quad \psi_0(x,y|\theta_1) = \theta_1 f_0(x,y)\,, \\
	f_2 \psi_0 = f_1 \theta_1 \psi_2\,.
\end{cases}  
\end{equation}
Substituting the first answer to the second equation we get:
\begin{equation}
	f_2 \theta_1 f_0 = f_1 \theta_1 \psi_2
\end{equation}
with the solution
\begin{equation}
\begin{cases}
	f_0 = f_1 p(x,y)\,, \\
	\psi_2 = f_2 p(x,y) + \theta_1 \chi(x,y)
\end{cases}
\end{equation}
with some polynomial $p(x,y)$. Thus, the zero mode is
\begin{equation}
	\Psi(x,y|\theta_1,\theta_2) = \theta_2 \theta_1 \chi(x,y) + D(p(x,y))\,,
\end{equation}
and the cohomology is $\IQ[x,y]\theta_1\theta_2/\langle f_1(x,y),\, f_2(x,y) \rangle$. Thus, in our case

\begin{equation}\label{1-un-ver-coh}
	\Psi_0 = \sum_{i,j=0}^{N-1} a_{ij} x^i y^j \theta_1 \theta_2\,,\quad \Psi_1 = \sum_{i=0}^{N-2} \sum_{j=0}^{N-1} b_{ij} x^i y^j \theta_1 \theta_2\,.
\end{equation}
Then, we find
\begin{equation}
	\chi_0(\Psi_0) = \sum_{i,j=0}^{N-1} a_{ij} x^i y^j \cdot \chi_0(\theta_1 \theta_2) = \sum_{i,j=0}^{N-1} a_{ij} x^i y^j \theta_1 \theta_2\,,
\end{equation}
and the resulting hypercube on cohomologies is
\begin{equation}
	\begin{array}{c}
		\begin{tikzpicture}[scale=0.7]
			\draw[thick,-stealth] (0.5,-0.5) to[out=90,in=270] (-0.5,0.5);
			\draw[thick, white, line width = 1.5mm] (-0.5,-0.5) to[out=90,in=270] (0.5,0.5);
			\draw[thick,-stealth] (-0.5,-0.5) to[out=90,in=270] (0.5,0.5);
			\draw[thick] (-0.5,0.5) to[out=90,in=90] (-0.8,0.5) -- (-0.8,-0.5) to[out=270,in=270] (-0.5,-0.5);
			\begin{scope}[xscale=-1]
				\draw[thick] (-0.5,0.5) to[out=90,in=90] (-0.8,0.5) -- (-0.8,-0.5) to[out=270,in=270] (-0.5,-0.5);
			\end{scope}
		\end{tikzpicture}
	\end{array}=\left[0\longrightarrow \Psi_0\overset{\epsilon_0\cdot 1}{\longrightarrow}\epsilon_0\,\Psi_1\longrightarrow 0\right]\,.
\end{equation}
The cohomology of the descended horizontal differential $\tilde \fD = \epsilon_0$ is
\begin{equation}
	\Psi = x^{N-1} \sum_{i=0}^{N-1} c_i \, y^i \theta_1 \theta_2
\end{equation}
so that the Poincaré polynomial is
\begin{equation}
	{\rm KhR}^{\bigcirc}(A,q,T) = q^{2(N-1)}\cdot q^{N-1}[N]\cdot q^{2(1-N)} = q^{N-1}[N]\,,
\end{equation}
and it is the right answer.

\subsection{Mirror 1-unknot}

For the unknot one-intersection resolutions~\eqref{unknot-ver-dif}, there is the following complex:
\begin{equation} 
	\begin{array}{c}
		\begin{tikzpicture}[scale=0.7, xscale=-1]
			\draw[thick,-stealth] (0.5,-0.5) to[out=90,in=270] (-0.5,0.5);
			\draw[thick,white, line width = 1.5mm] (-0.5,-0.5) to[out=90,in=270] (0.5,0.5);
			\draw[thick,-stealth] (-0.5,-0.5) to[out=90,in=270] (0.5,0.5);
			\draw[thick] (-0.5,0.5) to[out=90,in=90] (-0.8,0.5) -- (-0.8,-0.5) to[out=270,in=270] (-0.5,-0.5);
			\begin{scope}[xscale=-1]
				\draw[thick] (-0.5,0.5) to[out=90,in=90] (-0.8,0.5) -- (-0.8,-0.5) to[out=270,in=270] (-0.5,-0.5);
			\end{scope}
		\end{tikzpicture}
	\end{array}=\left[0\longrightarrow \Gamma_1\overset{\epsilon_1\chi_1}{\longrightarrow}\epsilon_1\,\Gamma_0\longrightarrow 0\right]
\end{equation}
with the vertical cohomologies~\eqref{1-un-ver-coh}. The action of the horizontal morphism is
\begin{equation}
	\chi_1(\Psi_1) = \sum_{i=0}^{N-2} \sum_{j=0}^{N-1} b_{ij} x^i y^j \cdot \chi_1(\theta_1\theta_2) = (x-y)\cdot \sum_{i=0}^{N-2} \sum_{j=0}^{N-1} b_{ij} x^i y^j\theta_1\theta_2
\end{equation} 
so that the descended horizontal differential is $\tilde \fD = \epsilon_1 (x-y)\,$, and the resulting complex  
is
\begin{equation} 
	\begin{array}{c}
		\begin{tikzpicture}[scale=0.7, xscale=-1]
			\draw[thick,-stealth] (0.5,-0.5) to[out=90,in=270] (-0.5,0.5);
			\draw[thick,white, line width = 1.5mm] (-0.5,-0.5) to[out=90,in=270] (0.5,0.5);
			\draw[thick,-stealth] (-0.5,-0.5) to[out=90,in=270] (0.5,0.5);
			\draw[thick] (-0.5,0.5) to[out=90,in=90] (-0.8,0.5) -- (-0.8,-0.5) to[out=270,in=270] (-0.5,-0.5);
			\begin{scope}[xscale=-1]
				\draw[thick] (-0.5,0.5) to[out=90,in=90] (-0.8,0.5) -- (-0.8,-0.5) to[out=270,in=270] (-0.5,-0.5);
			\end{scope}
		\end{tikzpicture}
	\end{array}=\left[0\longrightarrow \Psi_1\overset{\epsilon_1 \cdot (x-y)}{\longrightarrow}\epsilon_1\,\Psi_0\longrightarrow 0\right]\,.
\end{equation}
The horizontal cohomology is
\begin{equation}
	\Psi = \epsilon_1 \sum_{i=0}^{N-1} c_i \, y^i \theta_1 \theta_2
\end{equation}
so that the Poincaré polynomial is
\begin{equation}
	{\rm KhR}^{\overline{\bigcirc}}(A,q,T) = q^{N-1}[N]\cdot q^{1-N} \cdot q^{3-N} = q^{3-N}[N]\,,
\end{equation}
and it is the right answer.

\subsection{Parallel Hopf link}\label{sec:par-Hopf-ex}

In this section, we present in details an exercise of computation of the Poincar\'e polynomial for the Hopf link with a generic $N$.
We decompose the Hopf link into MOY diagrams:
\begin{equation}
	\begin{array}{c}
		\begin{tikzpicture}[scale=0.7]
			\begin{scope}[shift={(0,1)}]
				\draw[thick] (0.5,-0.5) to[out=90,in=270] (-0.5,0.5);
				\draw[white, line width = 1.5mm] (-0.5,-0.5) to[out=90,in=270] (0.5,0.5);
				\draw[thick] (-0.5,-0.5) to[out=90,in=270] (0.5,0.5);
			\end{scope}
			\draw[thick,-stealth] (0.5,-0.5) to[out=90,in=270] (-0.5,0.5) -- (-0.5,0.6);
			\draw[white, line width = 1.5mm] (-0.5,-0.5) to[out=90,in=270] (0.5,0.5);
			\draw[thick,-stealth] (-0.5,-0.5) to[out=90,in=270] (0.5,0.5) -- (0.5,0.6);
			\draw[thick] (-0.5,1.5) to[out=90,in=90] (-1.2,1.5) -- (-1.2,-0.5) to[out=270,in=270] (-0.5,-0.5);
			\begin{scope}[xscale=-1]
				\draw[thick] (-0.5,1.5) to[out=90,in=90] (-1.2,1.5) -- (-1.2,-0.5) to[out=270,in=270] (-0.5,-0.5);
			\end{scope}
			\node[left] at (-0.5,-0.5) {$\scriptstyle x_1$};
			\node[right] at (0.5,-0.5) {$\scriptstyle x_2$};
			\node[left] at (-0.5,0.5) {$\scriptstyle x_3$};
			\node[right] at (0.5,0.5) {$\scriptstyle x_4$};
			\node[left] at (-0.5,1.5) {$\scriptstyle x_1$};
			\node[right] at (0.5,1.5) {$\scriptstyle x_2$};
			\node[above] at (0,0) {$\scriptstyle 1$};
			\node[above] at (0,1) {$\scriptstyle 2$};
			\node[left] at (-0.3,0) {$\scriptstyle \theta_1$};
			\node[right] at (0.3,0) {$\scriptstyle \theta_2$};
			\node[left] at (-0.3,1) {$\scriptstyle \theta_3$};
			\node[right] at (0.3,1) {$\scriptstyle \theta_4$};
		\end{tikzpicture}
	\end{array}=\left[\begin{array}{c}
		\begin{tikzpicture}
			\node(A) at (0,0) {$\MOY_{00}$};
			\node(B) at (3,0.7) {$\epsilon_0^{(1)}\MOY_{10}$};
			\node(C) at (3,-0.7) {$\epsilon_0^{(2)}\MOY_{01}$};
			\node(D) at (6,0) {$\epsilon_0^{(1)}\epsilon_0^{(2)}\MOY_{11}$};
			\path (A) edge[-stealth] node[above left] {{\scriptsize $\epsilon_0^{(1)}\chi_0\left[\begin{array}{cc|c} x_3 & x_4 & \theta_1 \\ x_1 & x_2 & \theta_2 \end{array}\right]$}} (B) (A) edge[-stealth] node[below left] {{\scriptsize $\epsilon_0^{(2)}\chi_0\left[\begin{array}{cc|c} x_1 & x_2 & \theta_3 \\ x_3 & x_4 & \theta_4 \end{array}\right]$}} (C) (B) edge[-stealth] node[above right] {{\scriptsize $\epsilon_0^{(2)}\chi_0\left[\begin{array}{cc|c} x_1 & x_2 & \theta_3 \\ x_3 & x_4 & \theta_4 \end{array}\right]$}} (D) (C) edge[-stealth] node[below right] {{\scriptsize $\epsilon_0^{(1)}\chi_0\left[\begin{array}{cc|c} x_3 & x_4 & \theta_1 \\ x_1 & x_2 & \theta_2 \end{array}\right]$}} (D);
		\end{tikzpicture}
	\end{array}\right]\,,
\end{equation}
where the following MOY diagrams are used:
\begin{equation}
	\begin{aligned}
		& \MOY_{00}=\begin{array}{c}
			\begin{tikzpicture}[scale=0.55]
				\draw[thick] (0.5,-0.5) to[out=90,in=270] (0.3,0) to[out=90,in=270] (0.5,0.5) -- (0.5,0.6);
				\draw[thick] (-0.5,-0.5) to[out=90,in=270] (-0.3,0) to[out=90,in=270] (-0.5,0.5) -- (-0.5,0.6);
				\begin{scope}[shift={(0,1)}]
					\draw[thick] (0.5,-0.5) to[out=90,in=270] (0.3,0) to[out=90,in=270] (0.5,0.5);
					\draw[thick] (-0.5,-0.5) to[out=90,in=270] (-0.3,0) to[out=90,in=270] (-0.5,0.5);
				\end{scope}
				\draw[thick, postaction={decorate},decoration={markings, mark= at position 0.6 with {\arrow{stealth}}}] (-0.5,1.5) to[out=90,in=90] (-1.2,1.5) -- (-1.2,-0.5) to[out=270,in=270] (-0.5,-0.5);
				\begin{scope}[xscale=-1]
					\draw[thick, postaction={decorate},decoration={markings, mark= at position 0.6 with {\arrow{stealth}}}] (-0.5,1.5) to[out=90,in=90] (-1.2,1.5) -- (-1.2,-0.5) to[out=270,in=270] (-0.5,-0.5);
				\end{scope}
			\end{tikzpicture}
		\end{array},\quad \MOY_{10}=\begin{array}{c}
			\begin{tikzpicture}[scale=0.55]
				\draw[thick] (0.5,-0.5) to[out=90,in=270] (-0.5,0.5);
				\draw[thick] (-0.5,-0.5) to[out=90,in=270] (0.5,0.5);
				\begin{scope}[shift={(0,1)}]
					\draw[thick] (0.5,-0.5) to[out=90,in=270] (0.3,0) to[out=90,in=270] (0.5,0.5);
					\draw[thick] (-0.5,-0.5) to[out=90,in=270] (-0.3,0) to[out=90,in=270] (-0.5,0.5);
				\end{scope}
				\draw[thick, postaction={decorate},decoration={markings, mark= at position 0.6 with {\arrow{stealth}}}] (-0.5,1.5) to[out=90,in=90] (-1.2,1.5) -- (-1.2,-0.5) to[out=270,in=270] (-0.5,-0.5);
				\begin{scope}[xscale=-1]
					\draw[thick, postaction={decorate},decoration={markings, mark= at position 0.6 with {\arrow{stealth}}}] (-0.5,1.5) to[out=90,in=90] (-1.2,1.5) -- (-1.2,-0.5) to[out=270,in=270] (-0.5,-0.5);
				\end{scope}
				\draw[fill=\myred] (0,0) circle (0.13);
			\end{tikzpicture}
		\end{array},\quad\MOY_{01}=\begin{array}{c}
			\begin{tikzpicture}[scale=0.55]
				\draw[thick] (0.5,-0.5) to[out=90,in=270] (0.3,0) to[out=90,in=270] (0.5,0.5);
				\draw[thick] (-0.5,-0.5) to[out=90,in=270] (-0.3,0) to[out=90,in=270] (-0.5,0.5);
				\begin{scope}[shift={(0,1)}]
					\draw[thick] (0.5,-0.5) to[out=90,in=270] (-0.5,0.5);
					\draw[thick] (-0.5,-0.5) to[out=90,in=270] (0.5,0.5);
				\end{scope}
				\draw[thick, postaction={decorate},decoration={markings, mark= at position 0.6 with {\arrow{stealth}}}] (-0.5,1.5) to[out=90,in=90] (-1.2,1.5) -- (-1.2,-0.5) to[out=270,in=270] (-0.5,-0.5);
				\begin{scope}[xscale=-1]
					\draw[thick, postaction={decorate},decoration={markings, mark= at position 0.6 with {\arrow{stealth}}}] (-0.5,1.5) to[out=90,in=90] (-1.2,1.5) -- (-1.2,-0.5) to[out=270,in=270] (-0.5,-0.5);
				\end{scope}
				\draw[fill=\myred] (0,1) circle (0.13);
			\end{tikzpicture}
		\end{array},\quad\MOY_{11}=\begin{array}{c}
			\begin{tikzpicture}[scale=0.55]
				\draw[thick,-stealth] (0.5,-0.5) to[out=90,in=270] (-0.5,0.5) -- (-0.5,0.6);
				\draw[thick,-stealth] (-0.5,-0.5) to[out=90,in=270] (0.5,0.5) -- (0.5,0.6);
				\begin{scope}[shift={(0,1)}]
					\draw[thick] (0.5,-0.5) to[out=90,in=270] (-0.5,0.5);
					\draw[thick] (-0.5,-0.5) to[out=90,in=270] (0.5,0.5);
				\end{scope}
				\draw[thick, postaction={decorate},decoration={markings, mark= at position 0.6 with {\arrow{stealth}}}] (-0.5,1.5) to[out=90,in=90] (-1.2,1.5) -- (-1.2,-0.5) to[out=270,in=270] (-0.5,-0.5);
				\begin{scope}[xscale=-1]
					\draw[thick, postaction={decorate},decoration={markings, mark= at position 0.6 with {\arrow{stealth}}}] (-0.5,1.5) to[out=90,in=90] (-1.2,1.5) -- (-1.2,-0.5) to[out=270,in=270] (-0.5,-0.5);
				\end{scope}
				\draw[fill=\myred] (0,0) circle (0.13) (0,1) circle (0.13);
			\end{tikzpicture}
		\end{array}\,.
	\end{aligned}
\end{equation}
There are the following vertical differentials:
\begin{equation}
	\begin{aligned}
		D_{00} &=\underbrace{\pi_{31}\theta_1+(x_3-x_1)\theta_1^\dagger +\pi_{42}\theta_2+(x_4-x_2)\theta_2^\dagger}_{d_{00}} +\pi_{13}\theta_3+(x_1-x_3) \theta_3^\dagger +\pi_{24}\theta_4+(x_2-x_4) \theta_4^\dagger \,,\\
		D_{10} &=\underbrace{u_1\left[\begin{array}{cc}
				x_3 & x_4 \\
				x_1 & x_2 \\
			\end{array}\right] \theta_1-s_1\theta_1^\dagger +u_2\left[\begin{array}{cc}
				x_3 & x_4 \\
				x_1 & x_2 \\
			\end{array}\right] \theta_2-s_2\theta_2^\dagger}_{d_{10}} +\pi_{13}\theta_3+(x_1-x_3) \theta_3^\dagger +\pi_{24}\theta_4+(x_2-x_4) \theta_4^\dagger \,,\\
		D_{01} &=\pi_{31}\theta_1+(x_3-x_1) \theta_1^\dagger +\pi_{42}\theta_2+(x_4-x_2) \theta_2^\dagger +\underbrace{u_1\left[\begin{array}{cc}
			x_1 & x_2 \\
			x_3 & x_4 \\
		\end{array}\right]\theta_3+s_1 \theta_3^\dagger + u_2\left[\begin{array}{cc}
			x_1 & x_2 \\
			x_3 & x_4 \\
		\end{array}\right]\theta_4+s_2 \theta_4^\dagger}_{d_{01}} \,,\\
		D_{11} & = \underbrace{u_1\left[\begin{array}{cc}
			x_3 & x_4 \\
			x_1 & x_2 \\
		\end{array}\right]\theta_1-s_1 \theta_1^\dagger +u_2\left[\begin{array}{cc}
			x_3 & x_4 \\
			x_1 & x_2 \\
		\end{array}\right]\theta_2-s_2 \theta_2^\dagger}_{d_{11}} +u_1\left[\begin{array}{cc}
			x_1 & x_2 \\
			x_3 & x_4 \\
		\end{array}\right]\theta_3+s_1 \theta_3^\dagger +u_2\left[\begin{array}{cc}
			x_1 & x_2 \\
			x_3 & x_4 \\
		\end{array}\right]\theta_4+s_2 \theta_4^\dagger 
	\end{aligned}
\end{equation}
where $s_1 = x_1+x_2-x_3-x_4$, $s_2 = x_1 x_2-x_3 x_4$. Now we would like to solve equations for zero modes of these operators applying techniques of Section~\ref{sec:lin_red} and of Section~\ref{sec:quad_red}.

\paragraph{Cohomology $\Psi_{00}$.} We can apply linear reduction twice with respect to $(x_1-x_3)\theta_3^\dag$ and $(x_2-x_4)\theta_4^\dag$, so that we now deal with the reduced operator 
\begin{equation}
	D_{00}\big|_{\substack{x_3=x_1,\;x_4=x_2\\
			\theta_3=0,\;\theta_4=0}} = (N+1)\,x_1^N\theta_1 + (N+1)\,x_2^N\theta_2\,.
\end{equation}
Its cohomology is
\begin{equation}
	\Psi_{00}\big|_{\substack{x_3=x_1,\;x_4=x_2\\
			\theta_3=0,\;\theta_4=0}} = \sum_{i,j=0}^{N-1} a_{ij}x_1^i x_2^j \theta_1 \theta_2\,.
\end{equation}
Then, we restore the suppressed variables:
\begin{equation}
	\Psi_{00}\big|_{\substack{x_4=x_2\\
			\theta_4=0}} = \left( 1 - \theta_3 \frac{d_{00}\big|_{x_4=x_2} - d_{00}\big|_{x_3=x_1,\;x_4=x_2}}{x_1 - x_3} \right)\Psi_{00}\big|_{\substack{x_3=x_1,\;x_4=x_2\\
			\theta_3=0,\;\theta_4=0}} = \sum_{i,j=0}^{N-1} a_{ij}x_1^i x_2^j (\theta_1 \theta_2 + \theta_3 \theta_2)\,,
\end{equation}
and finally,
\begin{equation}
	\Psi_{00} = \left( 1 - \theta_4 \frac{d_{00} - d_{00}\big|_{x_4=x_2}}{x_2 - x_4} \right)\Psi_{00}\big|_{\substack{x_4=x_2\\
			\theta_4=0}} = \sum_{i,j=0}^{N-1} a_{ij}x_1^i x_2^j \Omega_{00}
\end{equation}
with
\begin{equation}
	\Omega_{00} = \theta_1 \theta_2 + \theta_3 \theta_2 + \theta_1 \theta_4 + \theta_3 \theta_4\,.
\end{equation}

\paragraph{Cohomology $\Psi_{10}$.} 
Again, we reduce by $x_3 = x_1$ and $x_4 = x_2$ and first find the cohomology
\begin{equation}
    \Psi_{10}\big|_{\substack{x_3=x_1,\;x_4=x_2\\
    \theta_3=0,\;\theta_4=0}}=\sum\lm_{i=0}^{N-1}\sum\lm_{j=0}^{N-2}b_{ij}x_1^ix_2^j\theta_1\theta_2
\end{equation}
of the reduced operator\footnote{For details, see Section~\ref{sec:MOY-I}.}
\begin{equation}
	D_{10}\big|_{\substack{x_3=x_1,\;x_4=x_2\\
			\theta_3=0,\;\theta_4=0}}=(N+1)\left(x_1^N +x_1^{N-1}x_2+\ldots+x_2^{N}\right)\hat\theta_1-(N+1)\left(x_1^{N-1}+x_1^{N-2}x_2+\ldots+x_2^{N-1}\right)\hat\theta_2\,.
\end{equation}
In order to find the full cohomology, we should use formula~\eqref{zerom}. First, we find
\begin{equation}
\begin{aligned}
    \Psi_{10}\big|_{\substack{x_4=x_2\\
    \theta_4=0}} &= \left(1 - \theta_3 \frac{d_{10}\big|_{x_4=x_2}-d_{10}\big|_{x_3=x_1,\;x_4=x_2}}{x_1 - x_3}\right) \Psi_{10}\big|_{\substack{x_3=x_1,\;x_4=x_2\\
    \theta_3=0,\;\theta_4=0}}
    &= \sum\lm_{i=0}^{N-1}\sum\lm_{j=0}^{N-2}b_{ij}x_1^ix_2^j\left(\theta_1\theta_2 -\theta_2 \theta_3 + x_2 \theta_1 \theta_3 \right)\,.
\end{aligned}
\end{equation}
Second,
\begin{equation}
\begin{aligned}
    \Psi_{10} = \left(1 - \theta_4 \frac{d_{10} - d_{10}\big|_{x_4=x_2}}{x_2 - x_4}\right) \Psi_{10}\big|_{\substack{x_4=x_2\\
    \theta_4=0}} = \sum\lm_{i=0}^{N-1}\sum\lm_{j=0}^{N-2}b_{ij}x_1^ix_2^j \Omega_{10}
\end{aligned}
\end{equation}
with
\begin{equation}
\begin{aligned}
    \Omega_{10}&=\theta_1\theta_2+x_2\theta_1\theta_3-\theta_2\theta_3+x_3\theta_1\theta_4-\theta_2\theta_4+\left(x_3-x_2\right)\theta_3\theta_4+\\
	&+\frac{\left(u_1\left[\begin{array}{cc}
			x_3 & x_4 \\
			x_1 & x_2 \\
		\end{array}\right]+x_2 u_2\left[\begin{array}{cc}
		x_3 & x_4 \\
		x_1 & x_2 \\
		\end{array}\right]\right)-\left(u_1\left[\begin{array}{cc}
		x_3 & x_2 \\
		x_1 & x_2 \\
		\end{array}\right]+x_2 u_2\left[\begin{array}{cc}
		x_3 & x_2 \\
		x_1 & x_2 \\
		\end{array}\right]\right)}{x_4-x_2}\,\theta_1\theta_2\theta_3\theta_4\,.
\end{aligned}
\end{equation}

\paragraph{Cohomology $\Psi_{01}$.} Here, we reduce with respect to $(x_3-x_1) \theta_1^\dagger$ and  $(x_4-x_2) \theta_2^\dagger$ and get
\begin{equation}
	D_{01}\big|_{\substack{x_3=x_1,\;x_4=x_2\\
			\theta_1=0,\;\theta_2=0}} = u_1\left[\begin{array}{cc}
			x_1 & x_2 \\
			x_1 & x_2 \\
		\end{array}\right]\theta_3 + u_2\left[\begin{array}{cc}
		x_1 & x_2 \\
		x_1 & x_2 \\
	\end{array}\right]\theta_4\,.
\end{equation}
Its cohomology is
\begin{equation}
	\Psi_{01}\big|_{\substack{x_3=x_1,\;x_4=x_2\\
			\theta_1=0,\;\theta_2=0}} = \sum_{i=0}^{N-1} \sum_{j=0}^{N-2} c_{ij} x_1^i x_2^j \theta_3 \theta_4\,. 
\end{equation}
According to~\eqref{zerom}, we ``dress'' this wave function, so that
\begin{equation}
	\Psi_{01}\big|_{\substack{x_4=x_2\\
			\theta_2=0}} = \left( 1 - \theta_1 \frac{d_{01}\big|_{x_4=x_2} - d_{01}\big|_{x_3=x_1,\;x_4=x_2}}{x_3 - x_1} \right)\Psi_{01}\big|_{\substack{x_3=x_1,\;x_4=x_2\\
			\theta_1=0,\;\theta_2=0}} = \sum_{i=0}^{N-1} \sum_{j=0}^{N-2} c_{ij} x_1^i x_2^j (\theta_3 \theta_4 + \theta_1 \theta_4 - x_2 \theta_1 \theta_3 )\,,
\end{equation}
and
\begin{equation}
	\Psi_{01} = \left( 1 - \theta_2 \frac{d_{01} - d_{01}\big|_{x_4=x_2}}{x_4 - x_2} \right)\Psi_{01}\big|_{\substack{x_4=x_2\\
			\theta_2=0}} = \sum_{i=0}^{N-1} \sum_{j=0}^{N-2} c_{ij} x_1^i x_2^j \Omega_{01}
\end{equation}
where
\begin{equation}
\begin{aligned}
	\Omega_{01} &= \theta_3 \theta_4 + \theta_1 \theta_4 - x_2 \theta_1 \theta_3 + \theta_2 \theta_4 + (x_3 - x_2) \theta_1 \theta_2 - x_3 \theta_2 \theta_3 + \\
	&+\frac{\left(u_1\left[\begin{array}{cc}
			x_1 & x_2 \\
			x_3 & x_4 \\
		\end{array}\right]+x_2 u_2\left[\begin{array}{cc}
		x_1 & x_2 \\
		x_3 & x_4 \\
		\end{array}\right]\right)-\left(u_1\left[\begin{array}{cc}
		x_1 & x_2 \\
		x_3 & x_2 \\
		\end{array}\right]+x_2 u_2\left[\begin{array}{cc}
		x_1 & x_2 \\
		x_3 & x_2 \\
		\end{array}\right]\right)}{x_2-x_4}\,\theta_1\theta_2\theta_3\theta_4\,.
\end{aligned}
\end{equation}

\paragraph{Cohomology $\Psi_{11}$.}  Now, we proceed to $D_{11}$. The cohomology of this operator is given by $f(x_1,x_2)+(x_3-x_4)g(x_1,x_2)$, see~\eqref{degen-psi0}. Still, in order to get $\Psi_{11}$ we use formulas from Section~\ref{sec:quad_red}. In our case, we have
\begin{equation}
    \Psi(\Gamma_{11}) = \psi_0 + \theta_3 \psi_3 + \theta_4 \psi_4 + \theta_3 \theta_4 \psi_{34}
\end{equation}
with
\begin{equation}
    \psi_0 = \sum\lm_{i=0}^{N-1}\sum\lm_{j=0}^{N-2}\left(d_{ij}x_1^ix_2^j+(x_3-x_4)e_{ij}x_1^ix_2^j\right)\theta_1 \theta_2\,.
\end{equation}
In order to find other summands of $\Psi(\Gamma_{11})$, we need
\begin{equation}
    d_{11} \psi_0 = \sum\lm_{i=0}^{N-1}\sum\lm_{j=0}^{N-2}\left(d_{ij}x_1^ix_2^j+(x_3-x_4)e_{ij}x_1^ix_2^j\right)\left( -s_1\theta_2 + s_2 \theta_1 \right)\,.
\end{equation}
Then, according to formulas~\eqref{D+-psi+-=0},~\eqref{psi12}:
{\small \begin{equation}
\begin{aligned}
    \psi_3 &= -v_1(d_{11} \psi_0) = \sum\lm_{i=0}^{N-1}\sum\lm_{j=0}^{N-2}\left(d_{ij}x_1^ix_2^j+(x_3-x_4)e_{ij}x_1^ix_2^j\right) \theta_2\,, \\
    \psi_4 &= -v_2(d_{11} \psi_0) = -\sum\lm_{i=0}^{N-1}\sum\lm_{j=0}^{N-2}\left(d_{ij}x_1^ix_2^j+(x_3-x_4)e_{ij}x_1^ix_2^j\right) \theta_1\,, \\
    \psi_{34} &= \frac{u_1\left[\begin{array}{cc}
    		x_1 & x_2 \\
    		x_3 & x_4 \\
    	\end{array}\right] \psi_0 - d_{11}\psi_3}{s_2} = \sum\lm_{i=0}^{N-1}\sum\lm_{j=0}^{N-2}\left(d_{ij}x_1^ix_2^j+(x_3-x_4)e_{ij}x_1^ix_2^j\right) \left(\frac{u_1\left[\begin{array}{cc}
    	x_1 & x_2 \\
    	x_3 & x_4 \\
    	\end{array}\right]-u_1\left[\begin{array}{cc}
    	x_3 & x_4 \\
    	x_1 & x_2 \\
    	\end{array}\right]}{s_2}\theta_1 \theta_2 + 1\right)\,,
\end{aligned}
\end{equation}}
and in total we obtain
\begin{equation}\label{Psi11-Hopf}
	\begin{aligned}
	&\Psi(\Gamma_{11})=\sum\lm_{i=0}^{N-1}\sum\lm_{j=0}^{N-2}\left(d_{ij}x_1^ix_2^j+(x_3-x_4)e_{ij}x_1^ix_2^j\right)\Omega_{11}\,,\\
	&\Omega_{11}=\left(\theta_1\theta_2-\theta_2\theta_3+\theta_1\theta_4+\theta_3\theta_4+\frac{u_1\left[\begin{array}{cc}
			x_1 & x_2 \\
			x_3 & x_4 \\
		\end{array}\right]-u_1\left[\begin{array}{cc}
		x_3 & x_4 \\
		x_1 & x_2 \\
		\end{array}\right]}{x_1x_2-x_3x_4}\,\theta_1\theta_2\theta_3\theta_4\right)\,.
	\end{aligned}
\end{equation}

\paragraph{Khovanov--Rozansky polynomial.} For the action of horizontal morphisms on forms we find:
\begin{equation}
	\begin{aligned}
		&\chi_0\left[\begin{array}{cc|c} x_3 & x_4 & \theta_1 \\ x_1 & x_2 & \theta_2 \end{array}\right](\Omega_{00})=\Omega_{10}\,,\\
		&\chi_0\left[\begin{array}{cc|c} x_1 & x_2 & \theta_3 \\ x_3 & x_4 & \theta_4 \end{array}\right](\Omega_{00})=\Omega_{01}+\MF_{01}(\theta_1\theta_2\theta_3)\,,\\
		&\chi_0\left[\begin{array}{cc|c} x_1 & x_2 & \theta_3 \\ x_3 & x_4 & \theta_4 \end{array}\right](\Omega_{10})=\frac{1}{2}(x_1-x_2+x_3-x_4)\Omega_{11}+\frac{1}{2}\MF_{11}(\theta_1\theta_2\theta_3)+\frac{1}{2}\MF_{11}(\theta_1\theta_3\theta_4)\,,\\
		&\chi_0\left[\begin{array}{cc|c} x_3 & x_4 & \theta_1 \\ x_1 & x_2 & \theta_2 \end{array}\right](\Omega_{01})=\frac{1}{2}(x_1-x_2+x_3-x_4)\Omega_{11}-\frac{1}{2}\MF_{11}(\theta_1\theta_2\theta_3)+\frac{1}{2}\MF_{11}(\theta_1\theta_3\theta_4)\,.
	\end{aligned}	
\end{equation}
Such morphisms are differential operators only in odd variables, thus, their action extends on the cohomologies, and we obtain the following complex:
\begin{equation}
	\begin{array}{c}
		\begin{tikzpicture}[scale=0.7]
			\begin{scope}[shift={(0,1)}]
				\draw[thick] (0.5,-0.5) to[out=90,in=270] (-0.5,0.5);
				\draw[white, line width = 1.5mm] (-0.5,-0.5) to[out=90,in=270] (0.5,0.5);
				\draw[thick] (-0.5,-0.5) to[out=90,in=270] (0.5,0.5);
			\end{scope}
			\draw[thick,-stealth] (0.5,-0.5) to[out=90,in=270] (-0.5,0.5) -- (-0.5,0.6);
			\draw[white, line width = 1.5mm] (-0.5,-0.5) to[out=90,in=270] (0.5,0.5);
			\draw[thick,-stealth] (-0.5,-0.5) to[out=90,in=270] (0.5,0.5) -- (0.5,0.6);
			\draw[thick] (-0.5,1.5) to[out=90,in=90] (-1.2,1.5) -- (-1.2,-0.5) to[out=270,in=270] (-0.5,-0.5);
			\begin{scope}[xscale=-1]
				\draw[thick] (-0.5,1.5) to[out=90,in=90] (-1.2,1.5) -- (-1.2,-0.5) to[out=270,in=270] (-0.5,-0.5);
			\end{scope}
			\node[left] at (-0.5,-0.5) {$\scriptstyle x_1$};
			\node[right] at (0.5,-0.5) {$\scriptstyle x_2$};
			\node[left] at (-0.5,0.5) {$\scriptstyle x_3$};
			\node[right] at (0.5,0.5) {$\scriptstyle x_4$};
			\node[left] at (-0.5,1.5) {$\scriptstyle x_1$};
			\node[right] at (0.5,1.5) {$\scriptstyle x_2$};
			\node[above] at (0,0) {$\scriptstyle 1$};
			\node[above] at (0,1) {$\scriptstyle 2$};
			\node[left] at (-0.3,0) {$\scriptstyle \theta_1$};
			\node[right] at (0.3,0) {$\scriptstyle \theta_2$};
			\node[left] at (-0.3,1) {$\scriptstyle \theta_3$};
			\node[right] at (0.3,1) {$\scriptstyle \theta_4$};
		\end{tikzpicture}
	\end{array}=\left[\begin{array}{c}
		\begin{tikzpicture}
			\node(A) at (0,0) {$\Psi_{00}$};
			\node(B) at (3,0.7) {$\epsilon_0^{(1)}\Psi_{10}$};
			\node(C) at (3,-0.7) {$\epsilon_0^{(2)}\Psi_{01}$};
			\node(D) at (6,0) {$\epsilon_0^{(1)}\epsilon_0^{(2)}\Psi_{11}$};
			\path (A) edge[-stealth] node[above left] {{\scriptsize $\epsilon_0^{(1)}\cdot 1$}} (B) (A) edge[-stealth] node[below left] {{\scriptsize $\epsilon_0^{(2)}\cdot 1$}} (C) (B) edge[-stealth] node[above right] {{\scriptsize $\epsilon_0^{(2)}\cdot \frac{1}{2}(x_1 - x_2 + x_3 - x_4) $}} (D) (C) edge[-stealth] node[below right] {{\scriptsize $\epsilon_0^{(1)}\cdot \frac{1}{2}(x_1 - x_2 + x_3 - x_4)$}} (D);
		\end{tikzpicture}
	\end{array}\right]\,,
\end{equation}
or accurately,
\begin{equation}
	\begin{array}{c}
		\begin{tikzpicture}[scale=0.7]
			\begin{scope}[shift={(0,1)}]
				\draw[thick] (0.5,-0.5) to[out=90,in=270] (-0.5,0.5);
				\draw[white, line width = 1.5mm] (-0.5,-0.5) to[out=90,in=270] (0.5,0.5);
				\draw[thick] (-0.5,-0.5) to[out=90,in=270] (0.5,0.5);
			\end{scope}
			\draw[thick,-stealth] (0.5,-0.5) to[out=90,in=270] (-0.5,0.5) -- (-0.5,0.6);
			\draw[white, line width = 1.5mm] (-0.5,-0.5) to[out=90,in=270] (0.5,0.5);
			\draw[thick,-stealth] (-0.5,-0.5) to[out=90,in=270] (0.5,0.5) -- (0.5,0.6);
			\draw[thick] (-0.5,1.5) to[out=90,in=90] (-1.2,1.5) -- (-1.2,-0.5) to[out=270,in=270] (-0.5,-0.5);
			\begin{scope}[xscale=-1]
				\draw[thick] (-0.5,1.5) to[out=90,in=90] (-1.2,1.5) -- (-1.2,-0.5) to[out=270,in=270] (-0.5,-0.5);
			\end{scope}
			\node[left] at (-0.5,-0.5) {$\scriptstyle x_1$};
			\node[right] at (0.5,-0.5) {$\scriptstyle x_2$};
			\node[left] at (-0.5,0.5) {$\scriptstyle x_3$};
			\node[right] at (0.5,0.5) {$\scriptstyle x_4$};
			\node[left] at (-0.5,1.5) {$\scriptstyle x_1$};
			\node[right] at (0.5,1.5) {$\scriptstyle x_2$};
			\node[above] at (0,0) {$\scriptstyle 1$};
			\node[above] at (0,1) {$\scriptstyle 2$};
			\node[left] at (-0.3,0) {$\scriptstyle \theta_1$};
			\node[right] at (0.3,0) {$\scriptstyle \theta_2$};
			\node[left] at (-0.3,1) {$\scriptstyle \theta_3$};
			\node[right] at (0.3,1) {$\scriptstyle \theta_4$};
		\end{tikzpicture}
	\end{array}=\left[\begin{array}{c}
		{\small \begin{tikzpicture}
			\node(A) at (0,0) {$\left(\IQ[x_1,x_2]/\langle x_1^N,\,x_2^N \rangle \right)\Omega_{00}$};
			\node(B) at (4.5,0.7) {$\epsilon_0^{(1)}\left(\IQ[x_1,x_2]/\langle \tilde u_1,\, \tilde u_2 \rangle \right)\Omega_{10}$};
			\node(C) at (4.5,-0.7) {$\epsilon_0^{(2)}\left(\IQ[x_1,x_2]/\langle \tilde u_1,\, \tilde u_2 \rangle \right)\Omega_{01}$};
			\node(D) at (9.5,0) {$\epsilon_0^{(1)}\epsilon_0^{(2)}\left((\IQ[x_1,x_2]\oplus (x_3 - x_4)\IQ[x_1,x_2])/\langle \tilde u_1,\, \tilde u_2 \rangle \right)\Omega_{11}$};
			\path (A) edge[-stealth] node[above left] {{\scriptsize $\epsilon_0^{(1)}\cdot 1$}} (B) (A) edge[-stealth] node[below left] {{\scriptsize $\epsilon_0^{(2)}\cdot 1$}} (C) (B) edge[-stealth] node[above right] {{\scriptsize $\epsilon_0^{(2)}\cdot \frac{1}{2}(x_1 - x_2 + x_3 - x_4) $}} (D) (C) edge[-stealth] node[below right] {{\scriptsize $\epsilon_0^{(1)}\cdot \frac{1}{2}(x_1 - x_2 + x_3 - x_4)$}} (D);
		\end{tikzpicture}}
	\end{array}\right]
\end{equation}
where $\tilde u_1 = u_1\left[\begin{array}{cc}
	x_1 & x_2 \\
	x_1 & x_2 \\
\end{array}\right]$, $\tilde u_2 = u_2\left[\begin{array}{cc}
x_1 & x_2 \\
x_1 & x_2 \\
\end{array}\right]$.

Now, consider an example of $N=2$. In this case, $\tilde u_1 = 3 x_1^2 + 3 x_1 x_2 + 3 x_2^2$, $\tilde u_2 = -3 x_1 - 3 x_2$. The zeroth differential is $\tilde \fD_0 = (1\cdot) + (1\cdot)$, and the elements are mapped as follows
\begin{equation}
\begin{array}{ccc}
	a_{00} & \rightarrow & (a_{00},\, a_{00})\,, \\
	a_{10}x_1 & \rightarrow & (a_{10}x_1,\, a_{10}x_1)\,, \\
	a_{01}x_2 & \rightarrow & (-a_{01}x_1,\, -a_{01}x_1)\,, \\
	a_{11}x_1x_2 & \rightarrow & (0,0)\,.
\end{array}
\end{equation}
Thus,
\begin{equation}\label{Ker-Im-D0}
	\Ker(\tilde \fD_0) \cong x_2^{N-1} \IQ[x_1]/\langle x_1^N \rangle \,,\quad \Im(\tilde \fD_0) = \IQ[x_1,x_2]/\langle \tilde u_1,\, \tilde u_2 \rangle\,.
\end{equation}
The first differential is $\tilde \fD_1 = \frac{1}{2}(x_1 - x_2 + x_3 - x_4)((1\cdot) + (-1\cdot))$. The corresponding action on the elements is
\begin{equation}
\begin{array}{ccc}
	(b_{00},\,c_{00}) & \rightarrow & \left(\frac{1}{2}(2x_1 + x_3 - x_4)b_{00},\, -\frac{1}{2}(2x_1 + x_3 - x_4)c_{00} \right)\,, \\ \\
	(b_{10}x_1,\,c_{10}x_1) & \rightarrow & \left( \frac{1}{2}x_1(x_3-x_4)b_{10},\, -\frac{1}{2}x_1(x_3-x_4)c_{10} \right)\,.
\end{array}
\end{equation} 
Therefore, 
\begin{equation}\label{Ker-Im-D1}
	\Ker(\tilde \fD_1) \cong \IQ[x_1,x_2]/\langle \tilde u_1,\, \tilde u_2 \rangle\,,\quad \Im(\tilde \fD_1) \cong (x_3 - x_4)(\IQ[x_1,x_2]/\langle \tilde u_1,\, \tilde u_2 \rangle)\,.
\end{equation}
The cohomology is obtained by the proper multiplication by odd variables:
\begin{equation}\label{Psi-Hopf}
	\Psi = x_2^{N-1} \sum_{i=0}^{N-1} a_i x_1^i \Omega_{00} + \epsilon_0^{(1)} \epsilon_0^{(2)} \sum_{i=0}^{N-1}\sum_{j=0}^{N-2} x_1^i x_2^j \Omega_{11}\,.
\end{equation}
One can make sure that equations~\eqref{Ker-Im-D0},~\eqref{Ker-Im-D1},~\eqref{Psi-Hopf} hold for any $N$. Thus, the Poincar\'e polynomial is
\begin{equation}
	\KhR(A,q,T)= q^{N-1}[N] + q T^2 [N][N-1]\,,
\end{equation}
and it coincides with the known expression from the literature~\cite{Carqueville:2011zea}.

\subsection{Antiparallel Hopf link}

We have already provided the detailed derivation of the cohomology and the KhR polynomial in the previous subsection. So here, we just present final answers. They are again obtained with the use of linear and quadratic reductions.   

Decompose the antiparallel Hopf link into MOY diagrams:
\begin{equation}
	\begin{array}{c}
		\begin{tikzpicture}[scale=0.75]
			\begin{scope}[shift={(0,1)}]
				\draw[thick,stealth-] (-0.5,-0.5) to[out=90,in=270] (0.5,0.5);
				\draw[white, line width = 1.5mm] (0.5,-0.5) to[out=90,in=270] (-0.5,0.5);
				\draw[thick] (0.5,-0.5) to[out=90,in=270] (-0.5,0.5);
			\end{scope}
			\draw[thick,-stealth] (-0.5,-0.5) to[out=90,in=270] (0.5,0.5) -- (0.5,0.6);
			\draw[white, line width = 1.5mm] (0.5,-0.5) to[out=90,in=270] (-0.5,0.5);
			\draw[thick] (0.5,-0.5) to[out=90,in=270] (-0.5,0.5) -- (-0.5,0.6);
			\draw[thick] (-0.5,1.5) to[out=90,in=90] (-1.2,1.5) -- (-1.2,-0.5) to[out=270,in=270] (-0.5,-0.5);
			\begin{scope}[xscale=-1]
				\draw[thick] (-0.5,1.5) to[out=90,in=90] (-1.2,1.5) -- (-1.2,-0.5) to[out=270,in=270] (-0.5,-0.5);
			\end{scope}
			\node[left] at (-0.5,-0.5) {$\scriptstyle x_2$};
			\node[right] at (0.5,-0.5) {$\scriptstyle x_3$};
			\node[left] at (-0.5,0.5) {$\scriptstyle x_1$};
			\node[right] at (0.5,0.5) {$\scriptstyle x_4$};
			\node[left] at (-0.5,1.5) {$\scriptstyle x_2$};
			\node[right] at (0.5,1.5) {$\scriptstyle x_3$};
			\node[below] at (0,0) {$\scriptstyle 1$};
			\node[above] at (0,1) {$\scriptstyle 2$};
			\node[below] at (0,-0.3) {$\scriptstyle \theta_1$};
			\node[above] at (0,0) {$\scriptstyle \theta_2$};
			\node[below] at (0,1) {$\scriptstyle \theta_3$};
			\node[above] at (0,1.35) {$\scriptstyle \theta_4$};
		\end{tikzpicture}
	\end{array}=\left[\begin{array}{c}
		\begin{tikzpicture}
			\node(A) at (0,0) {$\MOY_{00}$};
			\node(B) at (3,0.7) {$\epsilon_0^{(1)}\MOY_{10}$};
			\node(C) at (3,-0.7) {$\epsilon_0^{(2)}\MOY_{01}$};
			\node(D) at (6,0) {$\epsilon_0^{(1)}\epsilon_0^{(2)}\MOY_{11}$};
			\path (A) edge[-stealth] node[above left] {{\scriptsize $\epsilon_0^{(1)}\chi_0\left[\begin{array}{cc|c} x_4 & x_3 & \theta_2 \\ x_1 & x_2 & \theta_1 \end{array}\right]$}} (B) (A) edge[-stealth] node[below left] {{\scriptsize $\epsilon_0^{(2)}\chi_0\left[\begin{array}{cc|c} x_1 & x_2 & \theta_3 \\ x_4 & x_3 & \theta_4 \end{array}\right]$}} (C) (B) edge[-stealth] node[above right] {{\scriptsize $\epsilon_0^{(2)}\chi_0\left[\begin{array}{cc|c} x_1 & x_2 & \theta_3 \\ x_4 & x_3 & \theta_4 \end{array}\right]$}} (D) (C) edge[-stealth] node[below right] {{\scriptsize $\epsilon_0^{(1)}\chi_0\left[\begin{array}{cc|c} x_4 & x_3 & \theta_2 \\ x_1 & x_2 & \theta_1 \end{array}\right]$}} (D);
		\end{tikzpicture}
	\end{array}\right]\,.
\end{equation}
The resolutions are as follows:
	\begin{figure}[h!]
		\centering
		\includegraphics[width=1.0\linewidth]{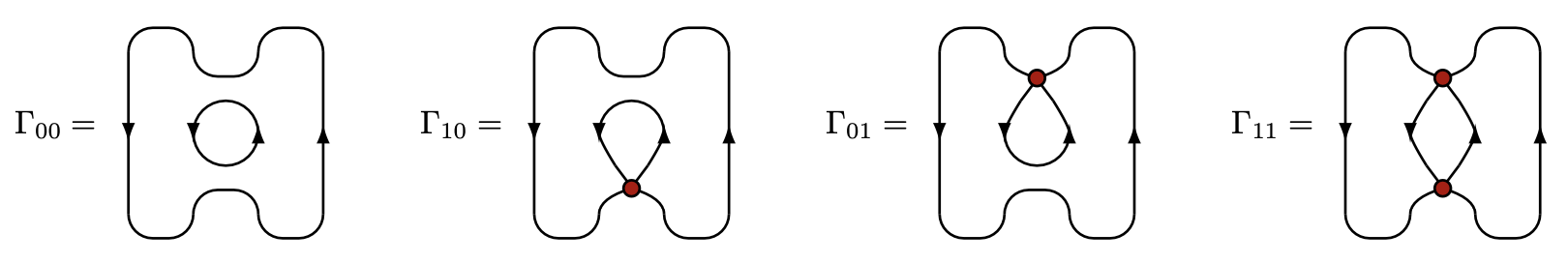}
	\end{figure}

\noindent The vertical operators are
\begin{equation}
	\begin{aligned}
		D_{00} &= \pi_{23} \theta_1 + (x_3 - x_2) \theta_1^\dag + \pi_{14} \theta_2 + (x_4 - x_1) \theta_2^\dag + \pi_{14} \theta_3 + (x_1 - x_4) \theta_3^\dag + \pi_{23} \theta_4 + (x_2 - x_3) \theta_4^\dag\,, \\
		D_{10} &= u_1\left[\begin{array}{cc}
			x_4 & x_3 \\
			x_1 & x_2 \\
		\end{array}\right] \theta_2 + (x_3 + x_4 - x_1 - x_2)\theta_2^\dag + u_2\left[\begin{array}{cc}
			x_4 & x_3 \\
			x_1 & x_2 \\
		\end{array}\right] \theta_1 + (x_3 x_4 - x_1 x_2)\theta_1^\dag + \\
		&+ \pi_{14} \theta_3 + (x_1 - x_4)\theta_3^\dag + \pi_{23} \theta_4 + (x_2 - x_3) \theta_4^\dag\,, \\
		D_{01} &= \pi_{23} \theta_1 + (x_3 - x_2) \theta_1^\dag + \pi_{14} \theta_2 + (x_4 - x_1) \theta_2^\dag + \\
		&+ u_1\left[\begin{array}{cc}
			x_1 & x_2 \\
			x_4 & x_3 \\
		\end{array}\right] \theta_3 + (x_1 + x_2 - x_3 - x_4)\theta_3^\dag + u_2\left[\begin{array}{cc}
			x_1 & x_2 \\
			x_4 & x_3 \\
		\end{array}\right] \theta_4 + (x_1 x_2 - x_3 x_4)\theta_4^\dag\,, \\
		D_{11} &= u_1\left[\begin{array}{cc}
			x_4 & x_3 \\
			x_1 & x_2 \\
		\end{array}\right] \theta_2 + (x_3 + x_4 - x_1 - x_2)\theta_2^\dag + u_2\left[\begin{array}{cc}
			x_4 & x_3 \\
			x_1 & x_2 \\
		\end{array}\right] \theta_1 + (x_3 x_4 - x_1 x_2)\theta_1^\dag + \\
		&+ u_1\left[\begin{array}{cc}
			x_1 & x_2 \\
			x_4 & x_3 \\
		\end{array}\right] \theta_3 + (x_1 + x_2 - x_3 - x_4)\theta_3^\dag + u_2\left[\begin{array}{cc}
			x_1 & x_2 \\
			x_4 & x_3 \\
		\end{array}\right] \theta_4 + (x_1 x_2 - x_3 x_4)\theta_4^\dag\,.
	\end{aligned}
\end{equation}
The wave functions (cohomologies) are
\begin{equation}
	\begin{aligned}
		\Psi_{00} &= \left( 1 - \theta_4 \frac{S_{00}^{(4)}}{x_2 - x_3} \right)\left( 1 - \theta_3 \frac{S_{00}^{(3)}}{x_1 - x_4} \right)\theta_1 \theta_2 x_1^k x_2^{k'}\,, \\
		\Psi_{10} &= \left( 1 - \theta_4 \frac{S_{10}^{(4)}}{x_2 - x_3} \right)\left( 1 - \theta_3 \frac{S_{10}^{(3)}}{x_1 - x_4} \right)\theta_1 \theta_2 x_1^j x_2^{k'}\,, \\
		\Psi_{01} &= \left( 1 - \theta_3 v_1 D_{01}\big|_{\theta_3=0,\,\theta_4=0} - \theta_4 v_2 D_{01}\big|_{\theta_3=0,\,\theta_4=0} - \theta_3 \theta_4 \frac{u_2\left[\begin{array}{cc}
				x_1 & x_2 \\
				x_4 & x_3 \\
			\end{array}\right] + D_{01}\big|_{\theta_3=0,\,\theta_4=0} v_2 D_{01}\big|_{\theta_3=0,\,\theta_4=0}}{x_1 + x_2 - x_3 - x_4} \right)\theta_1 \theta_2 x_1^j x_2^{k'}\,, \\
		\Psi_{11} &= \left( 1 - \theta_3 v_1 D_{11}\big|_{\theta_3=0,\,\theta_4=0} - \theta_4 v_2 D_{11}\big|_{\theta_3=0,\,\theta_4=0} - \theta_3 \theta_4 \frac{u_2\left[\begin{array}{cc}
				x_1 & x_2 \\
				x_4 & x_3 \\
			\end{array}\right] + D_{11}\big|_{\theta_3=0,\,\theta_4=0} v_2 D_{11}\big|_{\theta_3=0,\,\theta_4=0}}{x_1 + x_2 - x_3 - x_4} \right) \times \\
		&\times (\theta_1 \theta_2 x_1^j x_2^{k'} + (x_3 - x_4) \theta_1 \theta_2 x_1^j x_2^{k'})
	\end{aligned}
\end{equation}
where $0 \leq k,\, k' \leq n-1$, $0 \leq j \leq n-1$ and
\begin{equation}
	\begin{aligned}
		S_{00}^{(4)} &= (\pi_{23} - \pi_{22}) \theta_1 + (x_3 - x_2)\theta_1^\dag\,,\\
		S_{00}^{(3)} &= (\pi_{14} - \pi_{11}) \theta_2 + (x_4 - x_1)\theta_2^\dag\,, \\
		S_{10}^{(4)} &= \left( u_1\left[\begin{array}{cc}
			x_4 & x_3 \\
			x_1 & x_2 \\
		\end{array}\right] - u_1\left[\begin{array}{cc}
			x_4 & x_2 \\
			x_1 & x_2 \\
		\end{array}\right] \right)\theta_2 + (x_3 - x_2)\theta_2^\dag + \left( u_2\left[\begin{array}{cc}
			x_4 & x_3 \\
			x_1 & x_2 \\
		\end{array}\right] - u_2\left[\begin{array}{cc}
			x_4 & x_2 \\
			x_1 & x_2 \\
		\end{array}\right] \right)\theta_1 + x_4(x_3 - x_2)\theta_1^\dag\,, \\
		S_{10}^{(3)} &= \left( u_1\left[\begin{array}{cc}
			x_4 & x_2 \\
			x_1 & x_2 \\
		\end{array}\right] - u_1\left[\begin{array}{cc}
			x_1 & x_2 \\
			x_1 & x_2 \\
		\end{array}\right] \right)\theta_2 + (x_4 - x_1)\theta_2^\dag + \left( u_2\left[\begin{array}{cc}
			x_4 & x_2 \\
			x_1 & x_2 \\
		\end{array}\right] - u_2\left[\begin{array}{cc}
			x_1 & x_2 \\
			x_1 & x_2 \\
		\end{array}\right] \right)\theta_1 + x_2(x_4 - x_1)\theta_1^\dag\,.
	\end{aligned}
\end{equation}
The action of the horizontal morphisms on these wave functions is
\begin{equation}
	\begin{aligned}
		\chi_0\left[\begin{array}{cc|c}
			x_4 & x_3 & \theta_2 \\
			x_1 & x_2 & \theta_1
		\end{array}\right](\Psi_{00}) &= \Psi_{10}\,, \\
		\chi_0\left[\begin{array}{cc|c}
			x_1 & x_2 & \theta_2 \\
			x_4 & x_3 & \theta_1
		\end{array}\right](\Psi_{00}) &= \Psi_{01}\,, \\
		\chi_0\left[\begin{array}{cc|c}
			x_1 & x_2 & \theta_2 \\
			x_4 & x_3 & \theta_1
		\end{array}\right](\Psi_{10}) &= (x_2 - x_4)\Psi_{11} - D_{11}(\theta_1 \theta_2 \theta_3)\,, \\ 
		\chi_0\left[\begin{array}{cc|c}
			x_4 & x_3 & \theta_2 \\
			x_1 & x_2 & \theta_1
		\end{array}\right](\Psi_{01}) &= (x_2 - x_4)\Psi_{11} - D_{11}(\theta_1 \theta_2 \theta_3)\,.
	\end{aligned}
\end{equation}
The cohomology and the answer for the KhR polynomial of the Hopf link are the same as in Section~\ref{sec:par-Hopf-ex}, as it must be due to the fact that both the antiparallel and parallel diagrams correspond to the Hopf link.


\section{Reduction to Khovanov(-like) formalism}\label{sec:Kh-like}

In this section, we reduce the Khovanov--Rozansky operator formalism to the Khovanov(-like) technique. In Section~\ref{sec:N=2}, we show that KhR formalism in the $N=2$ case reduces to the usual Khovanov cycle calculus~\cite{khovanov2000categorification}. In Section~\ref{sec:bip-red} for bipartite links, we provide the reduction of the operator formalism for an arbitrary $N$ of double-complex to the single-complex cycle calculus on $N$-dimensional $q$-graded vector spaces with 3 universal morphisms.

\subsection{$N=2$ Khovanov--Rozansky reduction}\label{sec:N=2}

\noindent In~\cite{2508.05191}, we have shown the following quasi-isomorphism:
\begin{equation}\label{noMOY}
	\Gamma_1=\begin{array}{c}
		\begin{tikzpicture}[scale=1.2]
			\draw[thick, postaction={decorate},decoration={markings, mark= at position 0.75 with {\arrow{stealth}}, mark= at position 0.25 with {\arrow{stealth}}}] (-0.5,-0.5) -- (0.5,0.5) node[left,pos=0.3] {$\scriptstyle x_4$} node[right,pos=0.7] {$\scriptstyle x_2$};
			\draw[thick, postaction={decorate},decoration={markings, mark= at position 0.75 with {\arrow{stealth}}, mark= at position 0.25 with {\arrow{stealth}}}] (0.5,-0.5) -- (-0.5,0.5) node[right,pos=0.3] {$\scriptstyle x_3$} node[left,pos=0.7] {$\scriptstyle x_1$};
			\draw[fill=\myred] (0,0) circle (0.08);
			\begin{scope}[rotate=45]
				\node[\myblue] at (1.1,0) {$\scriptstyle G$};
				\draw[fill=\myblue, even odd rule] (0,0) circle (0.7)  (0,0) circle (0.9);
			\end{scope}
		\end{tikzpicture}
	\end{array}\cong\begin{array}{c}
	\begin{tikzpicture}[scale=0.7]
		\draw[thick] (-1,-0.7) to[out=60,in=120] (1,-0.7);
		\draw[thick] (1,0.7) to[out=240,in=300] (-1,0.7);
		\draw[fill=\myblue, even odd rule] (0,0) circle (1.2) (0,0) circle (1.5);
		\node[right] at (-1,0.7) {$\scriptstyle x_1$};
		\node[left] at (1,-0.7) {$\scriptstyle x_3$};
		\node[left] at (1,0.7) {$\scriptstyle x_2$};
		\node[right] at (-1,-0.7) {$\scriptstyle x_4$};
		\begin{scope}[rotate=45]
			\node[\myblue] at (1.8,0) {$\scriptstyle G$};
		\end{scope}
	\end{tikzpicture}
	\end{array}=\Gamma'_1
\end{equation}
by reducing the initial morphism
\begin{equation}
	D_{\Gamma_1}=u_1\left[\begin{array}{cc}
		x_1 & x_2 \\
		x_4 & x_3 \\
	\end{array}\right]\theta_1-3(x_3+x_4)\theta_2+(x_1+x_2-x_3-x_4)\theta_1^{\dagger}+(x_1x_2-x_3x_4)\theta_2^{\dagger}\,
\end{equation}
to the form
\begin{equation}\label{DD'}
	D_{\Gamma_1}=D_{\Gamma'_1}+\left[D_{\Gamma'_1},s\xi_1\xi_2\right]=e^{-s\xi_1\xi_2}D_{\Gamma'_1}e^{s\xi_1\xi_2}
\end{equation}
with
\begin{equation}
	D_{\Gamma'_1}:=\bar\pi_{12}\xi_1+\bar\pi_{34}\xi_2+\left(x_1+x_2\right)\xi_1^{\dagger}-\left(x_3+x_4\right)\xi_2^{\dagger}\,,
\end{equation}
where the following substitutions are implied
\begin{equation}
	\xi_1=\theta_1-\frac{1}{3}\theta_2^{\dagger},\quad \xi_2=\frac{1}{3}\theta_2^{\dagger},\quad \xi_1^{\dagger}=\theta_1^{\dagger},\quad \xi_2^{\dagger}=\theta_1^{\dagger}+3\theta_2\,,
\end{equation}
\begin{equation}
	s:=x_1+x_2+x_3+x_4\,,\quad \bar \pi_{ij} = \frac{x_i^3 + x_j^3}{x_i + x_j} = x_i^2 - x_i x_j + x_j^2\,.
\end{equation}
We will now show in details that the maps between $\Gamma'_1$ and
\begin{equation}\label{G0G1}
	\begin{aligned}
		&\Gamma_0=\begin{array}{c}
			\begin{tikzpicture}[scale=1.2]
				\draw[thick, postaction={decorate},decoration={markings, mark= at position 0.75 with {\arrow{stealth}}, mark= at position 0.25 with {\arrow{stealth}}}] (-0.5,-0.5) to[out=45,in=315] (-0.5,0.5);
				\draw[thick, postaction={decorate},decoration={markings, mark= at position 0.75 with {\arrow{stealth}}, mark= at position 0.25 with {\arrow{stealth}}}] (0.5,-0.5) to[out=135,in=225] (0.5,0.5);
				\node at (-0.2,0.45) {$\scriptstyle x_1$};
				\node at (-0.2,-0.45) {$\scriptstyle x_4$};
				\node at (0.2,0.45) {$\scriptstyle x_2$};
				\node at (0.2,-0.45) {$\scriptstyle x_3$};
				\begin{scope}[rotate=45]
					\node at (1.1,0) {$\scriptstyle G$};
				\end{scope}
				\draw[fill=\myblue, even odd rule] (0,0) circle (0.7)  (0,0) circle (0.9);
			\end{tikzpicture}
		\end{array} \quad \quad D_{\Gamma_0} = \pi_{14}\theta_1 + \pi_{23} \theta_2 + (x_1 - x_4)\theta_1^\dag + (x_2 - x_3)\theta_2^\dag
	\end{aligned}
\end{equation}
descend to the usual Khovanov $m$ and $\Delta$ after the closure of open ends of smoothings. The calculations of morphisms in the KhR cycle technique for bipartite links in Section~\ref{sec:bip-red} are analogous.

We know that the following commutation relation must hold:
\begin{equation}
	\chi_0 D_{\Gamma_0} = D_{\Gamma_1} \chi_0 = e^{-s\xi_1\xi_2}D_{\Gamma'_1}e^{s\xi_1\xi_2} \chi_0\,.
\end{equation}
Our new morphism $\chi'_0$ must commute with the proper $D$-morphisms too:
\begin{equation}
	\chi'_0 D_{\Gamma_0} = D_{\Gamma'_1} \chi'_0\,.
\end{equation}
Thus,
\begin{equation}\label{chi'}
	\boxed{\chi'_0 = e^{s\xi_1\xi_2} \chi_0=\chi_0 - \frac{1}{3}s\, \xi_1 \xi_2 \xi_2^\dagger \xi_1^\dagger + s\, \xi_1 \xi_2\,.}
\end{equation}
Let us rewrite the morphism $D_{\Gamma_0}$ in $\xi$-variables:
\begin{equation}
	D_{\Gamma_0} = \pi_{14} \xi_1 + \left( \pi_{14} + 3(x_2-x_3) \right)\xi_2 + \left( x_1-x_4-\frac{1}{3}\pi_{23} \right)\xi_1^\dagger + \frac{1}{3}\pi_{23}\xi_2^\dagger\,.
\end{equation}

\paragraph{Obtaining co-product.} Inside a knot complex, all open ends of smoothings are closed. Consider the case when $x_2 = - x_1$ and $x_4 = - x_3$. Then, there is the following part of the complex:
\begin{equation}
	\bigcirc \quad \overset{\chi'_0}{\longrightarrow} \quad \bigcirc \bigcirc\,,
\end{equation}
and the map $\chi'_0$ acting from the cohomology of $D_{\Gamma_0}$ to the cohomology of $D_{\Gamma'_1}$ must be Khovanov co-product $\Delta$. In this case, the cohomology of $D_{\Gamma_0}$ is spanned by $1+3\xi_1\xi_2$, and the cohomology of $D_{\Gamma'_1}$ is spanned by $\xi_1\xi_2$, and
\begin{equation}
	\begin{aligned}
		&\chi'_0(x_1,-x_1,x_3,-x_3) = \chi_0(x_1,-x_1,x_3,-x_3) = (x_1 - x_3)\xi_1^\dagger \xi_1 \xi_2 (\xi_2^\dagger - \xi_1^\dagger) + (\xi_1 + \xi_2)\xi_1^\dagger \xi_2^\dagger \xi_2 -\\
		&-\frac{1}{3}(2 x_1 - x_3)(\xi_1 + \xi_2)(\xi_2^\dagger - \xi_1^\dagger) + x_1 \xi_1 \xi_1^\dagger \xi_2 \xi_2^\dagger - 3 x_3 \xi_1 \xi_2 - \frac{1}{3}\xi_2^\dagger \xi_1^\dagger + \xi_1^\dagger (\xi_1+\xi_2)(\xi_2^\dagger-\xi_1^\dagger)\xi_2\,.
	\end{aligned}
\end{equation}
Then,
\begin{equation}
	\chi'_0(x_1,-x_1,x_3,-x_3)(1+3\xi_1\xi_2) = 3(x_1 - x_3)\xi_1 \xi_2\,.
\end{equation}
Thus, we obtain the following mapping of the cohomologies:
\begin{equation}
	\langle 1,\,x_1 = -x_3 \rangle \quad \overset{x_1 - x_3}{\longrightarrow} \quad \langle 1,\, x_1 \rangle \otimes \langle 1,\, x_3 \rangle\,.
\end{equation}
The explicit action is:
\begin{equation}
	x_1 - x_3: \quad 1 \; \mapsto \; x_1 \cdot 1 + 1 \cdot (-x_3)\,, \quad x_1 \; \mapsto \; x_1 \cdot (-x_3)\,,
\end{equation}
and after the grading shift, it exactly coincides with the action of the $\Delta$ operator from~\cite{dolotin2013introduction}:
\begin{equation}
	\Delta = (\vartheta_2^{(1)} \vartheta_1^{(3)} + \vartheta_1^{(1)} \vartheta_2^{(3)})\frac{\partial}{\partial\vartheta_1} + \vartheta_2^{(1)} \vartheta_2^{(3)} \frac{\partial}{\partial\vartheta_2}\,: \quad \vartheta_1 \; \mapsto \; \vartheta_2^{(1)} \vartheta_1^{(3)} + \vartheta_1^{(1)} \vartheta_2^{(3)}\,, \quad \vartheta_2 \; \mapsto \;  \vartheta_2^{(1)} \vartheta_2^{(3)}\,.
\end{equation}

\paragraph{Obtaining multiplication operator.} Now, let us perform another closure when $x_1 = x_4$ and $x_2 = x_3$. Then, $\Gamma_0$ becomes a pair of cycles and $\Gamma'_1$ turns into a cycle:
\begin{equation}
	\bigcirc\bigcirc \quad \overset{\chi'_0}{\longrightarrow} \quad \bigcirc\,,
\end{equation}
and the map $\chi'_0$ acting from the cohomology of $D_{\Gamma_0}$ to the cohomology of $D_{\Gamma'_1}$ must be Khovanov multiplication $m$. In this case, both the cohomologies of $D_{\Gamma_0}$ and of $D_{\Gamma'_1}$ are spanned by $\xi_1+\xi_2$, and
\begin{equation}
	\begin{aligned}
		\chi_0(x_1,x_2,x_2,x_1) &= (x_1 - x_2)\xi_1^\dagger \xi_1 \xi_2 (\xi_2^\dagger - \xi_1^\dagger) + (\xi_1 + \xi_2)\xi_1^\dagger \xi_2^\dagger \xi_2 - x_2 (\xi_1 + \xi_2)(\xi_2^\dagger - \xi_1^\dagger) + x_1 \xi_1 \xi_1^\dagger \xi_2 \xi_2^\dagger - \\
		&- 3 x_2 \xi_1 \xi_2 - \frac{1}{3}\xi_2^\dagger \xi_1^\dagger + \xi_1^\dagger (\xi_1 + \xi_2)(\xi_2^\dagger - \xi_1^\dagger)\xi_2\,.
	\end{aligned}
\end{equation}
Then,
\begin{equation}
	\chi'_0(x_1,x_2,x_2,x_1)(\xi_1 + \xi_2) = \chi_0(x_1,x_2,x_2,x_1)(\xi_1 + \xi_2) = \xi_1 + \xi_2\,.
\end{equation}
Thus, we obtain the following mapping of the cohomologies:
\begin{equation}
	\langle 1,\, x_1 \rangle \otimes \langle 1,\, x_2 \rangle \quad \overset{1}{\longrightarrow}\quad \langle 1,\, x_1 = - x_2 \rangle\,.
\end{equation}
The explicit action is:
\begin{equation}
	1:\quad 1\cdot 1 \; \mapsto \; 1\,,\quad x_1 \cdot 1 \; \mapsto \; x_1\,,\quad 1\cdot (-x_2) \; \mapsto \; -x_2=x_1\,,\quad x_1 \cdot x_2 \; \mapsto \; 0\,,
\end{equation}
and after the grading shift, it exactly coincides with the action of the $m$ operator from~\cite{dolotin2013introduction}:
\begin{equation}
	m = \vth_1 \frac{\p^2}{\p \vth_1^{(1)} \p \vth_1^{(2)}} + \vth_2 \left( \frac{\p^2}{\p \vth_1^{(1)} \p \vth_2^{(2)}}+\frac{\p^2}{\p \vth_2^{(1)} \p \vth_1^{(2)}} \right): \; \vth_1^{(1)}\vth_1^{(2)} \, \mapsto \, \vth_1\,, \; \vth_1^{(2)}\vth_2^{(1)} \, \mapsto \, \vth_2\,, \; \vth_2^{(2)}\vth_1^{(1)} \, \mapsto \, \vth_2\,, \; \vth_2^{(1)}\vth_2^{(2)} \, \mapsto \, 0\,.
\end{equation}

\paragraph{Generic case.} In general, in a knot complex, we have full smoothings glued from several $\Gamma_0$ and $\Gamma'_1$. The cut-and-join operator $\chi'_0$ acts inside local regions of these full smoothings, when one $\Gamma_0$ resolution transforms to $\Gamma'_1$ resolution, and depends locally only on two respective fermionic variables.

Our goal is to get rid of odd variables in order to reduce the problem to an already analyzed case. We do it by variables changes and conjugations by exponents. When $D$-operators are conjugated, $\chi'_0$ gets conjugated by the same exponents. However, these exponents commute with $\chi'_0$ because we get rid of ``far'' variables and make changes of Grassmann variables not coincident with $\chi'_0$ fermions. Proceeding this way, we transfer to new simple $D$-operators, whose cohomologies are tensor products of 2-dimensional spaces, and to morphisms~\eqref{chi'} giving rise to Khovanov $m$ and $\Delta$ operators.

\subsection{Bipartite cycle calculus}\label{sec:bip-red}

In our recent paper~\cite{2508.05191}, we have proved the bipartite vertex reduction of the Khovanov--Rozansky complex:

\begin{equation}\label{bip-KhR-complex}
	\begin{array}{c}
		\begin{tikzpicture}[scale=0.7]
			\draw[thick, postaction={decorate},decoration={markings, mark= at position 0.9 with {\arrow{stealth}}}] (0,0.4) to[out=0,in=120] (1,-0.7);
			\draw[thick, postaction={decorate},decoration={markings, mark= at position 0.9 with {\arrow{stealth}}}] (0,-0.4) to[out=180,in=300] (-1,0.7);
			\draw[white, line width = 1.2mm] (-1,-0.7) to[out=60,in=180] (0,0.4);
			\draw[thick, -stealth] (-1,-0.7) to[out=60,in=180] (0,0.4) -- (0.1,0.4);
			\draw[white, line width = 1.2mm] (1,0.7) to[out=240,in=0] (0,-0.4);
			\draw[thick, -stealth] (1,0.7) to[out=240,in=0] (0,-0.4) -- (-0.1,-0.4);
			\draw[fill=\myblue, even odd rule] (0,0) circle (1.2) (0,0) circle (1.5);
			\node[right] at (-1,0.7) {$\scriptstyle x_1$};
			\node[left] at (1,-0.7) {$\scriptstyle x_2$};
			\node[left] at (1,0.7) {$\scriptstyle x_3$};
			\node[right] at (-1,-0.7) {$\scriptstyle x_4$};
			\node[above] at (0,0.4) {$\scriptstyle x_5$};
			\node[below] at (0,-0.4) {$\scriptstyle x_6$};
		\end{tikzpicture}
	\end{array}\cong \;\left[\underset{D_+}{\begin{array}{c}
	\begin{tikzpicture}[scale=0.7]
		\node[left] at (-1.5,0) {$\scriptstyle G$};
		\draw[thick, postaction={decorate},decoration={markings, mark= at position 0.9 with {\arrow{stealth}}}] (-1,-0.7) to[out=60,in=300] (-1,0.7);
		\draw[thick, postaction={decorate},decoration={markings, mark= at position 0.9 with {\arrow{stealth}}}] (1,0.7) to[out=240,in=120] (1,-0.7);
		\draw[fill=\myblue, even odd rule] (0,0) circle (1.2) (0,0) circle (1.5);
		\node[right] at (-1,0.7) {$\scriptstyle x_1$};
		\node[left] at (1,-0.7) {$\scriptstyle x_2$};
		\node[left] at (1,0.7) {$\scriptstyle x_2$};
		\node[right] at (-1,-0.7) {$\scriptstyle x_1$};
	\end{tikzpicture}
	\end{array}} \xrightarrow[]{x_1 - x_2} \underset{D_+}{\begin{array}{c}
	\begin{tikzpicture}[scale=0.7]
		\node[left] at (-1.5,0) {$\scriptstyle G$};
		\draw[thick, postaction={decorate},decoration={markings, mark= at position 0.9 with {\arrow{stealth}}}] (-1,-0.7) to[out=60,in=300] (-1,0.7);
		\draw[thick, postaction={decorate},decoration={markings, mark= at position 0.9 with {\arrow{stealth}}}] (1,0.7) to[out=240,in=120] (1,-0.7);
		\draw[fill=\myblue, even odd rule] (0,0) circle (1.2) (0,0) circle (1.5);
		\node[right] at (-1,0.7) {$\scriptstyle x_1$};
		\node[left] at (1,-0.7) {$\scriptstyle x_2$};
		\node[left] at (1,0.7) {$\scriptstyle x_2$};
		\node[right] at (-1,-0.7) {$\scriptstyle x_1$};
	\end{tikzpicture}
	\end{array}} \xrightarrow[]{S = \frac{D_- - D_+}{x_1 - x_2}}
	\underset{D_-}{\begin{array}{c}
		\begin{tikzpicture}[scale=0.7]
			\node[left] at (-1.5,0) {$\scriptstyle G$};
			\draw[thick, postaction={decorate},decoration={markings, mark= at position 0.9 with {\arrow{stealth}}}] (-1,-0.7) to[out=60,in=120] (1,-0.7);
			\draw[thick, postaction={decorate},decoration={markings, mark= at position 0.9 with {\arrow{stealth}}}] (1,0.7) to[out=240,in=300] (-1,0.7);
			\draw[fill=\myblue, even odd rule] (0,0) circle (1.2) (0,0) circle (1.5);
			\node[right] at (-1,0.7) {$\scriptstyle x_1$};
			\node[left] at (1,-0.7) {$\scriptstyle x_2$};
			\node[left] at (1,0.7) {$\scriptstyle x_1$};
			\node[right] at (-1,-0.7) {$\scriptstyle x_2$};
		\end{tikzpicture}
	\end{array}}\right]\,.
\end{equation}
Here, we show that this reduced complex allows to built the Sh, $m$ and $\Delta$ operators guessed in~\cite{2506.08721}. 

Gluing the upper and lower strands, we obtain:
\begin{equation}
\begin{aligned}
	\bigcirc \quad D_+ &= \pi_{x_1 x_2}(\theta_1 + \theta_2) + (x_2 - x_1)(\theta_1^\dag - \theta_2^\dag)\,, \\
	\bigcirc \bigcirc \quad D_- &= \pi_{x_1 x_1} \theta_1 + \pi_{x_2 x_2} \theta_2\,.
\end{aligned}
\end{equation}
The cohomology of the differential $D_+$ is spanned by $\theta_1 + \theta_2$ and the cohomology of the operator $D_-$ is spanned by $\theta_1 \theta_2$. Induce the action of the horizontal morphism $S$ on the cohomologies:
\begin{equation}
	S(\theta_1 + \theta_2) = \frac{D_-}{x_1 - x_2} = \frac{\pi_{x_1 x_1} - \pi_{x_2 x_2}}{x_1 - x_2} \theta_1 \theta_2\,.
\end{equation}
Thus, we have the descended complex:
\begin{equation}
	\bigcirc \xrightarrow{0} \bigcirc \xrightarrow{\frac{\pi_{x_1 x_1} - \pi_{x_2 x_2}}{x_1 - x_2}} \bigcirc \bigcirc\,.
\end{equation}
Note that the first differential is zero because it acts between cohomologies with the identified $x_1 = x_2$. The action of $\frac{\pi_{x_1 x_1} - \pi_{x_2 x_2}}{x_1 - x_2}$ exactly coincides with the action of the co-product operator from~\cite{2506.08721}: 
\begin{equation}
	\Delta = \sum_{i=1}^N\left(\sum_{j=0}^{N-i}\vartheta^a_{N-j}\vartheta^b_{i+j}\right)\frac{\partial}{\partial \vartheta^c_i}\,:\quad V_c \; \mapsto \; V_a \otimes V_b\,.
\end{equation}
Let us explicitly consider the simplest example of $N=2$. We have the following mapping of the cohomologies:
\begin{equation}
	\langle 1, \, x_1 = x_2 \rangle \xrightarrow{x_1 + x_2} \langle 1,\, x_1 \rangle \otimes \langle 1,\, x_2 \rangle\,.
\end{equation}
The explicit action is:
\begin{equation}
	x_1 + x_2: \quad 1 \; \mapsto \; x_1 + x_2\,, \quad x_1 \; \mapsto \; x_1 x_2\,.
\end{equation}
After the grading shift, it exactly coincides with the action of the $\Delta$ operator:
\begin{equation}
	\Delta = (\vartheta_2^{(1)} \vartheta_1^{(2)} + \vartheta_1^{(1)} \vartheta_2^{(2)})\frac{\partial}{\partial\vartheta_1} + \vartheta_2^{(1)} \vartheta_2^{(2)} \frac{\partial}{\partial\vartheta_2}\,: \quad \vartheta_1 \; \mapsto \; \vartheta_2^{(1)} \vartheta_1^{(2)} + \vartheta_1^{(1)} \vartheta_2^{(2)}\,, \quad \vartheta_2 \; \mapsto \;  \vartheta_2^{(1)} \vartheta_2^{(2)}\,.
\end{equation}
Now, consider gluing of right and left ends in~\eqref{bip-KhR-complex}. There are the following resolutions and corresponding vertical differentials:
\begin{equation}
\begin{aligned}
	\bigcirc \bigcirc \quad D_+ &= \pi_{x_1 x_1} \theta_1 + \pi_{x_2 x_2} \theta_2\,, \\
	\bigcirc \quad D_- &= \pi_{x_1 x_2}(\theta_1 + \theta_2) + (x_2 - x_1)(\theta_1^\dag - \theta_2^\dag)\,.
\end{aligned}
\end{equation}
The cohomology of $D_+$ is spanned by $\theta_1 \theta_2$, and the cohomology of $D_-$ is spanned by $\theta_1 + \theta_2$. We again induce the action of the horizontal differential $S$ on the cohomologies:
\begin{equation}
	S (\theta_1 \theta_2) = \frac{D_-}{x_1 - x_2} \theta_1 \theta_2 = -(\theta_1 + \theta_2)\,.
\end{equation}
Thus, the descended complex is
\begin{equation}
	\bigcirc \bigcirc \xrightarrow{x_1 - x_2} \bigcirc \bigcirc \xrightarrow{1} \bigcirc\,.
\end{equation}
The action of $x_1 - x_2$ exactly coincides with the action of the Sh operator from~\cite{2506.08721}: 
\begin{equation}
	{\rm Sh} = \sum_{i,j = 1}^{N-1}\left(\vartheta^{{\color{blue} a}}_{i+1}\vartheta^{{\color{blue} b}}_j-\vartheta^{{\color{blue} a}}_i\vartheta^{{\color{blue} b}}_{j+1}\right)\frac{\partial^2}{\partial\vartheta_i^a\partial\vartheta_j^b}+\sum_{i=1}^{N-1}\vartheta^{{\color{blue} a}}_{i+1}\vartheta^{{\color{blue} b}}_N\frac{\partial^2}{\partial\vartheta_i^a\partial\vartheta_N^b}-\sum_{j=1}^{N-1}\vartheta^{{\color{blue} a}}_N\vartheta^{{\color{blue} b}}_{j+1}\frac{\partial^2}{\partial\vartheta_N^a\partial\vartheta_j^b}: \; V_a \otimes V_b \; \mapsto \; V_{{\color{blue} a}} \otimes V_{{\color{blue} b}}\,,
\end{equation} 
while the action of the unity is equivalent to the action of the multiplication operator from~\cite{2506.08721}:
\begin{equation}
	m = 
	\sum_{\substack{1\le i,j\le N\\i+j \le N+1}}\vartheta_{i+j-1}^a\frac{\partial^2}{\partial\vartheta_i^b\partial\vartheta_j^c}\,:\quad V_b \otimes V_c \; \mapsto \; V_a\,.
\end{equation} 
Again, consider the example of $N=2$. The first mapping is:
\begin{equation}
	\langle 1,\, x_1 \rangle \otimes \langle 1,\, x_2 \rangle \xrightarrow{x_1 - x_2} \langle 1,\, x_1 \rangle \otimes \langle 1,\, x_2 \rangle\,.
\end{equation}
The explicit action is:
\begin{equation}
	x_1 - x_2:\quad 1 \; \mapsto \; x_1 - x_2\,,\quad x_1 \; \mapsto \; - x_1 x_2\,,\quad x_2 \; \mapsto \; x_1 x_2\,,\quad x_1 x_2 \; \mapsto \; 0\,.
\end{equation}
It coincides with the action of the Sh morphism:
\begin{equation}
	{\rm Sh} = (\vth_2^{{\color{blue} (1)}}\vth_1^{{\color{blue} (2)}} - \vth_1^{{\color{blue} (1)}}\vth_2^{{\color{blue} (2)}})\frac{\p^2}{\p \vth_1^{(1)} \p \vth_1^{(2)}} + \vth_2^{{\color{blue} (1)}}\vth_2^{{\color{blue} (2)}}\frac{\p^2}{\p \vth_1^{(1)} \p \vth_2^{(2)}} - \vth_2^{{\color{blue} (1)}}\vth_2^{{\color{blue} (2)}}\frac{\p^2}{\p \vth_2^{(1)} \p \vth_1^{(2)}}\,.
\end{equation}
The second mapping is:
\begin{equation}
	\langle 1,\, x_1 \rangle \otimes \langle 1,\, x_2 \rangle \xrightarrow{1} \langle 1,\, x_1=x_2 \rangle\,.
\end{equation}
The explicit action is:
\begin{equation}
	1: \quad 1 \; \mapsto \; 1\,, \quad x_1 \; \mapsto \; x_1\,, \quad x_2 \; \mapsto \; x_1\,, \quad x_1 x_2 \; \mapsto \; 0\,,
\end{equation}
and it coincides with the action of the $m$ operator:
\begin{equation}
	m = \vth_1 \frac{\p^2}{\p \vth_1^{(1)} \p \vth_1^{(2)}} + \vth_2 \left( \frac{\p^2}{\p \vth_1^{(1)} \p \vth_2^{(2)}}+\frac{\p^2}{\p \vth_2^{(1)} \p \vth_1^{(2)}} \right): \; \vth_1^{(1)}\vth_1^{(2)} \, \mapsto \, \vth_1\,, \; \vth_1^{(2)}\vth_2^{(1)} \, \mapsto \, \vth_2\,, \; \vth_2^{(2)}\vth_1^{(1)} \, \mapsto \, \vth_2\,, \; \vth_2^{(1)}\vth_2^{(2)} \, \mapsto \, 0\,.
\end{equation}
In general, cycles are glued from several elementary vertical and horizontal resolutions, and there appear more variables on edges. As it has been discussed at the end of Section~\ref{sec:N=2}, due to the locality of horizontal morphisms, one can reduce the number of variables and arrive to the already analyzed case. 

Thus, we have proved the equivalence of the Khovanov--Rozansky calculus and cycle calculus for bipartite links~\cite{2506.08721}.

\section{Conclusion}

In this paper, we have studied different reductions in Khovanov--Rozansky operator formalism~\cite{2508.05191}. In this formalism, we deal with a bicomplex with vertical and horizontal morphisms, see Section~\ref{sec:categorification}. First, vertical reductions (Section~\ref{sec:vert-red}) allow to simplify vertical cohomologies. Second, horizontal reductions (Section~\ref{sec:IR}) allow to simplify horizontal cohomologies and differentials induced to vertical cohomologies. Third, in some cases ($N=2$ and arbitrary $N$ bipartite case, Section~\ref{sec:Kh-like}), one can transfer from bicomplex to monocomlex constructed from universal vector spaces (living on cycles) and universal morphisms.

In this text, we have introduced the constructions of the HOMFLY and Khovanov--Rozansky polynomial (in the differential operators formalism) in Section~\ref{sec:prerequisites}. In Sections~\ref{sec:elem_red} and~\ref{sec:cat-MOY}, we have listed types of vertical reductions. Section~\ref{sec:examples} is devoted to examples of vertical reductions. We then prove elementary linear and quadratic reductions in Sections~\ref{sec:lin_red} and~\ref{sec:quad_red},~\ref{sec:quad-red-2} correspondingly. In Sections~\ref{sec:MOY-I},~\ref{sec:MOY-II}--\ref{sec:box-2}, we provide proofs of the categorified MOY relations~\eqref{MOYcat-loc_II}--\eqref{MOYcat-loc_V}. 

The resulting horizontal complex of vertical cohomologies can be split into subcomplexes. Some of these subcomplexes are exact and fo not contribute to the horizontal cohomologies. This procedure of such an extraction of exact subcomplexes is called horizontal reduction. It is demonstrated in Section~\ref{sec:IR}. 

In Section~\ref{sec:ex}, we provide examples of calculation of the Khovanov--Rozansky polynomials for simple links. The examples of diagrams of Hopf links illustrate the application of the linear and quadratic reductions.

We show in Section~\ref{sec:N=2} that the sophisticated Khovanov--Rozansky operator formalism in the $N=2$ case can be reduced to the simple Khovanov cycle calculus. Moreover, for an arbitrary $N$ but in the case of specific bipartite links, the operator formalism can be transformed to similar (Khovanov-like) technique based on just a monocomplex of $N$-dimensional $q$-graded vector spaces with 3 universal morphisms.

The problems expected to be solved in the future are listed below.
\begin{enumerate}
	\item It is tempting to provide the KhR differential operator construction for higher representations --- \\ (anti)symmetric, rectangular, generic. In particular, in the case of an arbitrary representation, it is interesting to find out if there are some reductions of the operator formalism for bipartite links. We hope there are because in~\cite{ALM2}, we have shown that the planar bipartite calculus still exist for the HOMFLY polynomials in (anti)symmetric representations.
	
	\item A detailed derivation of~\eqref{bip-KhR-complex} should be performed.
	
	\item When dealing with bipartite reduction, there is only one type of end-points of these reduction --- these are cycles and their tensor products. Given generic links, the question is whether there is a finite number of types of reductions end-points. Now, it seems that the infinite number of types of end-points arises. This is due to the complexity of quadratic reduction that leaves resolutions to be tied, so that all links give rise to infinite number of diagrams connected by different ways.
	 
	\item Another difficulty of the quadratic reduction is that it does not seem to split a cohomology into a direct sum of cohomologies as two constituents are restricted by the relation. A solution of this relation is not unique. Still, in all our examples it allowed for such a splitting into a direct sum. The question is if it is always the case, or there exist an example when these two cohomologies are indeed connected by a relation. The related question is why the simple quadratic reduction restriction at the level of the HOMFLY polynomial: $\CM^{ii}_{ii}\equiv 0$, is categorified in such a mysterious way.
	
	\item A very interesting question is how to make an explicit downgrade to the RT approach. In particular, how to extract $\cal R$-matrix from the KhR operator formalism, or at least, from the Khovanov cohomologies.
	
	\item We have proved explicitly only three MOY relations. So, the remaining task is to demonstrate the validity of the forth MOY relation~\eqref{MOYcat-loc_VI}.
	
	\item We have demonstrated the invariance of our KhR operator formalism under the first Reidemeister move. The problem is to show explicitly the invariance under the second and third Reidemeister moves.
	
	\item It would be interesting to construct the operator formalism for categorifications of Wilson loops in the Chern--Simons theory for other simple Lie algebras.
\end{enumerate}

\section*{Acknowledgments}

We would like to thank Radomir Stepanov for illuminating discussions. 

Funding for this publication was generously provided by the Priority 2030 Academic Leadership Initiative, contributing
to the educational work of "Universities for a New Generation of Leaders", a project within the framework of the federal Youth and Children program.

\appendix

%
%
%

\section{Functions decomposition}

In this Section, we introduce a function decomposition that allows to separate functions with equated variables. In particular, this allows to provide projection on quotient rings in quadratic reduction, see Sections~\ref{sec:quad_red},~\ref{sec:quad-red-2}.

\subsection{2-variables function decomposition}\label{sec:2f-deco}

First, we derive the arbitrary 2-variables function decomposition. Any polynomial in 2 variables can be decomposed as follows:
\begin{equation}
	f(x_1,x_2) = a(x_1) + (x_1 - x_2) b(x_1,x_2)
\end{equation}
with polynomial $a(x_1)$ and $b(x_1,x_2)$. At the point $x_1 = x_2$, we get:
\begin{equation}
	f(x_1,x_1) = a(x_1)\,,
\end{equation}
and the final decomposition is
\begin{equation}
	f(x_1,x_2) = f(x_1,x_1) + (x_1 - x_2)\, \frac{f(x_1,x_2) - f(x_1,x_1)}{x_1 - x_2}\,.
\end{equation}

\subsection{3-variables function decomposition}\label{sec:3f-deco}

Any polynomial of 3 variables $f(x_1,x_2,x_3)$ can be decomposed as follows:
\begin{equation}
	f(x_1,x_2,x_3) = a(x_1,x_2) + b(x_1,x_2)x_3 + (x_1 - x_3)(x_2 - x_3)\tilde v_2 (f)\,.
\end{equation}
At the points $x_3 = x_1$ and $x_3 = x_2$, we get
\begin{equation}
	f(x_1,x_2,x_1) = a + b x_1\,,\quad f(x_1,x_2,x_2) = a + b x_2
\end{equation}
respectively. Thus, we get the following decomposition:
\begin{equation}
	\boxed{f(x_1,x_2,x_3) = P_- f_-(x_1,x_2) + P_+ f_+(x_1,x_2) + (x_1 - x_3)(x_2 - x_3)\tilde v_2 (f)}
\end{equation}
with the functions 
\begin{equation}
	f_+(x_1,x_2) = f(x_1,x_2,x_2)\,, \quad f_-(x_1,x_2) = f(x_1,x_2,x_1)\,,
\end{equation}
and the projectors
\begin{equation}
	P_- = \frac{x_3 - x_2}{x_1 - x_2}\,, \quad P_+ = \frac{x_1 - x_3}{x_1 - x_2}\,.
\end{equation}
These are the projectors in the ring of remainders:
\begin{equation}
\begin{aligned}
	P_-^2 &= P_- + (x_1 - x_3)(x_2 - x_3)\cdot \frac{1}{(x_1 - x_2)^2} = P_- \;{\rm mod}\;\langle (x_1 - x_3)(x_2 - x_3)\rangle\,, \\
	P_+^2 &= P_+ + (x_1 - x_3)(x_2 - x_3)\cdot \frac{1}{(x_1 - x_2)^2} = P_+ \;{\rm mod}\;\langle (x_1 - x_3)(x_2 - x_3)\rangle\,, \\
	P_+P_-&=P_-P_+=0\;{\rm mod}\;\langle (x_1 - x_3)(x_2 - x_3) \rangle\,.
\end{aligned}
\end{equation}

\subsection{4-variables function decomposition}\label{sec:4f-deco} 

An arbitrary 4-variables function can be represented as follows:
\begin{equation}
	f(x_1,x_2,x_3,x_4) = a(x_1,x_2) + b(x_1,x_2)(x_3 - x_4) + s_1 v_1 (f) + s_2 v_2(f)\,,
\end{equation}
where
\begin{equation}
	s_1 = x_1 + x_2 - x_3 - x_4\,,\quad s_2 = x_1 x_2 - x_3 x_4\,.
\end{equation}
At the points $x_1 = x_3$, $x_2 = x_4$ and $x_2 = x_3$, $x_1 = x_4$ we get
\begin{equation}
\begin{aligned}
	f\big|_{\substack{x_3=x_1\\ x_4=x_2}} &= f_- = a(x_1,x_2) + b(x_1,x_2)(x_1 - x_2)\,, \\
	f\big|_{\substack{x_3=x_2\\ x_4=x_1}} &= f_+ = a(x_1,x_2) + b(x_1,x_2)(x_2 - x_1)
\end{aligned}
\end{equation}
respectively. Thus, we can express $a$ and $b$ polynomials:
\begin{equation}
	a = \frac{1}{2}(f_+ f_-)\,, \quad b = \frac{1}{2} \frac{f_+ - f_-}{x_2 - x_1}
\end{equation}
so that 
\begin{equation}
	\boxed{f(x_1,x_2,x_3,x_4) = P_+ f_+ + P_- f_- + s_1 v_1(f) + s_2 v_2(f)}
\end{equation}
with $P_{\pm}=\frac{1}{2}(1\pm h)$ and $h=\cfrac{x_3-x_4}{x_1-x_2}\,$. 

Let us note that:
\begin{equation}
	h^2=\left(\frac{x_3-x_4}{x_1-x_2}\right)^2=1-s_1\frac{x_1+x_2+x_3+x_4}{(x_1-x_2)^2}+s_2\frac{4}{(x_1-x_2)^2}=1\;{\rm mod}\;\langle s_1,s_2\rangle\,.
\end{equation}
Thus, we see that $P_{\pm}=\frac{1}{2}(1\pm h)$ are projectors to disjoint components:
\begin{equation}
	P_{\pm}^2=P_{\pm}\;{\rm mod}\;\langle s_1,s_2\rangle,\quad P_+P_-=P_-P_+=0\;{\rm mod}\;\langle s_1,s_2\rangle\,.
\end{equation}

\paragraph{Another way.} We derive the arbitrary function decomposition:
\begin{equation}\label{f-g}
	f(x_1,x_2,x_3,x_4)=\frac{1}{2}\underbrace{\left(f(x_1,x_2,x_3,x_4)+f(x_1,x_2,x_4,x_3)\right)}_{g_1(x_1,x_2,x_3+x_4,x_3x_4)}+\frac{1}{2}(x_3-x_4)\underbrace{\frac{f(x_1,x_2,x_3,x_4)-f(x_1,x_2,x_4,x_3)}{x_3-x_4}}_{g_2(x_1,x_2,x_3+x_4,x_3x_4)}\,.
\end{equation}
Then, we further decompose\footnote{Here we use that for an arbitrary function $g(x,y)$ we can decompose
	\begin{equation}
		g(x,y)-g(x',y')=g(x,y)-g(x',y)+g(x',y)-g(x',y')=(x-x')\frac{g(x,y)-g(x',y)}{x-x'}+(y-y')\frac{g(x',y)-g(x',y')}{y-y'}\,.
	\end{equation}
}
{\small \begin{equation}\label{g-deco}
		\begin{aligned}
			g_i(x_1,x_2,x_3+x_4,x_3 x_4) &= g_i(x_1, x_2, x_1+x_2, x_1 x_2) + \underbrace{(x_1 + x_2 - x_3 - x_4)}_{s_1} \,\underbrace{\frac{g_i(x_1,x_2,x_3+x_4,x_3 x_4) - g_i(x_1,x_2,x_1+x_2,x_3 x_4)}{x_1 + x_2 - x_3 - x_4}}_{G_1^{(i)}(x_1,x_2,x_3+x_4,x_1+x_2,x_3 x_4)} \\
			&+ \underbrace{(x_1 x_2 - x_3 x_4)}_{s_2} \,\underbrace{\frac{g_i(x_1,x_2,x_1+x_2,x_3 x_4) - g_i(x_1,x_2,x_1+x_2,x_1 x_2)}{x_1 x_2 - x_3 x_4}}_{G_2^{(i)}(x_1,x_2,x_1+x_2,x_3 x_4,x_1 x_2)}
		\end{aligned}
\end{equation}}
where
\begin{equation}\label{g-coinc}
	\begin{aligned}
		g_1(x_1,x_2,x_1 + x_2, x_1 x_2) &= f(x_1,x_2,x_1,x_2) + f(x_1,x_2,x_2,x_1)\,, \\
		g_2(x_1,x_2,x_1 + x_2, x_1 x_2) &= \frac{f(x_1,x_2,x_1,x_2) - f(x_1,x_2,x_2,x_1)}{x_1 - x_2}\,.
	\end{aligned}
\end{equation}
We further plug~\eqref{g-deco} and~\eqref{g-coinc} into~\eqref{f-g}:
\begin{equation}
	\begin{aligned}
		f(x_1,x_2,x_3,x_4) &= \frac{1}{2}\left(f(x_1,x_2,x_1,x_2) + f(x_1,x_2,x_2,x_1) + s_1 G_1^{(1)} + s_2 G_2^{(1)}\right) + \\
		&+ \frac{1}{2}(x_3-x_4)\left( \frac{f(x_1,x_2,x_1,x_2) - f(x_1,x_2,x_2,x_1)}{x_1-x_2} + s_1 G_1^{(2)} + s_2 G_2^{(2)} \right)\,.
	\end{aligned}
\end{equation}
So that we get the final decomposition:
\begin{equation}
	\boxed{f(x_1,x_2,x_3,x_4)=\frac{1}{2}(1+h)f\big|_{\substack{x_3=x_1\\ x_4=x_2}}+\frac{1}{2}(1-h)f\big|_{\substack{x_3=x_2\\ x_4=x_1}}+s_1v_1(f)+s_2v_2(f)\,}
\end{equation}
with 
\begin{equation}
	\begin{aligned}
		v_1(f) &= \frac{1}{2}G_1^{(1)}+\frac{1}{2}(x_3-x_4)G_1^{(2)}\,, 
		\\
		v_2(f) &= \frac{1}{2}G_2^{(1)}+\frac{1}{2}(x_3-x_4)G_2^{(2)}\,. 
	\end{aligned}
\end{equation}

\printbibliography

@article{MOY,
	title="{HOMFLY polynomial via an invariant of colored plane graphs}",
	author={Murakami, H. and Ohtsuki, T. and Yamada, S.},
	journal={Enseignement Math{\'e}matique},
	volume={44},
	pages={325--360},
	year={1998},
	publisher={SWETS \& ZEITLINGER}
}

@article{KRI,
	title={Matrix factorizations and link homology},
	author={Khovanov, M. and Rozansky, L.},
	journal={Fundamenta mathematicae},
	volume={199},
	number={1},
	pages={1--91},
	year={2008},
	publisher={Institute of Mathematics},
	archivePrefix = {arXiv},
	eprint = {math/0401268},
	primaryClass = {math.QA}
}

@article{Carqueville:2011zea,
	author = "Carqueville, N. and Murfet, D.",
	title = "{Computing Khovanov\textendash{}Rozansky homology and defect fusion}",
	eprint = "1108.1081",
	archivePrefix = "arXiv",
	primaryClass = "math.QA",
	journal = "Algebr. Geom. Topol.",
	volume = "14",
	number = "1",
	pages = "489--537",
	year = "2014"
}

@article{2506.08721,
	title={Khovanov--Rozansky cycle calculus for bipartite links},
	author={Anokhina, A. and Lanina, E. and Morozov, A.},
	journal={The European Physical Journal C},
	volume={85},
	number={10},
	pages={1--35},
	year={2025},
	publisher={Springer},
	eprint = "2506.08721",
	archivePrefix = "arXiv",
	primaryClass = "hep-th",
}

@article{2508.05191,
	title={Operator lift of the Reshetikhin-Turaev formalism to Khovanov-Rozansky topological quantum field theory},
	author={Galakhov, D. and Lanina, E. and Morozov, A.},
	journal={Physical Review D},
	volume={113},
	number={2},
	pages={026013},
	year={2026},
	publisher={APS},
	eprint = "2508.05191",
	archivePrefix = "arXiv",
	primaryClass = "hep-th"
}

@article{aganagic2015knot,
	title={Knot homology and refined Chern--Simons index},
	author={Aganagic, M. and Shakirov, S.},
	journal={Communications in Mathematical Physics},
	volume={333},
	number={1},
	pages={187--228},
	year={2015},
	publisher={Springer},
	eprint = "1105.5117",
	archivePrefix = "arXiv",
	primaryClass = "hep-th"
}

@article{gukov2005khovanov,
	title={Khovanov-Rozansky homology and topological strings},
	author={Gukov, S. and Schwarz, A. and Vafa, C.},
	journal={Letters in Mathematical Physics},
	volume={74},
	number={1},
	pages={53--74},
	year={2005},
	publisher={Springer},
	eprint = "hep-th/0412243",
	archivePrefix = "arXiv",
	primaryClass = "hep-th"
}

@article{witten2011fivebranes,
	title={Fivebranes and knots},
	author={Witten, E.},
	journal={Quantum Topology},
	volume={3},
	number={1},
	pages={1--137},
	year={2011},
	eprint = "hep-th/1101.3216",
	archivePrefix = "arXiv",
	primaryClass = "hep-th"
}

@article{witten2012khovanov,
	title={Khovanov homology and gauge theory},
	author={Witten, E.},
	journal={Proceedings of the Freedman Fest},
	volume={18},
	pages={291--308},
	year={2012},
	publisher={Geometry \& Topology Publications Coventry},
	eprint = "1108.3103",
	archivePrefix = "arXiv",
	primaryClass = "hep-th"
}

@article{gaiotto2012knot,
	title={Knot invariants from four-dimensional gauge theory},
	author={Gaiotto, D. and Witten, E.},
	journal={Advances in Theoretical and Mathematical Physics},
	volume={16},
	number={3},
	pages={935--1086},
	year={2012},
	publisher={International Press, Inc.},
	eprint = "hep-th/1106.4789",
	archivePrefix = "arXiv",
	primaryClass = "hep-th"
}

@article{Witten,
	title={Quantum field theory and the Jones polynomial},
	author={Witten, E.},
	journal={Communications in mathematical physics},
	volume={121},
	number={3},
	pages={351--399},
	year={1989},
	publisher={Springer}
}

@article{Reshetikhin,
	title={Invariants of tangles 1},
	author={Reshetikhin, N.},
	journal={unpublished preprint},
	volume={},
	pages={},
	year={1987},
	publisher={}
}

@article{guadagnini1990chern2,
	title={Chern--Simons holonomies and the appearance of quantum groups},
	author={Guadagnini, E. and Martellini, M. and Mintchev, M.},
	journal={Physics Letters B},
	volume={235},
	number={3-4},
	pages={275--281},
	year={1990},
	publisher={Elsevier}
}

@article{reshetikhin1990ribbon,
	title={Ribbon graphs and their invaraints derived from quantum groups},
	author={Reshetikhin, N. and Turaev, V.},
	journal={Communications in Mathematical Physics},
	volume={127},
	number={1},
	pages={1--26},
	year={1990},
	publisher={Springer}
}

@article{turaev1990yang,
	title={The Yang--Baxter equation and invariants of links},
	author={Turaev, V.},
	journal={New Developments in the Theory of Knots},
	volume={11},
	pages={175},
	year={1990},
	publisher={World Scientific}
}

@article{reshetikhin1991invariants,
	title={Invariants of 3-manifolds via link polynomials and quantum groups},
	author={Reshetikhin, N. and Turaev, V.},
	journal={Inventiones mathematicae},
	volume={103},
	number={1},
	pages={547--597},
	year={1991}
}

@article{Kauff,
	title={State models and the Jones polynomial},
	author={Kauffman, L.H.},
	journal={Topology},
	volume={26},
	number={3},
	pages={395--407},
	year={1987},
	publisher={Elsevier}
}

@article{Chern-Simons,
	author = "Chern, S.-S. and Simons, J.",
	title = "{Characteristic forms and geometric invariants}",
	doi = "",
	journal = "Annals Math.",
	volume = "99",
	pages = "48--69",
	year = "1974"
}

@article{dolotin2013introduction,
	title={Introduction to Khovanov homologies I. Unreduced Jones superpolynomial},
	author={Dolotin, V. and Morozov, A.},
	journal={Journal of High Energy Physics},
	volume={2013},
	number={1},
	pages={1--48},
	year={2013},
	publisher={Springer},
	eprint = "1208.4994",
	archivePrefix = "arXiv",
	primaryClass = "hep-th"
}

@article{freyd1985new,
	title={A new polynomial invariant of knots and links},
	author={Freyd, P. and Yetter, D. and Hoste, J. and Lickorish, W.B.R. and Millett, K. and Ocneanu, A.},
	journal={Bulletin (new series) of the American mathematical society},
	volume={12},
	number={2},
	pages={239--246},
	year={1985},
	publisher={American Mathematical Society}
}

@article{przytycki1988invariants,
	title={Invariants of links of Conway type},
	author={Przytycki, J.H. and Traczyk, K.P.},
	journal={Kobe Journal of Mathematics},
	volume={4},
	number={2},
	pages={115--139},
	year={1988},
	publisher={理学研究科数学専攻},
	eprint = "1610.06679",
	archivePrefix = "arXiv",
	primaryClass = "math.GT"
}

@article{ALM,
	title={Planar decomposition of the HOMFLY polynomial for bipartite knots and links},
	author={Anokhina, A. and Lanina, E. and Morozov, A.},
	journal={The European Physical Journal C},
	volume={84},
	number={9},
	pages={990},
	year={2024},
	publisher={Springer},
	eprint = {2407.08724},
	archivePrefix = {arXiv},
	primaryClass = {hep-th}
}

@article{jones2005jones,
	title={The Jones polynomial},
	author={Jones, V.F.R.},
	journal={Discrete Math},
	volume={294},
	pages={275--277},
	year={2005}
}

@incollection{jones1987hecke,
	title={Hecke algebra representations of braid groups and link polynomials},
	author={Jones, V.F.R.},
	booktitle={New developments in the theory of knots},
	pages={20--73},
	year={1987},
	publisher={World Scientific}
}

@article{jones1985polynomial,
	title={A polynomial invariant for knots via von Neumann algebras},
	author={Jones, V.F.R.},
	journal={Bulletin of the American Mathematical Society},
	volume={12},
	number={1},
	pages={103--111},
	year={1985}
}

@article{khovanov2000categorification,
	title={A categorification of the Jones polynomial},
	author={Khovanov, M.},
	journal={Duke Mathematical Journal},
	volume={101},
	number={3},
	pages={359--426},
	year={2000},
	publisher={Duke University Press},
	eprint = {math/9908171},
	archivePrefix = {arXiv},
	primaryClass = {math.QA}
}

@article{dolotin2014introduction,
	title={Introduction to Khovanov homologies. III. A new and simple tensor-algebra construction of Khovanov--Rozansky invariants},
	author={Dolotin, V. and Morozov, A.},
	journal={Nuclear Physics B},
	volume={878},
	pages={12--81},
	year={2014},
	publisher={Elsevier},
	eprint = {1308.5759},
	archivePrefix = {arXiv},
	primaryClass = {hep-th}
}

@article{ALM2,
	title={Planar decomposition of bipartite HOMFLY polynomials in symmetric representations},
	author={Anokhina, A. and Lanina, E. and Morozov, A.},
	journal={Physical Review D},
	volume={111},
	number={4},
	pages={046018},
	year={2025},
	publisher={APS},
	eprint = {2410.18525},
	archivePrefix = {arXiv},
	primaryClass = {hep-th}
}

@article{ALM3,
	title={Bipartite expansion beyond biparticity},
	volume={1014},
	journal={Nuclear Physics B},
	publisher={Elsevier BV},
	author={Anokhina, A. and Lanina, E. and Morozov, A.},
	year={2025},
	pages={116881},
	eprint = {2501.15467},
	archivePrefix = {arXiv},
	primaryClass = {hep-th}
}

@article{MMM2,
	title={Character expansion for HOMFLY polynomials. II. Fundamental representation. Up to five strands in braid},
	author={Mironov, A. and Morozov, A. and Morozov, And.},
	journal={Journal of High Energy Physics},
	volume={2012},
	number={3},
	pages={1--34},
	year={2012},
	publisher={Springer},
	eprint = {1112.2654},
	archivePrefix = {arXiv},
	primaryClass = {math.QA}
}

@article{AnoM,
	title={Towards R-matrix construction of Khovanov-Rozansky polynomials I. Primary T-deformation of HOMFLY},
	author={Anokhina, A. and Morozov, A.},
	journal={Journal of High Energy Physics},
	volume={2014},
	number={7},
	pages={1--183},
	year={2014},
	publisher={Springer},
	eprint = {1403.8087},
	archivePrefix = {arXiv},
	primaryClass = {hep-th}
}

@incollection{MMM1,
	title={Character expansion for HOMFLY polynomials I: Integrability and difference equations},
	author={Mironov, A. and Morozov, A. and Morozov, And.},
	booktitle={Strings, Gauge Fields, and the Geometry Behind: The Legacy of Maximilian Kreuzer},
	pages={101--118},
	year={2013},
	publisher={World Scientific},
	eprint = {1112.5754},
	archivePrefix = {arXiv},
	primaryClass = {hep-th}
}

\end{document}